\pdfoutput=1
\RequirePackage{ifpdf}

\documentclass{JINST}
\usepackage{graphicx}
\usepackage{amsmath}
\usepackage[font=it]{caption}
\usepackage{graphicx}
\usepackage{subfig}
\usepackage{amsmath}
\usepackage{url}
\usepackage{booktabs}
\usepackage{multirow} 
\usepackage{float}
\usepackage{heppennames2}
\usepackage{placeins}

\makeatletter
\renewcommand\@makefntext[1]{\leftskip=0em\hskip-0em\@makefnmark#1}
\makeatother

\title{Shower development of particles with momenta from 15\,GeV to 150\,GeV in the CALICE scintillator-tungsten hadronic calorimeter}

\author{\centering
\includegraphics[width=0.2\textwidth]{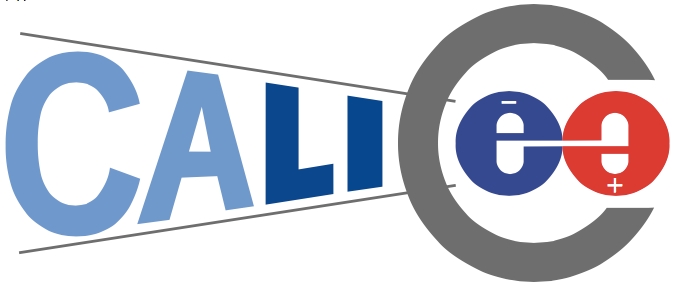}\\
\LARGE The CALICE collaboration 
}

\author{\centering
M.\,Chefdeville, 
Y.\,Karyotakis
\\ \it
Laboratoire d'Annecy-le-Vieux de Physique des Particules, Universit\'{e} de Savoie,
CNRS/IN2P3,
9 Chemin de Bellevue BP110, F-74941 Annecy-le-Vieux CEDEX, France
}

\author{\centering
J.\,Repond, 
J.\,Schlereth, 
L.\,Xia
\\ \it
Argonne National Laboratory,
9700 S.\ Cass Avenue,
Argonne, IL 60439-4815,
USA}

\author{\centering
G.\,Eigen 
\\ \it
University of Bergen, Inst.\, of Physics, Allegaten 55, N-5007 Bergen, Norway
}

\author{\centering 
J.\,S.\,Marshall, 
M.\,A.\,Thomson, 
D.\,R.\,Ward
\\ \it
University of Cambridge, Cavendish Laboratory, J J Thomson Avenue, CB3 0HE, UK
}

\author{\centering 
N.\,Alipour Tehrani,
J.\,Apostolakis, 
D.\,Dannheim, 
K.\,Elsener,
G.\,Folger, 
C.\,Grefe$^a$, 
V.\,Ivantchenko, 
M.\,Killenberg$^b$, 
W.\,Klempt, 
E.\,van der Kraaij, 
L.\,Linssen,
A.\,-I.\,Lucaci-Timoce, 
A.\,M\"unnich$^b$, 
S.\,Poss,
A.\,Ribon,
P.\,Roloff,
A.\,Sailer, 
D.\,Schlatter, 
E.\,Sicking$^\spadesuit$,
J.\,Strube$^c$, 
V.\,Uzhinskiy
 \\ \it 
CERN, 1211 Gen\`{e}ve 23, Switzerland
}

\author{\centering 
S.\,Chang, 
A.\,Khan, 
D.\,H.\,Kim,
D.\,J.\,Kong, 
Y.\,D.\,Oh
\\ \it
Department of Physics, Kyungpook National University, Daegu, 702-701,
Republic of Korea
}

\author{\centering
G.\,C.\,Blazey, 
A.\,Dyshkant, 
K.\,Francis, 
V.\,Zutshi
\\ \it
NICADD, Northern  Illinois University,
Department of Physics,
DeKalb, IL 60115,
USA
}

\author{\centering 
J.\,Giraud, 
D.\,Grondin, 
J.\,-Y.\,Hostachy
\\ \it
Laboratoire de Physique Subatomique et de Cosmologie - Universit\'{e}
Grenoble-Alpes, CNRS/IN2P3, Grenoble, France
}

\author{\centering 
E.\,Brianne,
U.\,Cornett, 
D.\,David, 
G.\,Falley, 
K.\,Gadow, 
P.\,G\"{o}ttlicher, 
C.\,G\"{u}nter, 
O.\,Hartbrich,
B.\,Hermberg, 
A.\,Irles,
S.\,Karstensen, 
F.\,Krivan, 
K.\,Kr\"{u}ger, 
J.\,Kvasnicka$^d$, 
S.\,Lu, 
B.\,Lutz, 
S.\,Morozov, 
V.\,Morgunov$^e$, 
C.\,Neub\"user, 
A.\,Provenza,
M.\,Reinecke,  
F.\,Sefkow, 
P.\,Smirnov, 
M.\,Terwort, 
H.L.\,Tran,
A.\,Vargas-Trevino
\\ \it
DESY, Notkestrasse 85,
D-22603 Hamburg, Germany
}

\author{\centering   
E.\,Garutti, 
S.\,Laurien,
M.\,Matysek
M.\,Ramilli, 
S.\,Schr\"oder
\\ \it
Univ. Hamburg,
Physics Department,
Institut f\"ur Experimentalphysik,
Luruper Chaussee 149,
22761 Hamburg, Germany
}

\author{\centering 
K.\,Briggl,
P.\,Eckert, 
T.\,Harion, 
Y.\,Munwes,
H.\,-Ch.\,Schultz-Coulon, 
W.\,Shen, 
R.\,Stamen
\\ \it
University of Heidelberg, Fakult\"at f\"ur Physik und Astronomie,
Albert Uberle Str. 3-5 2.OG Ost,
D-69120 Heidelberg, Germany
}

\author{\centering 
B.\,Bilki,
Y.\,Onel
\\ \it
University of Iowa, Dept. of Physics and Astronomy,
203 Van Allen Hall, Iowa City, IA 52242-1479, USA
}

\author{\centering 
K.\,Kawagoe,
H.\,Hirai, 
Y.\,Sudo, 
T.\,Suehara, 
H.\,Sumida, 
S.\,Takada,
T.\,Tomita,
T.\,Yoshioka
\\ \it
Department of Physics, Kyushu University, Fukuoka 812-8581, Japan
}

\author{\centering 
M.\,Wing
\\ \it
Department of Physics and Astronomy, University College London,
Gower Street,
London WC1E 6BT, UK
}

\author{\centering 
E.\,Calvo~Alamillo, 
M.-C.\,Fouz, 
J.\,Marin,
J.\,Puerta-Pelayo, 
A.\,Verdugo
\\ \it
CIEMAT, Centro de Investigaciones Energeticas, Medioambientales y Tecnologicas, Madrid, Spain 
}

\author{\centering 
B.\,Bobchenko$^f$, 
M.\,Chadeeva$^f$, 
M.\,Danilov$^f$, 
O.\,Markin$^f$, 
R.\,Mizuk$^f$, 
E.\,Novikov, 
V.\,Rusinov$^f$, 
E.\,Tarkovsky$^f$ 
\\ \it
Institute of Theoretical and Experimental Physics, B. Cheremushkinskaya ul. 25,
RU-117218 Moscow, Russia
}

\author{\centering 
N.\,Kirikova,  
V.\,Kozlov, 
P.\,Smirnov, 
Y.\,Soloviev 
\\ \it
P.\,N.\, Lebedev Physical Institute,
Russian Academy of Sciences,
117924 GSP-1 Moscow, B-333, Russia
}

\author{\centering 
D.\,Besson, 
P.\,Buzhan, 
E.\,Popova 
\\ \it
National Research Nuclear University 
MEPhI (Moscow Engineering Physics Institute)
31, Kashirskoye shosse,
115409 Moscow, Russia
}

\author{\centering 
M.\,Gabriel, 
C.\,Kiesling,
N.\,van\,der\,Kolk, 
K.\,Seidel,
F.\,Simon, 
C.\,Soldner, 
M.\,Szalay, 
M.\,Tesar, 
L.\,Weuste
\\ \it
Max Planck Inst. f\"ur Physik,
F\"ohringer Ring 6,
D-80805 Munich, Germany
}

\author{\centering 
M.\,S.\,Amjad, 
J.\,Bonis, 
P.\,Cornebise, 
F.\,Richard, 
R.\,P\"oschl, 
J.\,Rou\"en\'e, 
A.\,Thiebault
\\ \it
Laboratoire de L'acc\'elerateur Lin\'eaire,
Centre d'Orsay, Universit\'e de Paris-Sud XI,
BP 34, B\^atiment 200,
F-91898 Orsay CEDEX, France
}

\author{\centering 
M.\,Anduze,
V.\,Balagura,
V.\,Boudry, 
J-C.\,Brient, 
J-B.\,Cizel,
R.\,Cornat,
M.\,Frotin,
F.\,Gastaldi, 
Y.\,Haddad,
F.\,Magniette,
J.\,Nanni,
S.\,Pavy, 
M.\,Rubio-Roy,
K.\,Shpak,
T.H.\,Tran,
H.\,Videau,
D.\,Yu
\\ \it
Laboratoire Leprince-Ringuet (LLR)  -- \'{E}cole Polytechnique,
CNRS/IN2P3,
Palaiseau, F-91128 France
}

\author{\centering 
S.\,Callier,
S.\,Conforti di Lorenzo, 
F.\,Dulucq, 
J.\,Fleury,
G.\,Martin-Chassard, 
Ch.\,de la Taille, 
L.\,Raux, 
N.\,Seguin-Moreau
\\ \it
OMEGA Microelectronics
Ecole Polytechnique, CNRS/IN2P3
Avenue Gustave Coriolis
F-91128 Palaiseau CEDEX, France
}

\author{\centering 
J.\,Cvach, 
P.\,Gallus, 
M.\,Havranek, 
M.\,Janata, 
M.\,Kovalcuk,
J.\,Kvasnicka,
D.\,Lednicky,
M.\,Marcisovsky, 
I.\,Polak, 
J.\,Popule, 
L.\,Tomasek, 
M.\,Tomasek, 
P.\,Ruzicka, 
P.\,Sicho, 
J.\,Smolik, 
V.\,Vrba, 
J.\,Zalesak 
\\ \it
Institute of Physics, Academy of Sciences of the Czech Republic, Na Slovance 2,
CZ-18221 Prague 8, Czech Republic
}

\author{\centering              
S.\,Ieki,
Y.\,Kamiya,
W.\,Ootani, 
N.\,Shibata
\\ \it
ICEPP, The University of Tokyo, 7-3-1 Hongo, Bunkyo-ku, Tokyo
113-0033, Japan}

\author{\centering              
S.\,Chen, 
D.\,Jeans, 
S.\,Komamiya$^g$, 
C.\,Kozakai, 
H.\,Nakanishi
\\ \it
Department of Physics, Graduate School of Science, The University of
Tokyo, 7-3-1 Hongo, Bunkyo-ku, Tokyo 113-0033, Japan
}

\author{\centering 
M.\,G\"otze,  
J.\,Sauer, 
S.\,Weber, 
C.\,Zeitnitz
\\ \it
Bergische Universit\"{a}t Wuppertal
Fachbereich 8 Physik,
Gaussstrasse 20,
D-42097 Wuppertal, Germany
}

\author{
\it
$^\spadesuit$ Corresponding author\newline
E-mail: \email{eva.sicking@cern.ch}
}

\author{  \\
\llap{$^a$}Now at University of Bonn, Germany\\
\llap{$^b$}Now at DESY Hamburg, Germany\\
\llap{$^c$}Now at PNNL, USA\\
\llap{$^d$}Also at IPASCR Prague, Czech Republic\\
\llap{$^e$}Also at ITEP, Russia, deceased\\
\llap{$^f$}Also at NRNU MEPhI, Moscow, Russia\\
\llap{$^g$}Also at ICEPP, The University of Tokyo, Japan
}

\abstract{
We present a study of showers initiated by electrons, pions, kaons, and protons with momenta from 15\,GeV to 150\,GeV in the highly granular CALICE scintillator-tungsten analogue hadronic calorimeter.
The data were recorded at the CERN Super Proton Synchrotron in~2011. 
The analysis includes measurements of the calorimeter response to each particle type as well as
measurements of the energy resolution and studies of the longitudinal and radial shower development for selected particles. 
The results are compared to Geant4 simulations (version 9.6.p02).
In the study of the energy resolution we include previously published data with beam momenta from 1\,GeV to 10\,GeV recorded at the CERN Proton Synchrotron in 2010.
}


\begin{document}

\clearpage

\section{Introduction}
A sampling hadronic calorimeter with highly granular readout, as required for Particle Flow Analysis~(PFA)~\cite{Thomson:2009rp,Marshall:2012ry}, and with tungsten as absorber, is considered for experiments at future multi-TeV $\Pep\Pem$ colliders such as the Compact Linear Collider (CLIC)~\cite{CLIC_PhysDet_CDR}.
In these experiments the use of tungsten as dense absorber provides a calorimeter system sufficiently deep to contain jets of TeV energy, with a diameter fitting inside a solenoid of acceptable size.
\\
This paper presents the analysis of data obtained in test beam experiments of the \mbox{CALICE} tungsten analogue hadronic calorimeter prototype \mbox{(W-AHCAL)}~\cite{CaliceCollaboration,CALICE_AHCAL,LCD-Note-2012-002} using mixed beams containing electrons, muons, pions, kaons, and protons with a momentum range from 15\,GeV to 150\,GeV (in this paper, the natural system of units with $\hbar = c=1$ is used). 
The analysis includes measurements of the calorimeter response to each particle type, the energy resolution, and studies of the longitudinal and radial shower development. 
A purpose of the analysis is the comparison of data with Geant4~\cite{Agostinelli:2002hh, Allison:2006ve} simulation models. 
Based on the comparisons, the reliability of the models used in detector studies for future collider experiments can be assessed and the simulation models can be further improved.\\
Section \ref{sec:experimentalSetup} describes the experimental setup used in the test beam experiment.
The calibration procedure is discussed in section \ref{sec:calibration} and the data selection in section \ref{sec:DataSelection}.
The Monte Carlo (MC) simulation used in the comparison with data is introduced in section \ref{sec:simulation}.
Section \ref{sec:systematics} lists the systematic uncertainties relevant for the studied observables.
The analysis results for positrons are shown in section~\ref{sec:positron} and for hadrons in section~\ref{sec:hadron}.
In section \ref{sec:ComparisonOfResponse}, we compare the W-AHCAL response for different particle types, and in section \ref{sec:summary} we draw conclusions.

\section{Experimental setup}
\label{sec:experimentalSetup}
The \mbox{W-AHCAL} setup comprises a stack of 38 layers of absorber plates interleaved with 0.5\,cm thick scintillator tiles, read out by wavelength shifting fibres coupled to silicon photo-multipliers (SiPMs). 
Each absorber plate is 1\,cm thick and is made of a tungsten alloy consisting of 92.99\% tungsten, 5.25\% nickel, and 1.76\% copper, with a density of 17.8\,g/cm$^3$. 
The nuclear interaction length of this alloy is $\lambda_\mathrm{I} = 10.80$\,cm and the radiation length is $X_0=0.39$\,cm. 
The scintillator tiles of each layer are mounted into steel cassettes, with 0.2\,cm thick walls and are connected to printed circuit boards (PCB). 
The overall thickness of a calorimeter layer is approximately 2.5\,cm.
The combined structure of absorber, steel support, PCB, and scintillators has an effective interaction length of $\lambda_{\mathrm{I}} = 19.0$\,cm and radiation length of $X_0=0.88$\,cm such that one calorimeter layer corresponds in total to 0.13~$\lambda_{\mathrm{I}}$ and to 2.8~$X_0$.
The overall dimensions of the prototype are $0.9\times 0.9\times 0.95\;\mathrm{m}^3$, amounting to 4.9~$\lambda_{\mathrm{I}}$ and to 108~$X_0$. 
The high granularity of the detector is ensured by  the $3\times 3\; \mathrm{cm}^2$ tiles placed in the centre of each active plane,  surrounded by $6\times 6\,\mathrm{cm}^2$ and \mbox{$12\times 12\,\mathrm{cm}^2$} tiles at the edges.  
The last 8 layers of the stack are equipped with $6\times 6\,\mathrm{cm}^2$ tiles in the centre surrounded by $12\times 12\, \mathrm{cm}^2$ tiles.
In total the calorimeter is equipped with 7608 cells.
The temperature of each layer is monitored by five sensors in order to correct for the temperature dependence of the SiPM response.
The response of all SiPMs is calibrated and monitored with a LED system~\cite{CALICE_AHCAL}. 
\\
The data were taken in 2011 at the H8 beam line~\cite{SPS-H8-beamline,Coet&Dobel} of the CERN Super Proton Synchrotron (SPS).
The primary proton beam from the SPS hits a  target 605\,m upstream of the \mbox{W-AHCAL} prototype.
The secondary beam can then either be used for the experiment or it can hit a secondary target 477\,m upstream of the \mbox{W-AHCAL} prototype.
A momentum selection and focusing system is used to deliver the secondary or tertiary beam to the experiment. 
Either beams of $\Pepm$ with momenta between 10 and 40\,GeV or mixed beams of $\PGmpm$, $\PGppm$, $\PKpm$, and protons between 10 and 300\,GeV were used.
\\
The data analysis is restricted to $\Pep$ events with beam momenta from 15\,GeV to 40\,GeV as well as to positively-charged hadron events with beam momenta from 25\,GeV to 150\,GeV for which high selection purity can be achieved and leakage effects are limited (cf. section~\ref{sec:DataSelection}). 
Negative-polarity data have smaller statistics due to less allocated beam time, but they are included if necessary for the study of detector-related systematics.

\begin{figure}[t!]
\centering
\includegraphics[width=\textwidth]{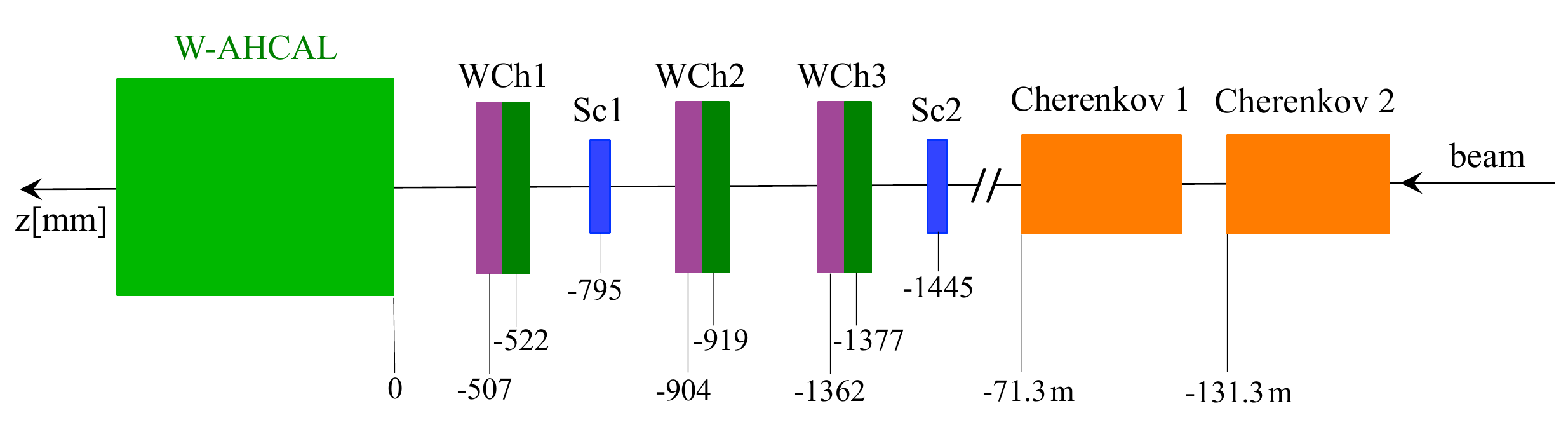}
\caption{Sketch of the CERN 2011 test beam line (not to scale), where $Sc$ stands for scintillator and $WCh$ for wire chamber. 
The beam enters from the right.
}
\label{fig:tbeam_H8}
\end{figure}

A sketch of the CERN SPS test beam setup is presented in figure~\ref{fig:tbeam_H8}.
A right handed coordinate system is used, with the $z$-axis given by the beam axis, the $x$-axis pointing horizontally and the \mbox{$y$-axis} pointing up. 
The beam passes through two Cherenkov threshold counters, two trigger scintillators and a tracking system of three delay wire chambers~\cite{Spanggaard1998} before reaching the W-AHCAL.
The Cherenkov counters are filled with $\mathrm{He}$ gas. 
The Cherenkov radiation is detected by photo-multiplier tubes that are operated with discriminators at fixed thresholds.
The Cherenkov signals are used offline for hadron identification, one Cherenkov counter being used for the separation of pions from kaons and the other one for the separation of kaon from protons.
The beam trigger is defined by the coincidence of two \mbox{$10\times10\times1$\,cm$^3$} scintillator counters. 
The information from the three $11\times11$\,cm$^2$ wire chambers is used offline to reconstruct the trajectory of the incident particle and predict its position on the calorimeter surface.
\\
In the analysis of the W-AHCAL energy resolution, data from the 2010 measurement campaign~\cite{CALICE-WAHCAL-2010} at the CERN Proton Synchrotron (PS) with beam momenta from 1\,GeV to 10\,GeV are also included. 
The experimental setup for these low energy measurements  was slightly different. 
The calorimeter was equipped only with the first 30 layers accounting for 3.9\,$\lambda_{\mathrm{I}}$ and 85\,$X_0$.
The Cherenkov counters were placed at 3.5\,m and 7\,m in front of the experimental setup and filled with $\text{CO}_2$ to better distinguish the lower momentum particles.

\clearpage
\section{Calibration}
\label{sec:calibration}
The response of all calorimeter cells is equalised to a common physics signal produced by minimum ionising particles (MIP), which is used throughout the paper as the energy scale, being the most natural one for the comparison with Monte Carlo simulations.
For this calibration, data from dedicated muon runs are used, as muons are considered to be minimum ionising particles. 
Several steps are necessary to translate signals recorded by the SiPM measured in ADC counts into deposited energy in units of MIP.
\\
First a track finding algorithm \cite{Adloff:2013vra} is used to reconstruct the muon track. 
The energy distribution in each cell for hits associated to a reconstructed muon track is fitted with a Landau function convolved with a Gaussian function.
The MIP calibration factor of a single cell $i$, $A_i^{\textrm{MIP}}$~[ADC] is given by the most probable value of the signal of a muon in that cell, expressed in ADC counts. 
The calibration of a single cell $i$ is then performed according to
\begin{equation}
\displaystyle E_i [\textrm{MIP}] = \frac{A_i[\textrm{ADC}]}{A_i^{\textrm{MIP}}[\textrm{ADC}]}\cdot f_{\textrm{resp}}(A_i[\textrm{pixel}]), \label{eq:calibFormula}
\end{equation}
where
\begin{itemize}
\item $A_i[\textrm{ADC}]$ is the pedestal subtracted amplitude registered in cell $i$, in units of ADC counts,
\item $A_i^{\textrm{MIP}}[\textrm{ADC}]$ is the MIP amplitude in cell $i$ in ADC counts, called the MIP calibration factor and extracted from the dedicated muon runs,
\item $f_{\textrm{resp}}(A_i[\textrm{pixel}])$ is the SiPM response function which corrects for the non-linearity of the SiPM response due to the limited number of pixels (1156) and the finite pixel recovery time (\mbox{$20-500$\,ns}).  
This function acts on the amplitude expressed in number of pixels, and returns the saturation correction factor to linearise the amplitude in MIPs. 
Its asymptotic behaviour is described using a double-exponential fit function and a linearised extrapolation for large values~\cite{CALICE_AHCAL_emPaper}.
\item The amplitude in pixels is obtained by dividing the amplitude of a cell by the corresponding SiPM gain $G_i[\textrm{ADC/pixel}]$
\begin{equation}
\label{eq:gain}
\displaystyle A_i[\textrm{pixel}] = \frac{A_i[\textrm{ADC}]}{G_i[\textrm{ADC/pixel}]}.
\end{equation}
The gain values $G_i[\textrm{ADC/pixel}]$ are obtained from fits of photo-electron spectra taken with low intensity LED light provided by the calibration and monitoring LED system~\cite{CALICE_AHCAL_emPaper}.
\end{itemize}
In the following data analysis, only cells with energy depositions above a threshold of 0.5\,MIP are considered.

\subsection{Saturation scaling factor of the SiPM response function}
\label{sec:calibration_saturation}
\begin{figure}[ht!]
\begin{minipage}[b]{0.49\textwidth}
\includegraphics[width=\textwidth]{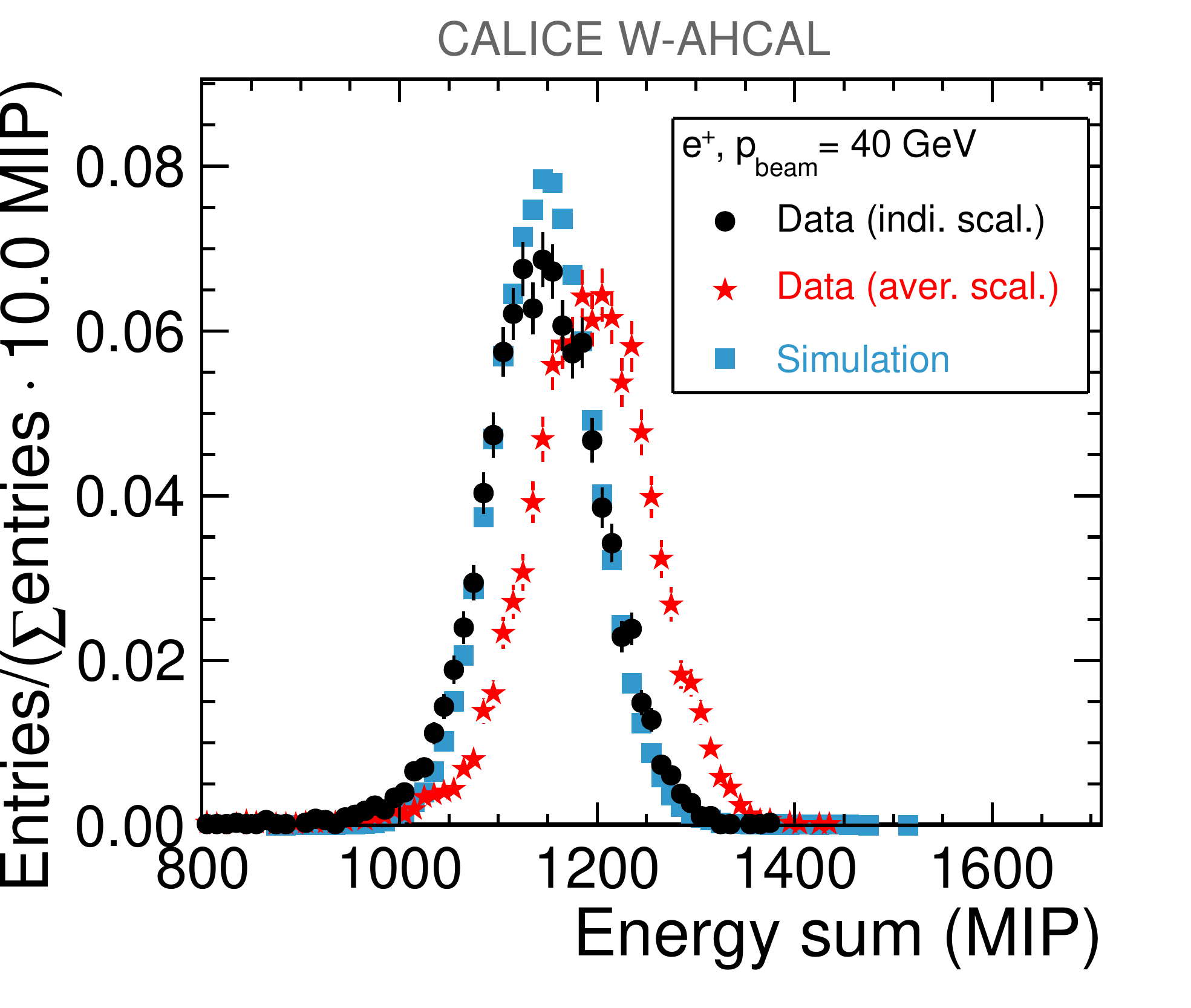}
\caption{Distribution of the energy sum over all cells for 40\,GeV $\Pep$ data. 
Data reconstructed using individual and average saturation scaling factors are compared to simulations.}
\label{fig:esum_scaling}
\end{minipage}
\hfill
\begin{minipage}[b]{0.49\textwidth}
\includegraphics[width=\textwidth]{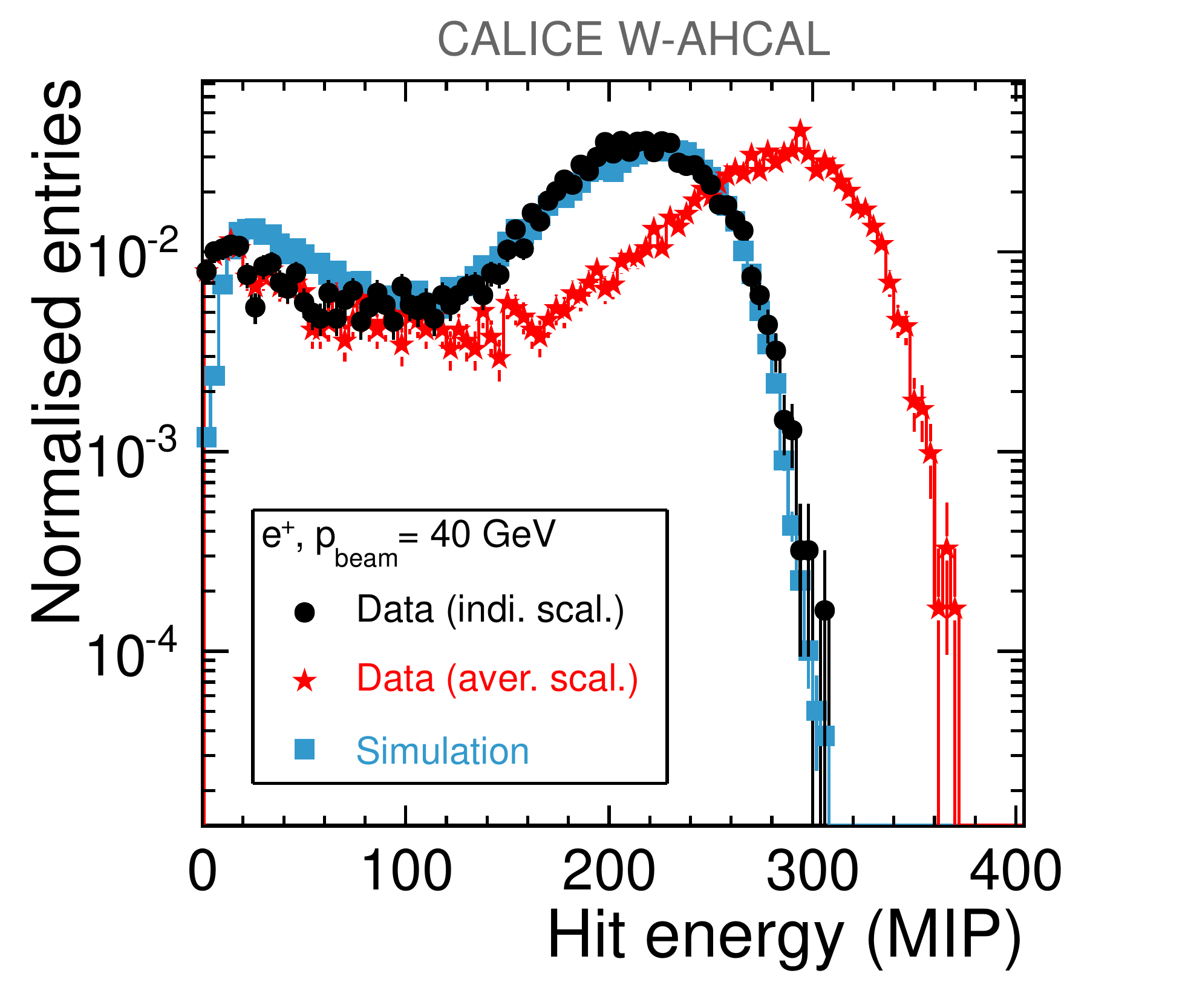}
\caption{Hit energy distribution for the most central cell with respect to the beam position in layer 3 for 40\,GeV $\Pep$. 
Data reconstructed using individual and average saturation scaling factors are compared to simulations.
}
\label{fig:ePlus40GeV_hitEnergy_sysScalingFactor}
\end{minipage}
\end{figure}

The response of all SiPMs was measured before mounting the sensors on the scintillator tiles~\cite{CALICE_AHCAL}. 
Due to geometric alignment effects, the effective maximum number of pixels for mounted SiPMs is on average 20\% lower than the value for bare SiPMs \cite{CALICE_AHCAL_emPaper}. 
For this reason, the response functions which correct for saturation in the SiPM response of all AHCAL cells are scaled by a factor $s$ which has been determined in situ using the LED calibration system. 
While for steel and low energy tungsten data the usage of an average factor of $s=0.8$ was found to be sufficient, the high energy tungsten data exhibit higher single cell energies, such that the usage of the exact scaling values is required for a correct reconstruction.
Hence, individual scale factors are used where available (in 60\% of the cells) and the average value of $s=0.8$ otherwise.
The impact of using individual or average scale factors is largest for the highest positron energy, as shown for the total energy sum distribution of the 40\,GeV $\Pep$ data in figure~\ref{fig:esum_scaling}. 
The data event selection and the simulation used in this study are introduced in sections \ref{sec:DataSelection} and \ref{sec:simulation}.
The effect is particularly striking in single-cell hit energy distributions of cells which have an individual scaling factor that differs significantly from the average scaling factor.
As an example, the hit energy distribution for 40\,GeV $\Pep$ for the most central cell with respect to the beam position in layer 3, as evaluated using the average (0.8) and individual (0.91) scaling factor is shown in figure~\ref{fig:ePlus40GeV_hitEnergy_sysScalingFactor}.
\\
Differences between the shower properties reconstructed using individual and average saturation scaling factors decreases towards lower beam momenta and are smaller for hadron showers than for electromagnetic showers.
For this reason, in the reconstruction of the PS data sets with beam momenta between 1\,GeV and 10\,GeV used in the study of the energy resolution in sections~\ref{sec:positron} and \ref{sec:hadron}, average scaling values as used in \cite{CALICE-WAHCAL-2010} were adopted.
In this momentum range, the visible energy and the energy resolution obtained using individual and average scaling agree with each other on the sub-percent level.

\subsection{Temperature correction}
\label{sect:tempCorrection}

The temperature inside the calorimeter is monitored using five sensors per layer. 
Details about the temperature measurements are given in~\cite{LCD-Note-2011-001}. 
Temperature variations of several degrees were observed with time and along the beam direction.
Due to the temperature dependence of the SiPM response, the MIP calibration factors need to be corrected for temperature variations.
The temperature dependence of the MIP calibration factor of a given cell $i$ can be described by the formula
\begin{equation}
\label{eq:mipTemp}
A_i^{\textrm{MIP}} (T_{\mathrm{run}}) =
A_{i}^{\textrm{MIP}}(T_0) + C \cdot (T_{\mathrm{run}} - T_0) ,
\end{equation}
where
\begin{itemize}
\item $A_i^{\textrm{MIP}}(T_{\mathrm{run}})$ is the MIP amplitude in cell $i$ in ADC counts at the temperature $T_{\mathrm{run}}$,
\item $A_{i}^{\textrm{MIP}} (T_0)$ is the MIP amplitude in cell $i$ in ADC counts, obtained at the reference temperature $T_{0}$, i.e.\ at the average temperature during the muon runs of the sensor closest to the cell,
\item $T_{\mathrm{run}}$ is the temperature of the sensor closest to the given cell in the given run in $^\circ$C,
\item $C$ is the MIP temperature coefficient determined for the layer of the given cell in ADC/$^\circ$C.
\end{itemize}

Due to limited statistics, the MIP temperature coefficients are not determined for each cell independently but estimated in average for all cells in one calorimeter layer, using the method described in~\cite{CALICE-WAHCAL-2010}. 
For a few calorimeter layers, no reliable estimations for the MIP temperature coefficient could be determined. 
For these specific layers, the average of all other coefficients is used. 
The average MIP temperature coefficient considering all layers is $-4.5\%/^\circ$C before the temperature correction, and better than $0.06\%/^\circ$C after.
A similar procedure is used to correct the temperature dependence of the gain calibration coefficients.

\section{Data selection}
\label{sec:DataSelection}
\begin{figure}[t!]
\centering
\includegraphics[width=0.8\textwidth]{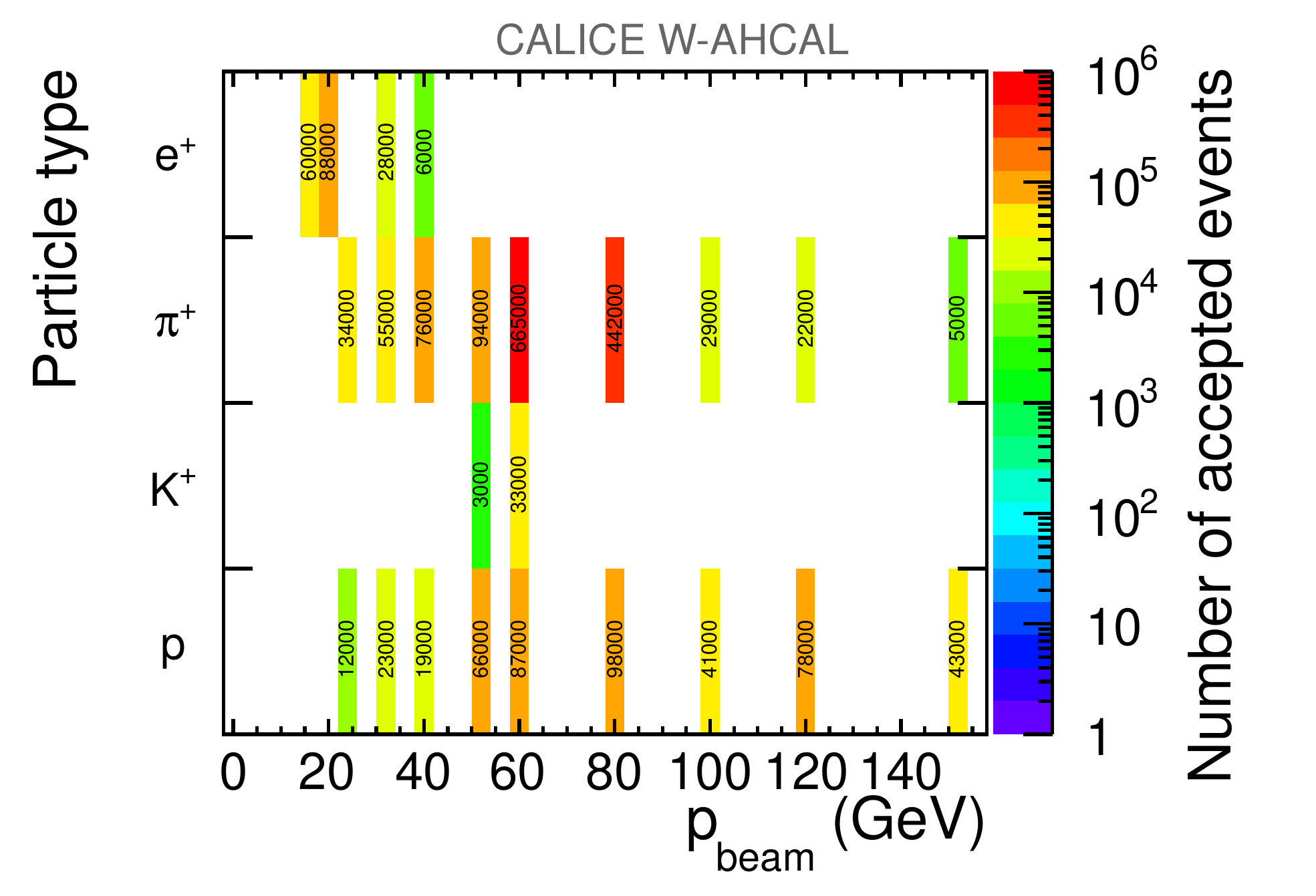} 
\caption{Number of events after selection for a given particle type.}
\label{fig:NumberOfEvents}
\end{figure}

The data analysis was performed for $\Pep$ events with beam momenta from 15\,GeV to 40\,GeV, $\PGpp$ and proton events with beam momenta from 25\,GeV to 150\,GeV, and $\PKp$ events at 50\,GeV and 60\,GeV.
For these data sets, high selection purity can be achieved~\cite{LCD-Note-2013-006} and leakage effects are limited (cf.\ section \ref{sec:HadronSelection}).
\\
For most of the particles and beam momenta, 20\,000--100\,000 events (figure \ref{fig:NumberOfEvents}) pass the selection steps introduced in the following.
For 60\,GeV and 80\,GeV, high statistics hadron data sets were recorded with approximately up to 700\,000 events after event selection.
Beam momenta with particularly low number of events after event selection are 40\,GeV $\Pep$ (6\,000 events), 50\,GeV $\PKp$ (3\,000~events), and 150\,GeV $\PGpp$ (5\,000 events).

\subsection{Electromagnetic showers}
\label{sec:emSelection}
\begin{figure}[t!]
\centering
\begin{minipage}{0.45\textwidth}
\centering
\includegraphics[height=\textwidth]{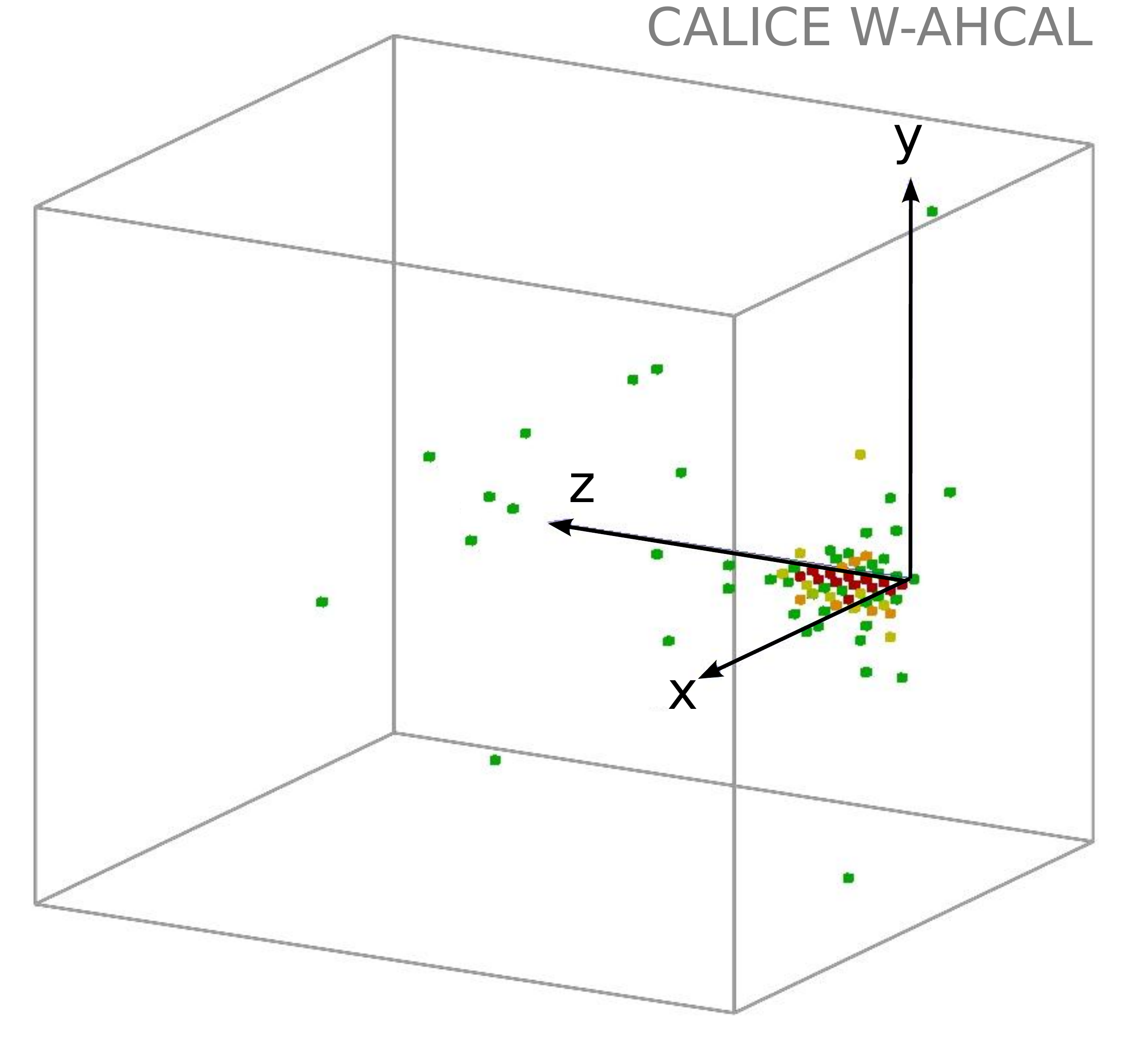}
\caption{Display of a 15\,GeV $\Pep$ event. 
The colours of the W-AHCAL cells indicate the deposited energy, red being the highest (more than 5.4\,MIP).
}
\label{fig:ePlus15GeV_eventDisplay}
\end{minipage}
\hfill
\begin{minipage}{0.45\textwidth}
\centering
\includegraphics[height=\textwidth]{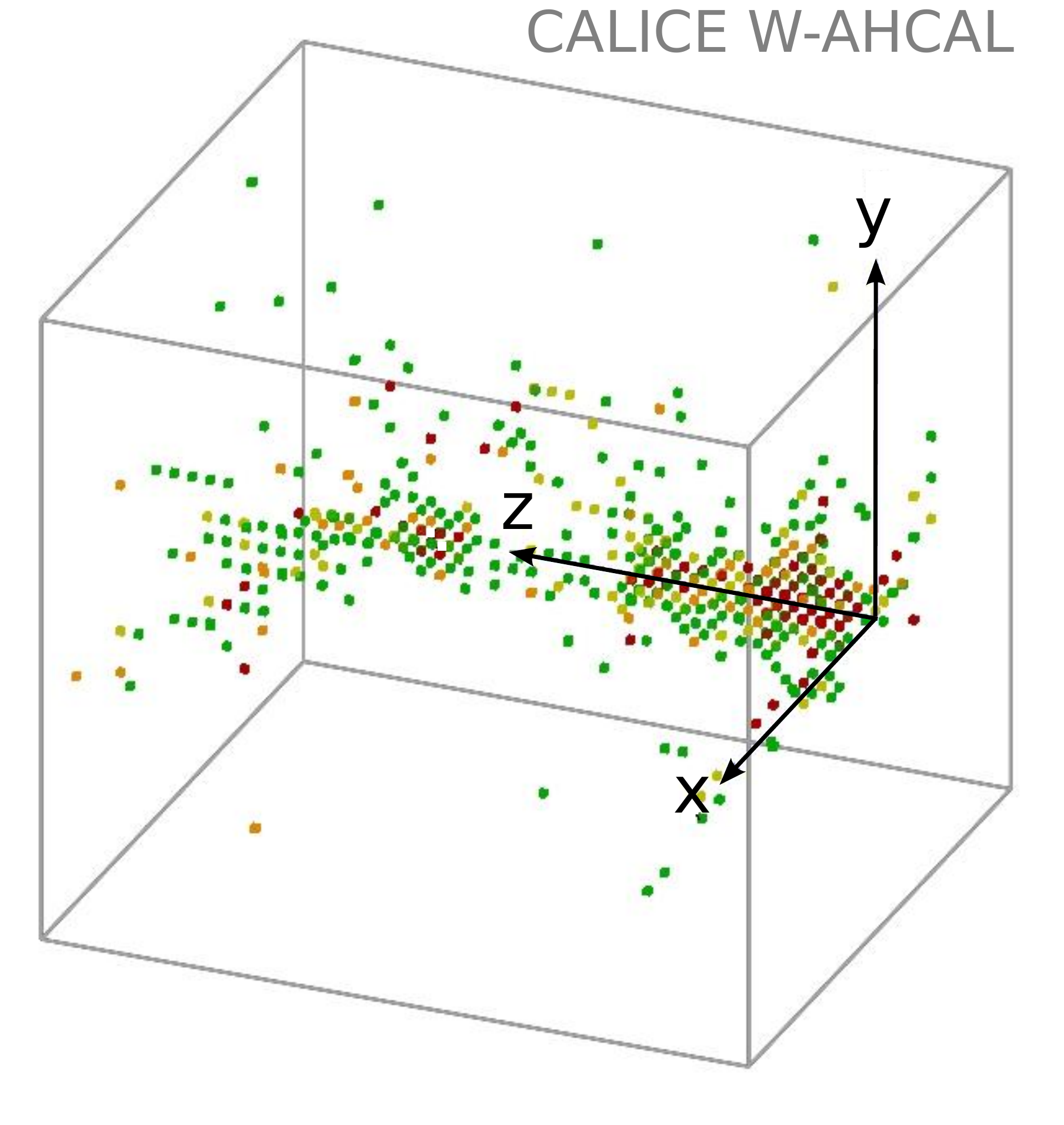}
\caption{Display of a 80\,GeV $\PGpp$ event. 
The colours of the W-AHCAL cells indicate the deposited energy, red being the highest (more than 5.4\,MIP).
}
\label{fig:pionEventDisplay}
\end{minipage}
\end{figure}

Positron events were recorded at the SPS with a dedicated positron beam that has a small contamination from muons.
To increase the purity of the $\Pep$ data, an optimised cluster reconstruction and subsequent event selection procedure is pursued.
\\
Three-dimensional clusters are identified using the nearest-neighbour algorithm developed in~\cite{Lutz2010}. 
Cells with an energy above a threshold of 2\,MIP are used as seeds for the clustering. 
All active cells which surround these seed cells and have an energy deposit above a threshold of 0.5\,MIP are added to the cluster.
\\
The reconstructed $\Pep$ event candidates are selected if they fulfil the following requirements.
First, the number of identified clusters in the whole calorimeter is required to be one. 
Electromagnetic showers form a cluster near the calorimeter front face, hence in the second step events are rejected if their centre-of-gravity in the $z$-direction ($z_{\text{cog}}$) exceeds 160\,mm, which rejects muon events.
Here, $z_{\text{cog}}$ is defined as
\begin{equation}
z_{\text{cog}}=\displaystyle \frac{\sum_{i=1}^{N_{\text{hits}}} E_i \cdot z_i}{\sum_{i=1}^{N_{\text{hits}}} E_i},
\label{eq:zcog}
\end{equation}
where $E_i$ is the visible energy deposited in the $i$-th active calorimeter cell, $z_i$ is its position in the reference frame of the calorimeter with the W-AHCAL front plane at \mbox{$z=0$\,mm}, and $N_{\text{hits}}$ is the number of W-AHCAL cells which have an energy deposition above the threshold of 0.5\,MIP.
In a third selection step, the events are required not to contain an identified track which could indicate a muon or a hadron event in which either the muon or the hadron themselves or particles inside the hadron shower traverse (parts of) the calorimeter stack as a minimum ionising particle.  
As a last requirement, $N_{\text{hits}}$ should be inside the range defined in table~\ref{tab:nHits}. 
This is done in order to reject the few events in which either the electron shower is initiated before entering the W-AHCAL or events that are particularly noisy.
A typical $\Pep$ event after all selection cuts is shown in figure~\ref{fig:ePlus15GeV_eventDisplay}.
Depending on the beam momentum, between 60\% and 80\% of all positron candidate events pass the described positron selection.

\begin{table}[t!]
\centering
\caption{Cuts on the minimum and the maximum number of hits applied for the $\Pep$ selection.}
\label{tab:nHits}
\begin{tabular}{lrr}
\toprule
$p_{\text{beam}}$ (GeV) & $N_{\mathrm{hits}}^{\mathrm{min}}$ & $ N_{\mathrm{hits}}^{\mathrm{max}}$\\
\midrule
15  &  75 & 110\\
20  &  80 & 130\\
30  &  95 & 140\\
40  & 110 & 170\\
\bottomrule
\end{tabular}
\end{table}

To reduce the influence of noise in the selected $\Pep$ events, only calorimeter cells within the first 20 calorimeter layers (corresponding to 56\,$X_0$) and within the central $10\times10$ tiles of $3\times3\;\textrm{cm}^2$ size (corresponding to more than 3.5 effective Moli\`{e}re radii) are used to calculate the positron shower properties discussed below.
This strategy is appropriate because $\Pep$ showers in the energy range studied here deposit energy mostly within the first 5 to 10 calorimeter layers, the beam is centred on the calorimeter centre, and the width of the beam profile is not more than three tiles.

\bigskip
\subsection{Hadronic showers}
\label{sec:HadronSelection}

\begin{figure}[t!]
\centering
\includegraphics[width=0.75\textwidth]{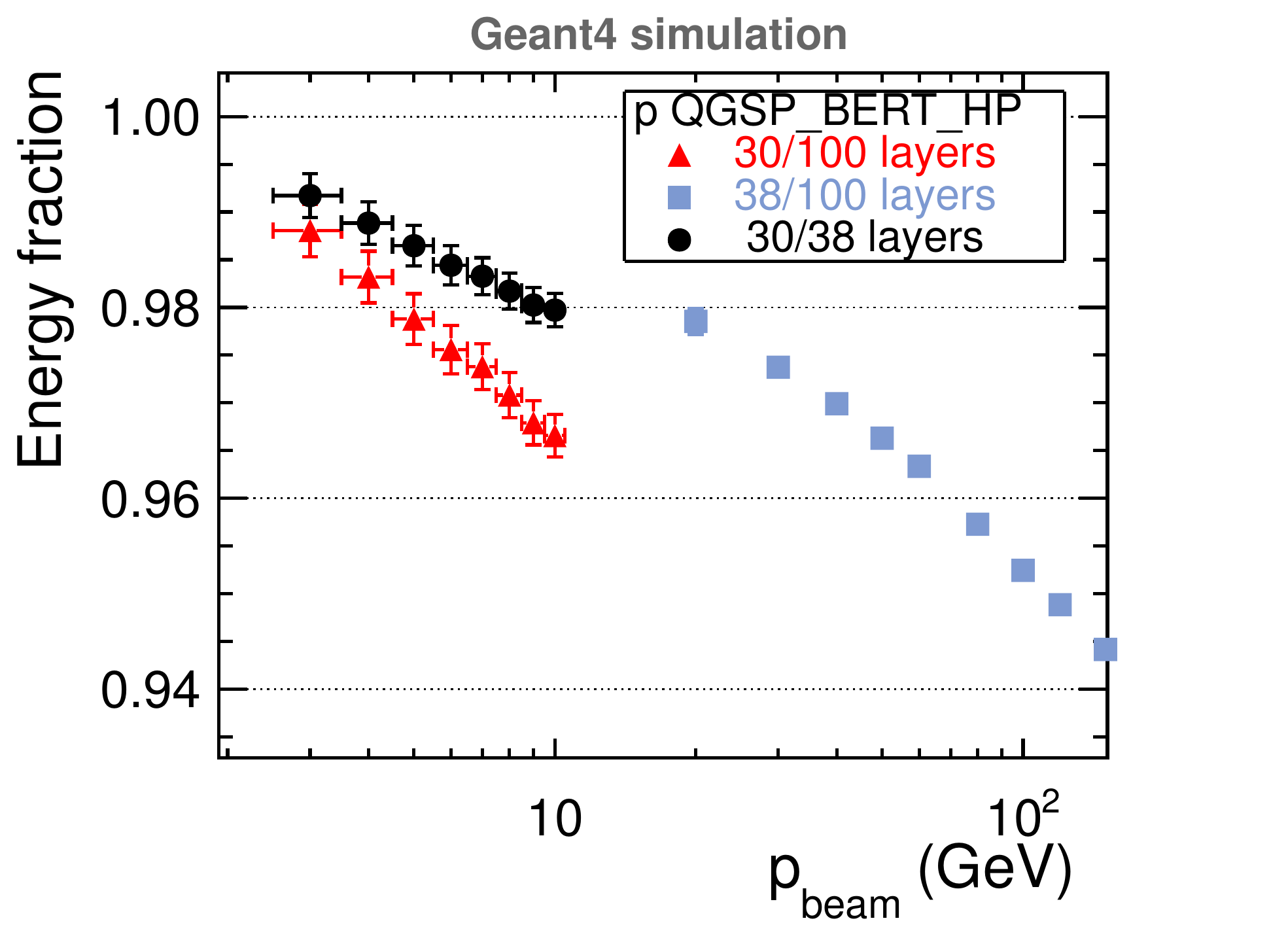}
\caption{
Energy fraction detected versus beam momentum of simulated proton (p) showers within 30 (38) layers of the 100-layer deep W-AHCAL at PS (SPS) energies and energy fraction detected within 30 layers of 38-layer deep W-AHCAL at PS energies.
The showers are required to start within the first three layers of the calorimeter.
}
\label{fig:leakage}
\end{figure}

Hadron events are recorded at the SPS with a mixed beam containing mostly $\PGpp$, $\PKp$, protons, and $\PGmp$ from in-flight decays.
Contamination from $\Pep$ was effectively suppressed by using a lead absorber in the beam line. 
The absorber had a thickness of 8\,mm for the 25\,GeV and 30\,GeV data sets and 18\,mm for higher beam momenta. 
Even for an absorber thickness of 8\,mm the effect of the resulting $\Pep$ impurity on the shower observables is negligible.
Contamination from $\PGmp$ from in-flight decays is rejected based on information from the W-AHCAL itself, as described below.
\\
The first step in the hadron selection is based on particle identification information obtained from the two Cherenkov threshold counters installed upstream of the \mbox{W-AHCAL} prototype as described in section~\ref{sec:experimentalSetup}. 
A detailed description of the Cherenkov selection strategy and results can be found elsewhere~\cite{LCD-Note-2013-006}. 
The event samples of pions, kaons, and protons selected based on the Cherenkov threshold counters contain between 85\% and 100\% of the selected hadron type.
\\
Events are then rejected if they do not contain a reconstructed cluster (cf.\ section \ref{sec:emSelection}).
After that, the layer of primary interaction of the reconstructed shower, i.e.\ the shower start, introduced in \cite{Adloff:2013mns}, must be in the first three layers of the calorimeter, except when measuring the longitudinal profiles where an alternative selection is used (see section \ref{sec:had:long}). 
The latter requirement allows for a selection of hadron showers that are mostly contained in the W-AHCAL stack as discussed below. 
Both requirements reject muon events.
The remaining muon contamination is reduced by exploiting the fact that muons show a different signature in the highly granular W-AHCAL from hadron events, i.e.\ they deposit energy only in comparatively few cells of the full W-AHCAL stack and the mean depth of all energy depositions given by the centre-of-gravity of all hits in the $z$-direction has a different distribution from that for hadrons.
In the ($N_{\text{hits}}:z_{\text{cog}}$)-plane, all events below a threshold given by
$N_{\text{hits}} + a(p_{\text{beam}})*z_{\text{cog}} < b(p_{\text{beam}})$
(where $a$ and $b$ depend linearly on $p_{\text{beam}}$) are rejected.
While removing few remaining muon events, the cut does not have any impact on the central part of the energy sum distribution of the hadron data.
An example $\PGpp$ event after all selection cuts is shown in figure~\ref{fig:pionEventDisplay}.
Depending on the beam momentum, between 5\% and 30\% of all hadron candidate events pass the described hadron selection, mostly influenced by the variation of the beam's muon content.
\\
The W-AHCAL prototypes used in the two test beam campaigns at the PS and SPS, with respective total lengths of 3.9\,$\lambda_{\text{I}}$ and 4.9\,$\lambda_{\text{I}}$, do not fully contain all hadronic showers at the studied beam momentum ranges of 1--10\,GeV and 15--150\,GeV, even if the showers start within the first three \mbox{W-AHCAL} layers.
The remaining longitudinal leakage of energy after applying the shower start cut has been studied using simulated hadron showers in a long W-AHCAL prototype of 100 layers corresponding to $13\,\lambda_{\text{I}}$. 
The energy fraction contained in 30 or 38 layers of this long prototype is shown in figure~\ref{fig:leakage} for simulated proton showers.
Events are selected by requiring that the shower starts within the first three layers of the calorimeter, as also required in the analysis of hadron data.
Similar results are found for different Geant4 physics lists and hadron types. 

As can be seen in figure~\ref{fig:leakage} for protons, 5.5\% of the shower energy measured in 100 layers is expected to leak out of a 38 layers calorimeter at 150\,GeV.
In order not to be dominated by leakage effects in the characterisation of the W-AHCAL, we limit the momentum range studied to this upper value of 150\,GeV.
We estimated that leakage has a 1.0\%--5.5\% effect on the W-AHCAL response (figure \ref{fig:leakage}), a \mbox{0\%--17\%} effect on the energy resolution for the studied momentum range and it has a negligible effect on shower shapes.
The detector simulations introduced in section \ref{sec:simulation} use the same \mbox{W-AHCAL} geometry as in data such that leakage effects are taken into account.

\subsection{Combination of data recorded at the CERN PS and SPS}
To better constrain the W-AHCAL energy resolution fit results discussed in sections \ref{sec:positron} and \ref{sec:hadron}, the \mbox{W-AHCAL} data of the test beam campaign at the CERN PS with beam momenta from 1\,GeV to 10\,GeV~\cite{CALICE-WAHCAL-2010} are included.
For this, the PS data were selected based on the SPS selection cuts described above.
The PS analysis results obtained in this way are slightly different from but consistent within the uncertainties to the previously published results~\cite{CALICE-WAHCAL-2010}.
In the combination of the PS and SPS data for the energy resolution, the difference in the energy resolution between 30 and 38 layers has been taken into account by introducing a systematic uncertainty of 2\% on the PS data points.
This uncertainty was estimated using hadron shower simulations at PS energies as introduced in section~\ref{sec:HadronSelection} by comparing the energy resolution of a W-AHCAL of 30 and 38 layers.
For SPS energies we observe that the difference in the energy resolution obtained with 30 layers and 38 layers in data is well described by MC.

\section{Monte Carlo simulation}
\label{sec:simulation}
The simulation includes the W-AHCAL, the two trigger scintillators and the three wire chambers.
Particles of a given type and energy are generated at \mbox{$z=-55$\,m}. 
The transverse beam profiles in the simulations are chosen such that they reflect the beam profiles measured in the corresponding data runs.
Interactions are modelled using the QGSP\_BERT\_HP and the FTFP\_BERT\_HP physics lists of Geant4 version 9.6.p02. 
These lists are described in~\cite{CALICE-WAHCAL-2010}.
The extension HP indicates that the physics lists use a data driven high precision neutron package for the transport of neutrons below 20\,MeV down to thermal energies. 
This is crucial for the simulation of the high-A absorber tungsten where many more neutrons due to spallation are produced than in steel absorbers \cite{Adloff:2014rya}.
The size of the Monte Carlo data sets are chosen such that they reflect the approximate number of events of the individual data sets used in the analysis. 
The same event selection as introduced in section~\ref{sec:DataSelection} for data is applied to the simulated data sets;
similar selection efficiencies in simulation and data are observed.
\\
For a valid comparison between simulation and data, realistic detector effects need to be considered at the generation and the digitisation level.
In the generation step, the signal shaping time of the readout electronics is emulated in the Monte Carlo simulation by accepting only hits generated within a time window of 150\,ns~\cite{Lutz2010}. 
In addition, saturation effects in plastic scintillators, described by Birks' law, are applied \cite{Birks:1951}.
In the digitisation, the same detector granularity, calibration values and dead or uncalibrated cells are considered as for the reconstruction of the experimental data.
The simulated energy is converted into MIP based on a conversion factor estimated in muon simulations~\cite{Clemens_Thesis}. 
Light leakage between neighbouring scintillator tiles is also taken into account~\cite{Clemens_Thesis}.
\\
During data taking, some events were recorded with random triggers in order to monitor the noise level in the calorimeter \cite{CALICE_AHCAL_emPaper}.
This noise is not subtracted from the data but it is accounted for by superimposing random trigger events from the relevant run period on simulations. 
\\
In contrast to previous AHCAL studies, in this study an additional treatment of saturation effects is included in the simulations.
In the data reconstruction, the asymptotic exponential form of the SiPM response function introduced in section~\ref{sec:calibration} is extrapolated with a linear function for very large signal amplitudes in order to avoid unphysical results.
For the large amplitudes occurring when using tungsten absorbers and at large beam momenta, this potentially leads to an under-correction of saturation effects in data.
In previous AHCAL studies, the same form of the SiPM response function was used in simulation and reconstruction such that the effect cancels out in the case of MC, but not for data.
By retaining the asymptotic form in the simulation while using the linearised function in the reconstruction, saturation effects observed in data are better reproduced by the simulations.

\section{Systematic uncertainties}
\label{sec:systematics}
The most relevant sources of systematic uncertainties for the W-AHCAL test beam data are
the limited knowledge of the saturation scaling factor of the SiPM response function,
the uncertainty in the MIP calibration,
the stability of the \mbox{W-AHCAL} response over time, and
the choice of the event selection based on the shower start layer.
For the Geant4 simulations, the most relevant systematic uncertainty originates from the simulation of the cross-talk between neighbouring scintillator tiles.
\\
The systematic uncertainties due to all relevant sources and all discussed observables are studied and quantified.
In the following, for brevity, only the systematic uncertainty on the reconstructed visible energy and, in specific cases, on the longitudinal or the radial energy profile are quoted.

\subsection{Saturation scaling factor of the SiPM response function}
\label{sec:scaling}
For electromagnetic showers in the W-AHCAL, due to the dense absorber with about 3~radiation lengths per layer, most of the shower energy is deposited in a few cells in the first calorimeter layers.  
Hence, uncertainties in the SiPM response function are expected to have significant impact on the reconstructed $\Pep$ shower properties.
\\
As discussed in section~\ref{sec:calibration_saturation}, the response functions for all SiPMs were measured before mounting the sensors on the scintillator tiles.
These are then scaled with a saturation scaling factor $s$, accounting for the imperfect mounting of the SiPMs to the wavelength shifting fibres of the scintillator tiles.
The distribution of the measured saturation scaling factors with a mean of $s=0.8$ has a standard deviation of 0.09, reflecting cell-to-cell variations and measurement uncertainties.
Individually measured values are available for 60\% of the cells.
The uncertainties on these factors are taken to be $\pm 0.5$~standard deviations of their distribution ($\pm 0.045$) for the individually measured factors and $\pm 1$~standard deviation ($\pm 0.09$) for the saturation scaling factors for the unmeasured cells.
\\
Systematic uncertainties in the reconstructed deposited energy are estimated by reconstructing the data using scaling factors varied by the full range of their respective uncertainties, denoted as high and low scaling.
An increase in the scaling factor results in a decrease of the reconstructed energy and vice versa. 

\begin{figure}[t!]
\centering
\begin{minipage}[c]{0.45\linewidth}
\centering
\includegraphics[width=\textwidth]{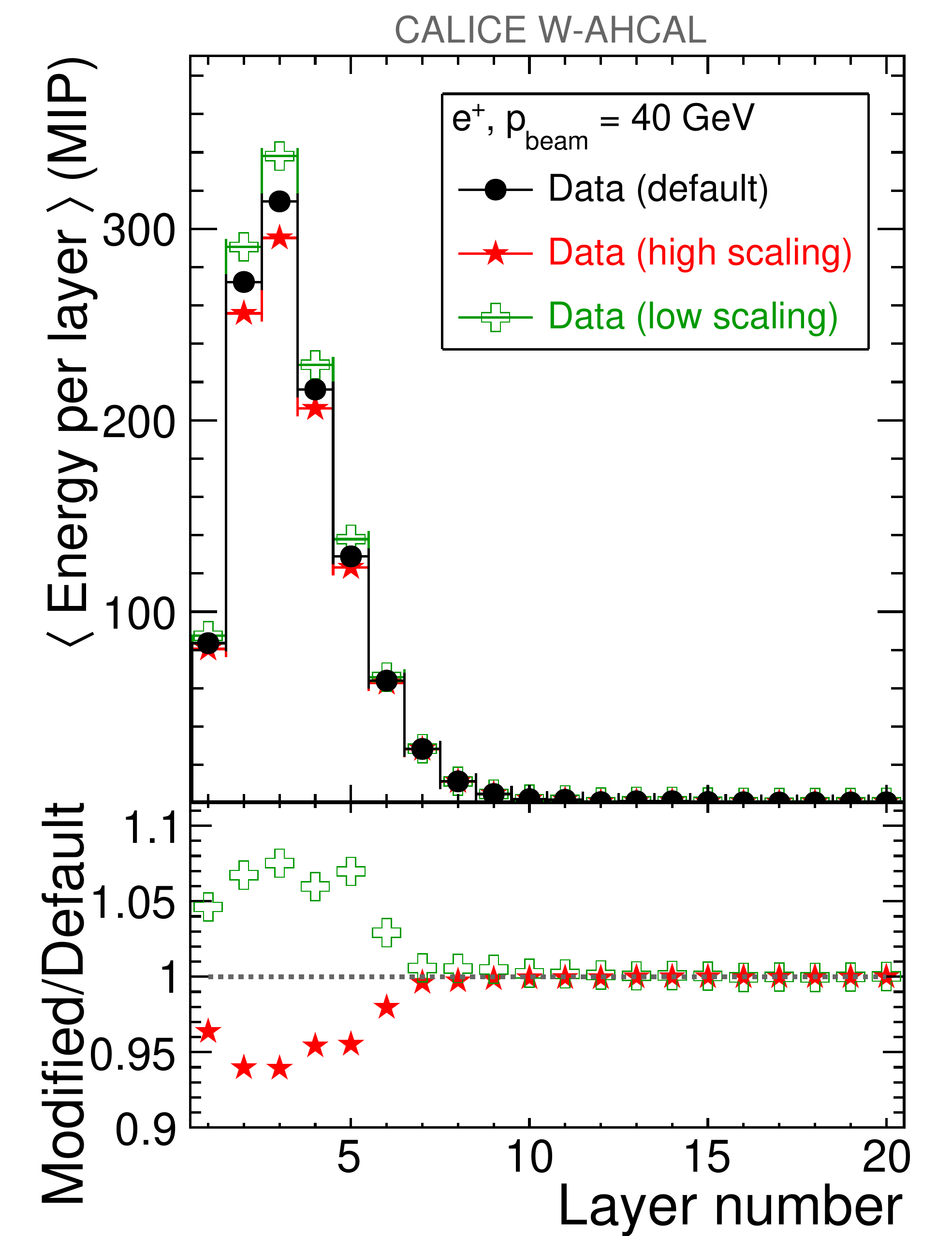}
\end{minipage}
\hfill
\begin{minipage}[c]{0.45\linewidth}
\centering
\includegraphics[width=\textwidth]{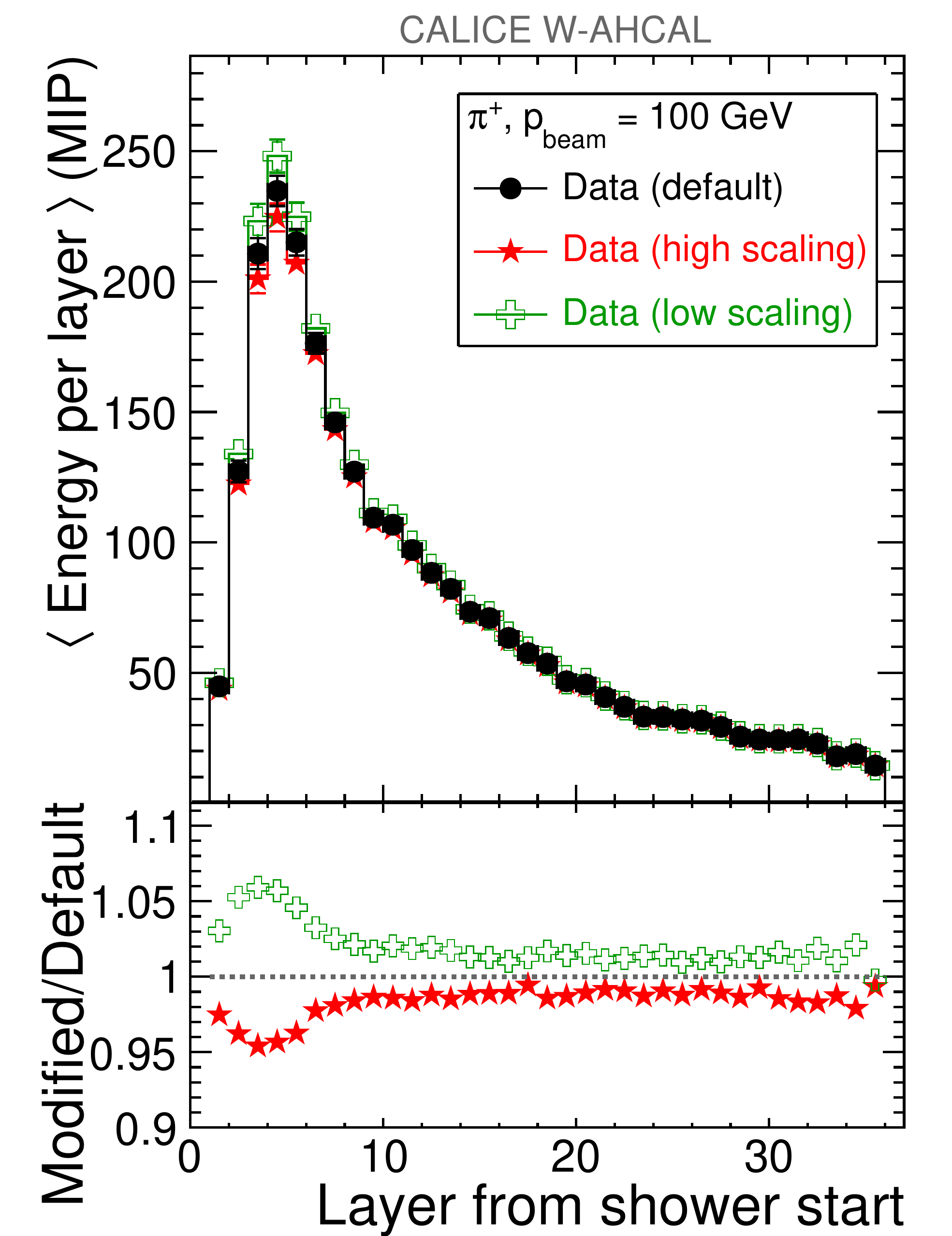}
\end{minipage}
\caption{
Comparison of the longitudinal energy profile for positron-induced showers at 40\,GeV (left) and pion-induced showers at 100\,GeV (right) when using SiPM scaling factors varied within the full range of their uncertainties.
The bottom panels (same for figures \protect\ref{fig:ePlus_mcComparison_long}, \protect\ref{fig:ePlus_linearity_resolution}, and \protect\ref{fig:piLinearityResolution}--\protect\ref{fig:proton_linearity_resolution}) show the ratio between the distributions for the different choices of scaling factor.
}
\label{fig:ScalingFactorSystematics}
\end{figure}

Figure \ref{fig:ScalingFactorSystematics} shows how the maximal modification of the saturation scaling factors within their uncertainties increases or decreases the reconstructed energy deposition per layer in longitudinal energy profiles for $\Pep$ (left) and $\PGpp$ (right) induced showers.
The observed variations, $\delta E_{\text{vis per layer}}$, are treated as the limits of a rectangular distribution and the resulting uncertainties are used as systematic uncertainties on the longitudinal energy profile.
As an example, the observed variations and the resulting systematic uncertainty on the layer-by-layer energy deposition for $\Pep$ showers for the layers with the maximal variation are listed in table \ref{tab:ePlus_longProfile_scaling}.
The systematic uncertainty increases as a function of the beam momentum, since saturation of single cells is reached sooner at higher momenta and saturation scaling factors play an increasingly important role in the cell-energy reconstruction.
The layer-by-layer uncertainties in hadron showers at the same beam momentum have lower values than in the denser $\Pep$ showers as more cells contribute to the energy sum of one layer and saturation is less common. 
For hadrons, the maximal layer-by-layer systematic uncertainties range from 1.0\% at 25\,GeV to 3.5\% at 150\,GeV.

\begin{table}[t!]
\centering
\captionof{table}{
Maximal impact of the SiPM saturation scaling factor modification on the longitudinal energy profiles for $\Pep$ showers and the resulting maximal systematic uncertainty on the layer-by-layer energy deposition. 
}
\label{tab:ePlus_longProfile_scaling}
\begin{tabular}{lrrr r}
\toprule
$p_{\mathrm{beam}}$ (GeV) & \multicolumn{2}{c}{$\delta E_{\text{vis per layer, max}}$ (\%)} & Syst.\ uncertainty (\%)\\
\midrule
 & Low scaling & High scaling \\
\midrule
15  & $+3.6$ & $-2.8$  & 1.8\\
20  & $+4.8$ & $-3.7$  & 2.5\\
30  & $+6.6$ & $-5.4$  & 3.5\\
40  & $+7.5$ & $-6.1$  & 3.9\\
\bottomrule
\end{tabular}
\end{table}

In the study of the detector response discussed in sections \ref{sec:positron} and \ref{sec:hadron}, the average visible energy in a particle shower measured with the W-AHCAL, $\langle E_{\text{vis}} \rangle$, for particle showers of one particle type and beam momentum is obtained by fitting the corresponding energy sum distribution to a Gaussian function. The event's energy sum is obtained by summing up all cell energies $E_i$ for all cells $i$ which have an energy above threshold.
\\
The systematic uncertainty $\delta E_{\text{vis}}$ on the average visible energy is approximated by the quadratic sum of the layer-by-layer response variation given by
\begin{equation}
 \delta E_{\text{vis}} = \frac{\sqrt{\sum_{\ell=1}^{\ell_{\text{max}}}(E_{\ell\text{,default}}-E_{\ell\text{,modified}})^2}}{E_{\text{vis}}}
\end{equation}
where $\ell$ is the index of the layer, $E_{\ell}$ is the visible energy in layer $\ell$ as reconstructed with default or modified saturation scaling values in all cells and $\ell_{\text{max}}$ is 20 for \Pep and 38 for hadrons.
The resulting systematic uncertainties on the visible energy for the analysed $\Pep$ energies are given in table~\ref{tab:ePlus_meanEnergy_scaling}.
For the visible energy of hadron showers, in which many more cells contribute to the full shower energy, the impact is reduced and ranges from 0.4\% at 25\,GeV up to 1.5\% for 150\,GeV.

\begin{table}[t!]
\centering
\caption{Systematic uncertainty on the average visible energy of the full $\Pep$ showers due to the uncertainty on the SiPM saturation scaling factor.
}
\label{tab:ePlus_meanEnergy_scaling}
\begin{tabular}{lrrr}
\toprule
$p_{\mathrm{beam}}$ (GeV) & \multicolumn{2}{c}{$\delta E_{\text{vis}}$ (\%)} \\
\midrule
 & Low scaling & High scaling \\
\midrule
15  & $+1.4$ & $-1.1$  \\
20  & $+1.9$ & $-1.4$  \\
30  & $+2.7$ & $-2.1$  \\
40  & $+3.0$ & $-2.5$  \\
\bottomrule
\end{tabular}
\end{table}

\subsection{MIP calibration}
The MIP calibration factors have two types of uncertainties: a statistical uncertainty on the measurement itself and a systematic uncertainty associated with the method.
The statistical uncertainties are uncorrelated between the different cells. 
In~\cite{CALICE_AHCAL_emPaper} a MIP systematic uncertainty of 2\% was found, due to imperfections of the parametrisation of the MIP line shape and other effects such as binning or bias in the muon track selection. 
This uncertainty affects all cells in the same way.
\\
Muon calibration runs, which are used for a cell-wise MIP calibration as introduced in section~\ref{sec:calibration}, were recorded twice, in summer and autumn 2011. 
By comparing the analysis results after MIP calibration, such as the visible energy, obtained with the MIP calibrations of the two independent periods, the systematic uncertainty of the MIP calibration can be extracted. 
As systematic uncertainty we use the width of the  difference in the analysis results obtained with the two independent calibrations.
This yields a variation of $2.6\%$, which is uncorrelated from one cell to the other\footnote{The uncertainty of 2.6\% includes a contribution from the uncorrelated cell-to-cell variations of the MIP temperature coefficient, since we use a
representative average coefficient per layer to correct for the temperature difference between the two periods. 
Hence, the overall uncertainty can be larger than the  statistical uncertainty extracted from the MIP fit.}. 
This uncertainty has a negligible impact on the total energy sum, because the effect is reduced by a factor $\sqrt{N}$, where $N$ is the number of cells contributing to the measurement. 
However, its effect on the longitudinal shower profiles is larger, because the energy deposition in an individual layer can be dominated by few cells only, as in the case of electromagnetic showers.
The uncertainties due to the MIP calibration factors are summarised in table~\ref{tab:mipSys}.

\begin{table}[t!]
\centering
\caption{Summary of uncertainties on the average visible energy and the average energy per layer due to the MIP calibration factors.}
\label{tab:mipSys}
\begin{tabular}{llll}
\toprule
Measurement & Source & Total & Comments \\
\midrule
Average visible & 2\% from correlated systematic & $\pm2\%$\\
energy         & uncertainties (fit) & \\
\midrule
Average energy &  2\% from correlated systematic & $\pm3.3\%$ & Added in quadrature\\
per layer & uncertainties and 2.6\% from  & & due to independent\\
&  uncorrelated statistical uncertainties & & origins \\
\bottomrule
\end{tabular}
\end{table}

The MIP calibration factors are corrected for temperature variations as discussed in section~\ref{sect:tempCorrection} using MIP temperature coefficients obtained by means of a linear fit. 
The average uncertainty on the MIP temperature coefficients is 2.3\%. 
Modifying the MIP temperature correction within this uncertainty results in a negligible change in the average visible energy.
\\
Similar to the MIP calibration factors, the SiPM gain factors are also corrected for temperature variations. 
Uncertainties on the temperature correction of the SiPM gain are negligible, first, because the gain variation itself is small $-1.7\%/^\circ$C as compared to $-4.5\%/^\circ$C for the MIP value, and second, as they enter only via the saturation corrections, where uncertainties on the total number of pixels enter in a similar way and are dominant. 

\subsection{Stability of the detector response in time}
\label{sec:stability}
\begin{figure}[t!]
\centering
\begin{minipage}[c]{0.49\linewidth}
\centering
\includegraphics[width=\textwidth]{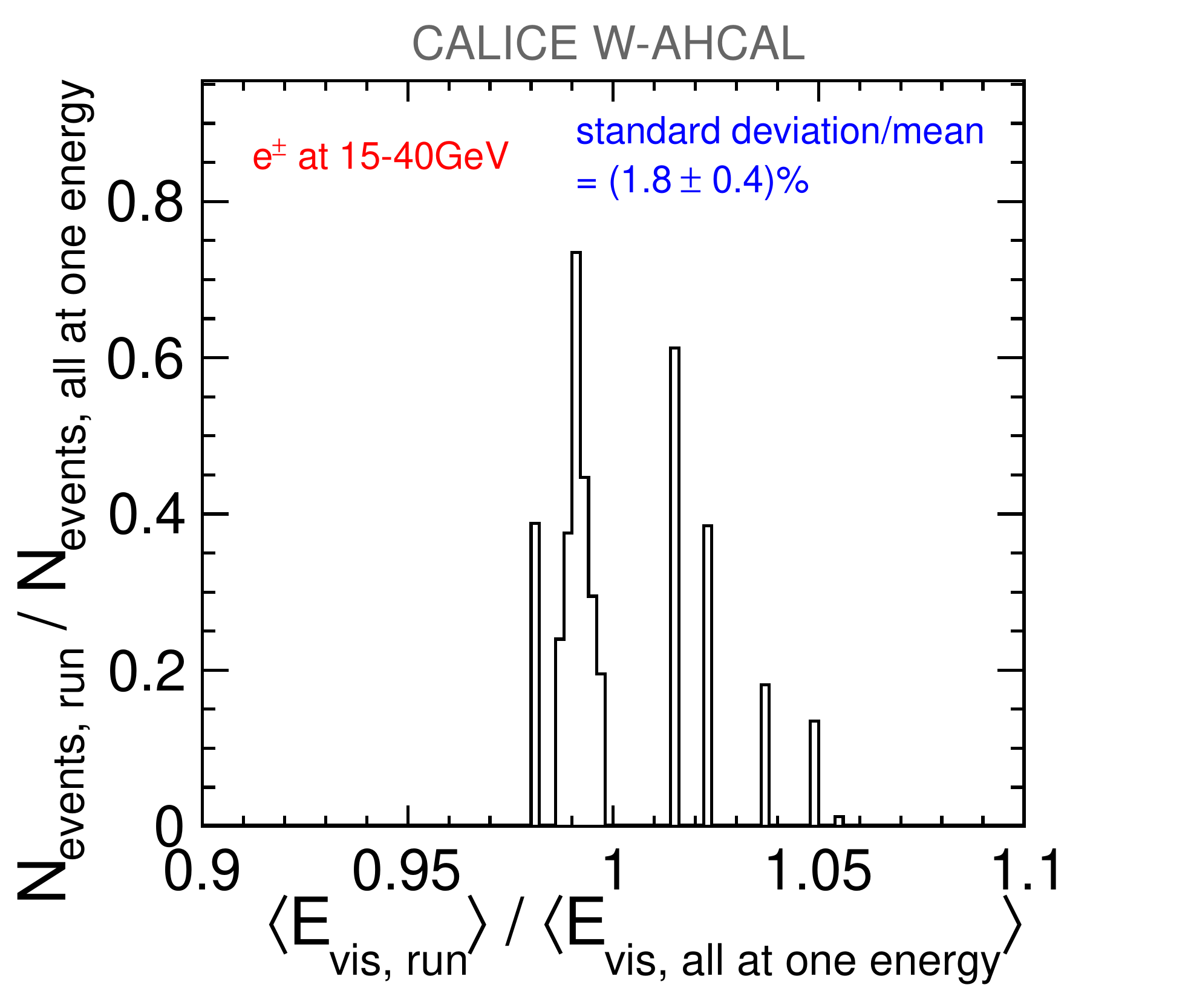}
\end{minipage}
\begin{minipage}[c]{0.49\linewidth}
\centering
\includegraphics[width=\textwidth]{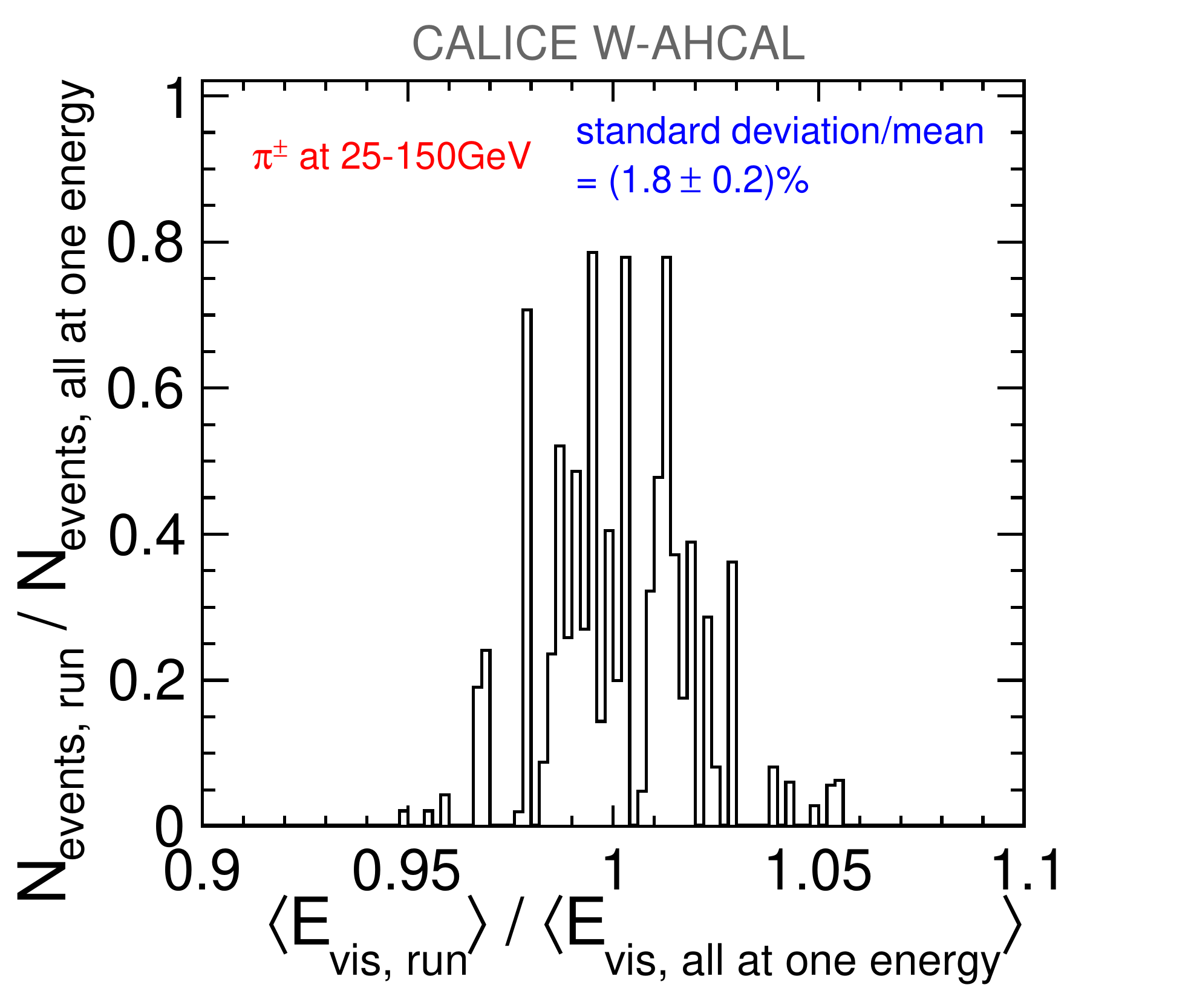}
\end{minipage}
\caption{
Histogram of the ratio of the visible energy per run over the visible energy for the full data set at one beam energy for $\Pepm$ (left panel) and $\PGppm$ (right panel).
The relative number of events per run with respect to the full data sets at one beam momentum is used as weights for the histogram.
}
\label{fig:hadronRatios}
\end{figure}

The test beam experiments of the W-AHCAL at the CERN SPS were performed in two separate data taking periods in summer and autumn separated by five months.
We observed that the detector response for data sets at the same beam energy in these two periods was slightly altered, in particular the data from negative particles were slightly shifted with respect to the data from positive particles, independently of beam momentum. 
This cannot be explained by changes in the beam properties, consequently we attribute it to a detector instability. 
The origin and full time dependence are unknown, since corrections for temperature variation have been applied, and since the shift is not reproduced in the Monte Carlo simulations, which do account, for example, for changes in the beam position, beam profile and dead cells. 
The magnitude of the variation is beyond the corresponding uncertainties.
\\
Therefore we adopted the following strategy to quantify the stability of the detector response in time: 
we study the variation of the average visible energy deposited in the calorimeter measured in one run, $\langle E_{\text{vis, run}}\rangle$, divided by the average visible energy measured using all runs at the same beam energy, $\langle E_{\text{vis, all at one energy}}\rangle$, including positive and negative polarity data. $\langle E_{\text{vis, run}}\rangle$ is obtained in all cases with a Gaussian fit function performed in the central region of the energy sum peak defined by \mbox{mean $\pm $1.5 $\sigma$}.
\\
The obtained distribution of the W-AHCAL response including runs with positive and negative polarity are shown in figure~\ref{fig:hadronRatios} for $\Pepm$ from 15\,GeV to 40\,GeV (left) and $\PGppm$ from 25\,GeV to 150\,GeV (right).
The corresponding distributions in simulations exhibit only variations of a few per mill. 
\\
In the following, the systematic uncertainty due to the observed variations of the calorimeter response is quantified based on the standard deviation of these distribution divided by the mean. For both \Pep and hadrons, the resulting systematic uncertainty due to the instability of the detector response is 1.8\%.
The uncertainties are applied to data. 

\subsection{Shower start layer reconstruction}
As discussed in section~\ref{sec:HadronSelection}, hadronic events are selected only if the shower starts within the first three \mbox{W-AHCAL} layers in order to reduce the effect of leakage.
The observables used in this paper are tested for all beam momenta with data and simulation for two different shower start selections: shower starts in layers 1--3 and in layers 2--4. 
Differences of a few percent are observed. 
In general, the simulation describes these effects to better than 1\% except for the energy resolution where the difference is 2.1\%. 
The observed disagreement between data and simulation is used as systematic uncertainty of the shower start selection. 
The systematic uncertainty is applied to data.

\clearpage
\subsection{Selection purity}
\label{sec:noise}
The analysed hadron data sets contain mixed muon, pion, kaon, and proton events. 
Events from a given particle type are selected using Cherenkov threshold counters and information based on shower shapes.
While the muon contamination is efficiently reduced using calorimeter information, the distinction between pions, kaons, and protons is solely based on Cherenkov counters.
The purity of the selected hadron samples was determined in~\cite{LCD-Note-2013-006}.
For this analysis, only runs which have a purity between 85\% and 100\% for a given hadron type are considered. 
An analysis of shower shapes in the Fe-AHCAL \cite{CAN040} showed that for a particle purity above 85\% the systematic uncertainty on the shower shapes is at most 1.5\% on the shower radius and 1\% on other observables.
Since the profiles in iron and tungsten are similar, one may assume that the systematic uncertainties due to the particle impurity are negligible in comparison to the already discussed systematic uncertainties.
Therefore no systematic uncertainties due to possible contaminations are considered.

\subsection{Systematic uncertainties of the simulations}
\paragraph{Optical cross-talk between tiles}
Due to the imperfect reflective coating of the scintillator tiles, light may leak between neighbouring calorimeter cells. 
This is taken into consideration in the simulation using the cross-talk factor, which is the fraction of light for which leakage will appear in neighbouring cells. 
Measurements of the cross-talk found values of 2.5\%~\cite{CALICE_AHCAL_emPaper}, and between 3.3\% and 4.6\% \cite{Clemens_Thesis} of the total energy per edge for the $3\times3$~cm$^2$ cells. 
\\
In the simulation, the cross-talk factor is relevant as it sets the energy scale. 
The higher this factor, the higher the final energy is on the MIP scale, since unlike for muon calibration events, in showers most of the signal leaks into neighbouring tiles which already have signals above threshold.
\\
To account for the imperfect knowledge of the cross-talk, simulations of positron- and hadron-induced showers are performed with cross-talk values corresponding to the lowest and the highest cross-talk measurements of 2.5\% and 4.6\%. 
The difference in the response divided by $\sqrt{12}$ is used as an estimate for the systematic uncertainty due to the cross-talk.
For both positrons and hadrons, a systematic uncertainty of 2.7\% was found for the average visible energy.
For the radial energy profiles studied in hadron-induced showers, a maximum systematic uncertainty due to cross-talk of $4.2\%$ was found.

\paragraph{Timing cut}
As introduced in section~\ref{sec:simulation}, a timing cut of 150\,ns is applied on simulated hits in order to emulate the signal
shaping time of the CALICE readout electronics. 
Due to differences in the SiPM production series used for the AHCAL, cell-by-cell differences in the signal shaping time of up to 15\,ns were observed when optimising the readout time window of single cells.
For the \mbox{Fe-AHCAL}, it was found that changing the timing cut within this observed variations of 15\,ns has very small effects (at the per mill level) on the reconstructed hadron shower energy~\cite{Lutz2010}.
Because tungsten is used as absorber, a rather large fraction of the signal is produced by late spallation neutrons~\cite{Adloff:2014rya}.
While most of the energy is deposited within the first nanoseconds, there are also significant late energy deposits. 
To estimate the effects due to the timing cut, variations of $\pm 30$\,ns were considered, resulting in variations of the measured energy in the simulation of $\pm0.3\%$, which is negligible.

\clearpage
\subsection{Summary of uncertainties}
In this paper, an extensive study of systematic uncertainties of data and simulations is performed and summarised for contributions to the average visible energy $\langle E_{\text{vis}}\rangle$ in table~\ref{tab:sys}.
The total uncertainty for a given data set is obtained by adding all relevant contributions in quadrature.

\begin{table}[t!]
\centering
\caption{Systematic uncertainties on the W-AHCAL average visible energy $\langle E_{\text{vis}}\rangle$ for the beam momentum range of 15\,GeV--40\,GeV for $\Pep$ and 25\,GeV--150\,GeV for $\PGpp$, $\PKp$, and protons.}
\label{tab:sys}
\begin{tabular}{lrrr}
\toprule
Source                 & \multicolumn{2}{c}{--- Systematic uncertainty on $\langle E_{\text{vis}}\rangle$ ---} & Assigned to  \\
                       &  for \Pep (\%)& for \PGpp, \PKp, and protons (\%)            &  \\
\midrule 
SiPM saturation scaling         & 1.4--3.0       &  0.4--1.5               & data \\
MIP constants          		& 2.0            &  2.0                    & data \\
Detector stability     		& 1.8            &  1.8                    & data \\
Shower start           		& -              & 0.1                     & data \\
Cross-talk             		& 2.7            & 2.7                     & MC   \\
\midrule
Quadratic sum for data 		& 3.1--4.0       & 2.7--3.1                       \\
MC   				& 2.7            & 2.7                            \\
\bottomrule
\end{tabular}
\end{table}
\section{Analysis of positron data}
\label{sec:positron}
In the following section, the analysis of positron-induced showers in the W-AHCAL is discussed.
As Geant4 simulations of electromagnetic showers are well established, the analysis of the $\Pep$ showers permits a validation of the detector calibration and implementation of the detector simulation.
Due to the limited penetration depth of positrons this is only possible for the first $\sim$10 layers of the hadronic calorimeter.
In addition, the W-AHCAL response to positrons is compared to the hadron responses, discussed in section \ref{sec:ComparisonOfResponse}, to study the degree of compensation of the calorimeter.

\begin{figure}[t!]
\centering
\includegraphics[width=0.46\textwidth]{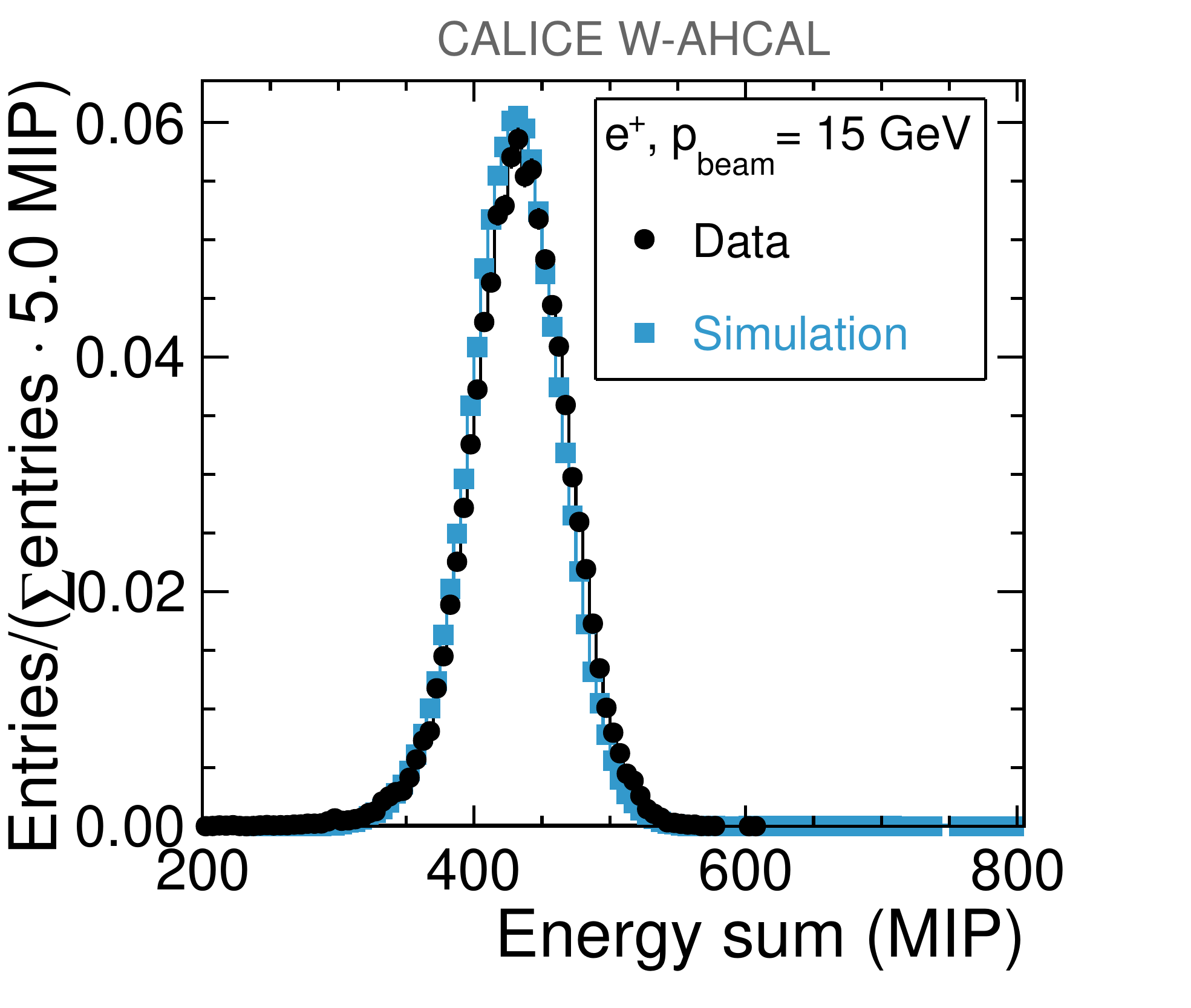}\hfill
\includegraphics[width=0.46\textwidth]{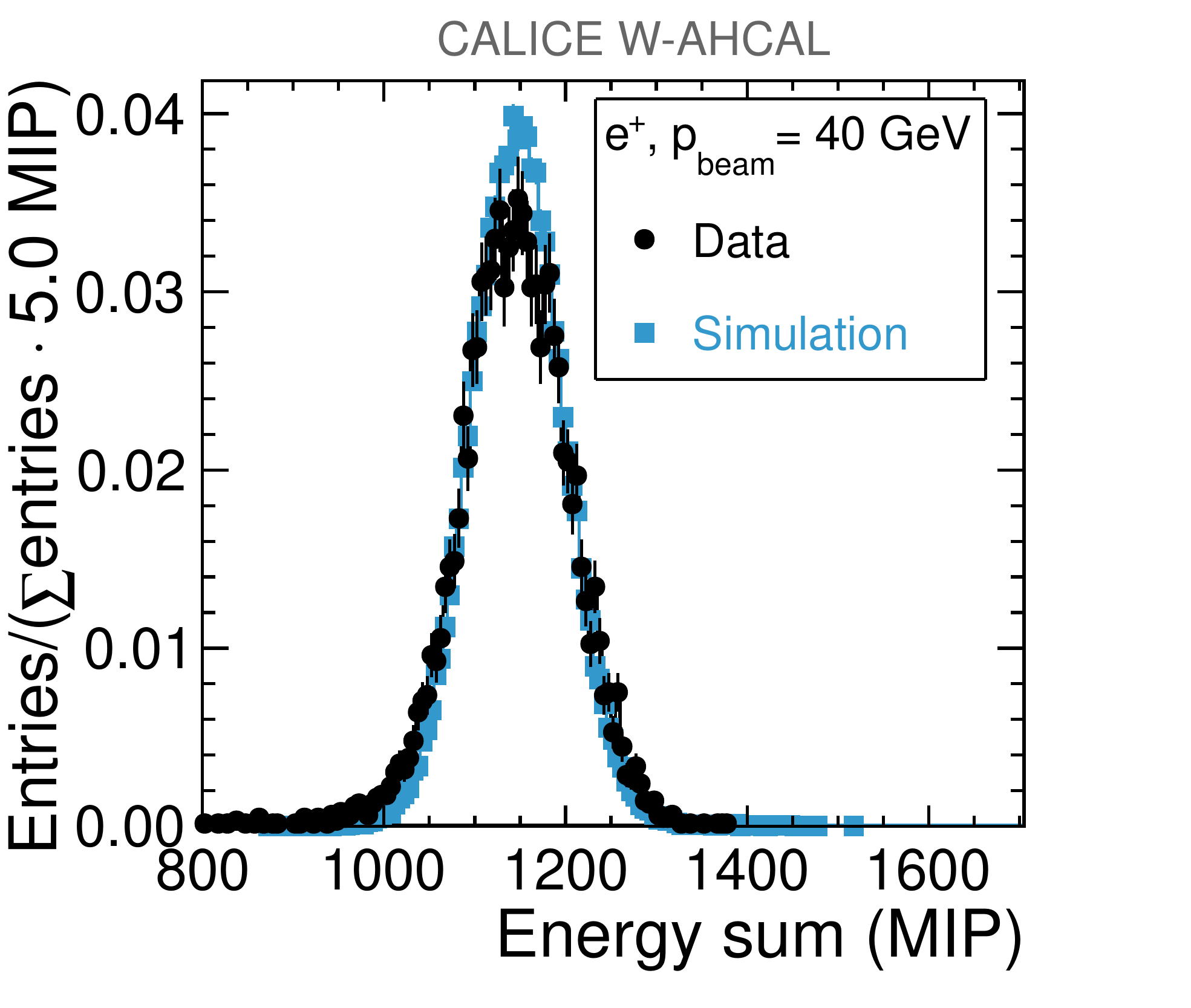}
\caption{Comparisons of the energy sum distribution of 15\,GeV (left) and 40\,GeV (right) $\Pep$ data with simulation.
}
\label{fig:ePlus_mcComparison}
\end{figure}

\begin{figure}[t!]
\centering
\includegraphics[width=0.46\textwidth]{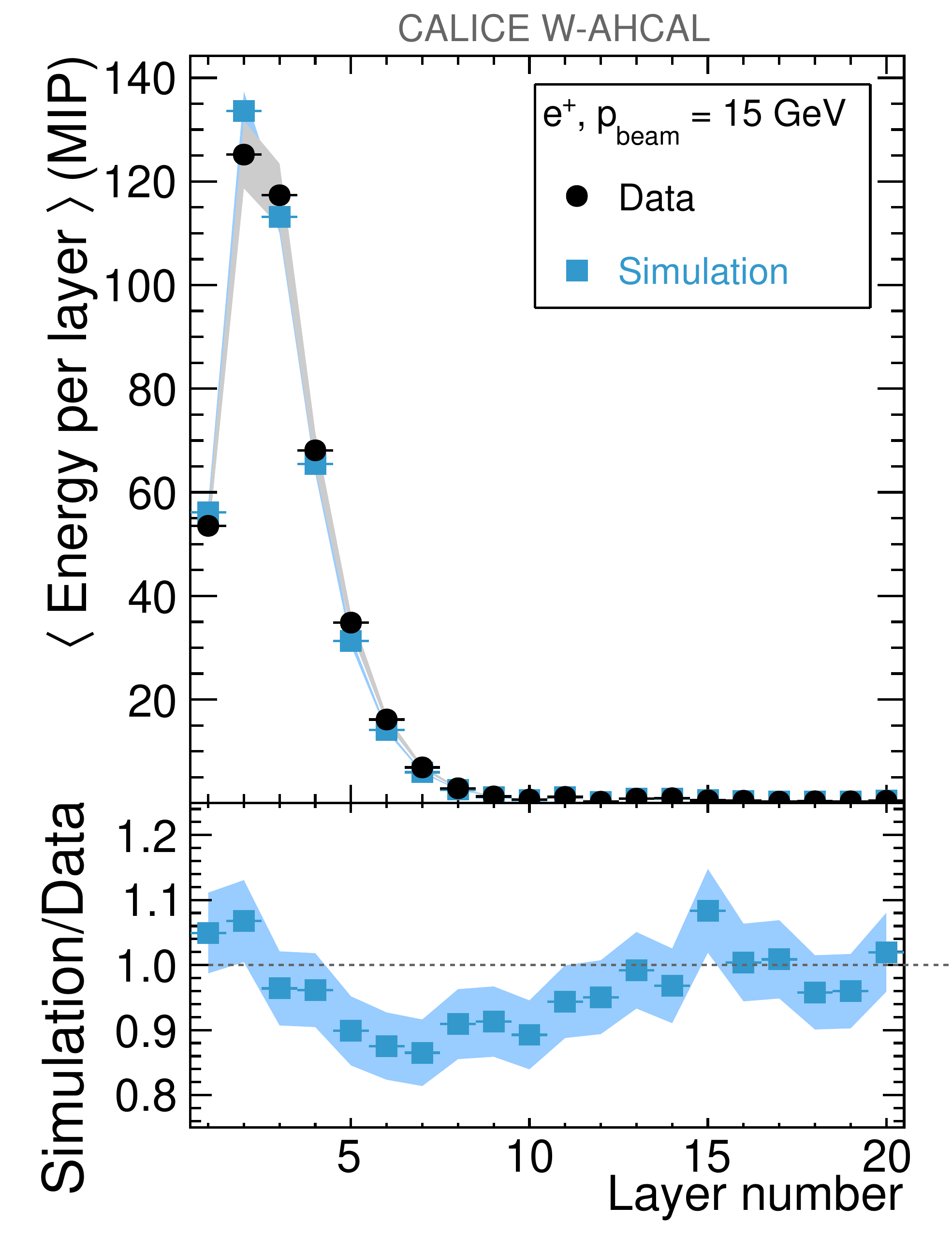}\hfill
\includegraphics[width=0.46\textwidth]{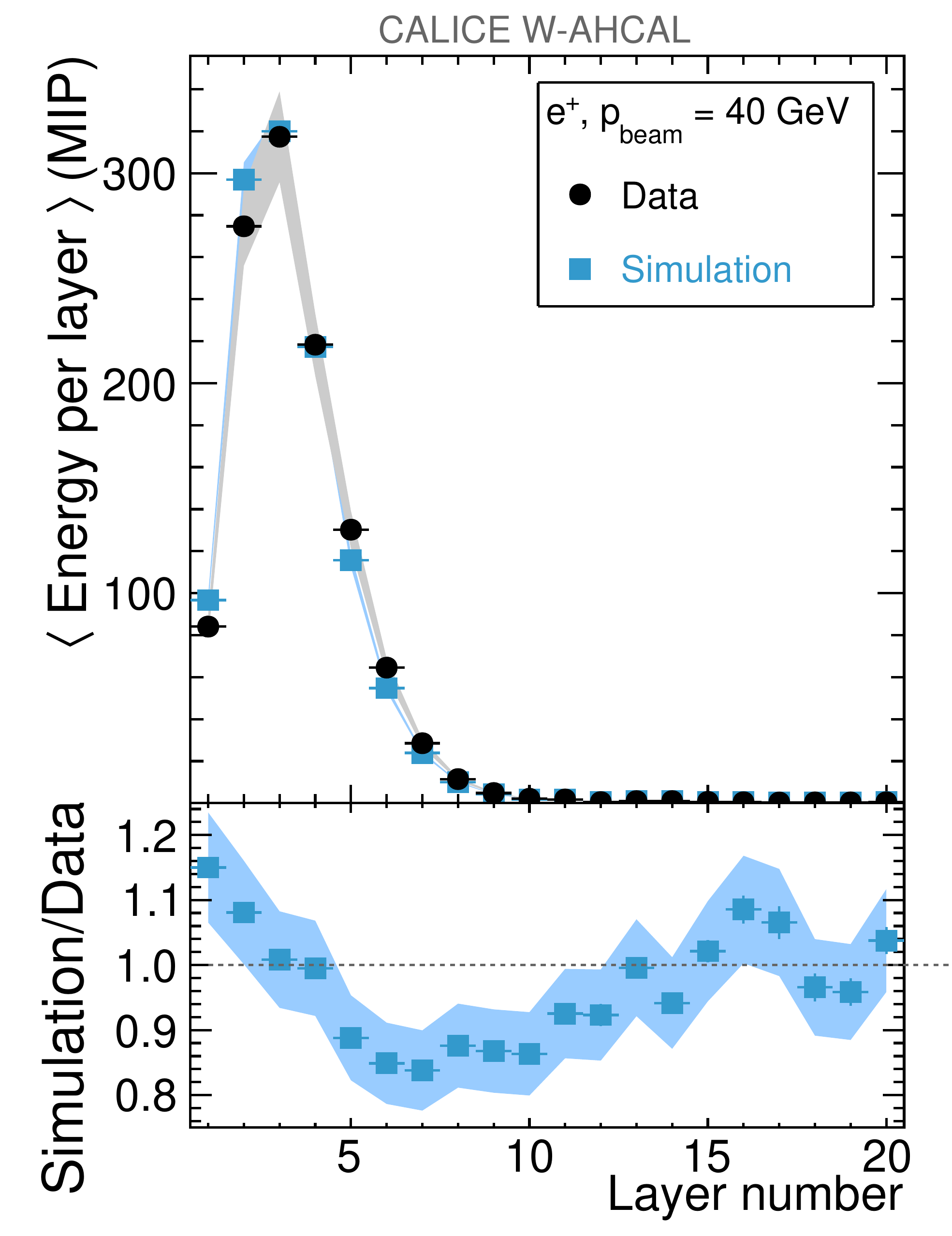}
\caption{Comparisons of the longitudinal energy profiles of 15\,GeV (left) and 40\,GeV (right) $\Pep$ data with simulation.
Here and in the following figures, the uncertainty bands include statistical and systematic uncertainties while the data points only show statistical uncertainties.
}
\label{fig:ePlus_mcComparison_long}
\end{figure}

Figures~\ref{fig:ePlus_mcComparison} and \ref{fig:ePlus_mcComparison_long} show the energy sum distributions and the longitudinal profiles (i.e., the distribution of the average energy deposited in a given calorimeter layer as a function of the layer number) for 15\,GeV and 40\,GeV positrons for data and simulations.
The energy sum distributions of data and simulation agree reasonably well with each other.
In the longitudinal profiles, discrepancies of up to 15\% between data and simulation are observed.
These are assumed to be due to missing individual measurements of the SiPM saturation scaling factors, an imperfect description of material in the beam line, and uncertainties in Geant4 modelling.

\begin{figure}[t!]
\begin{minipage}[t]{0.46\textwidth}
\includegraphics[width=\textwidth]{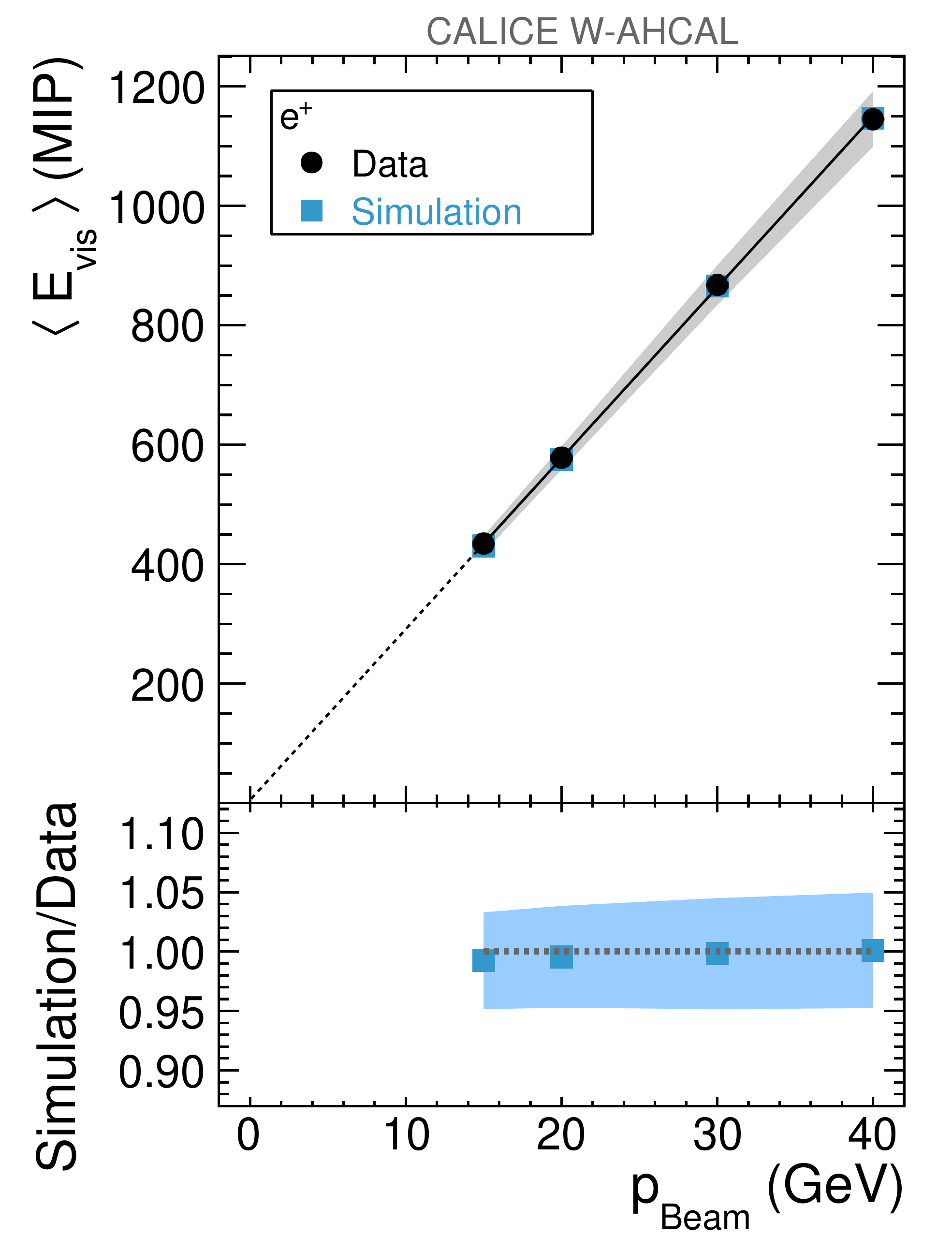}
\end{minipage}
\hfill
\begin{minipage}[t]{0.46\textwidth}
\includegraphics[width=\textwidth]{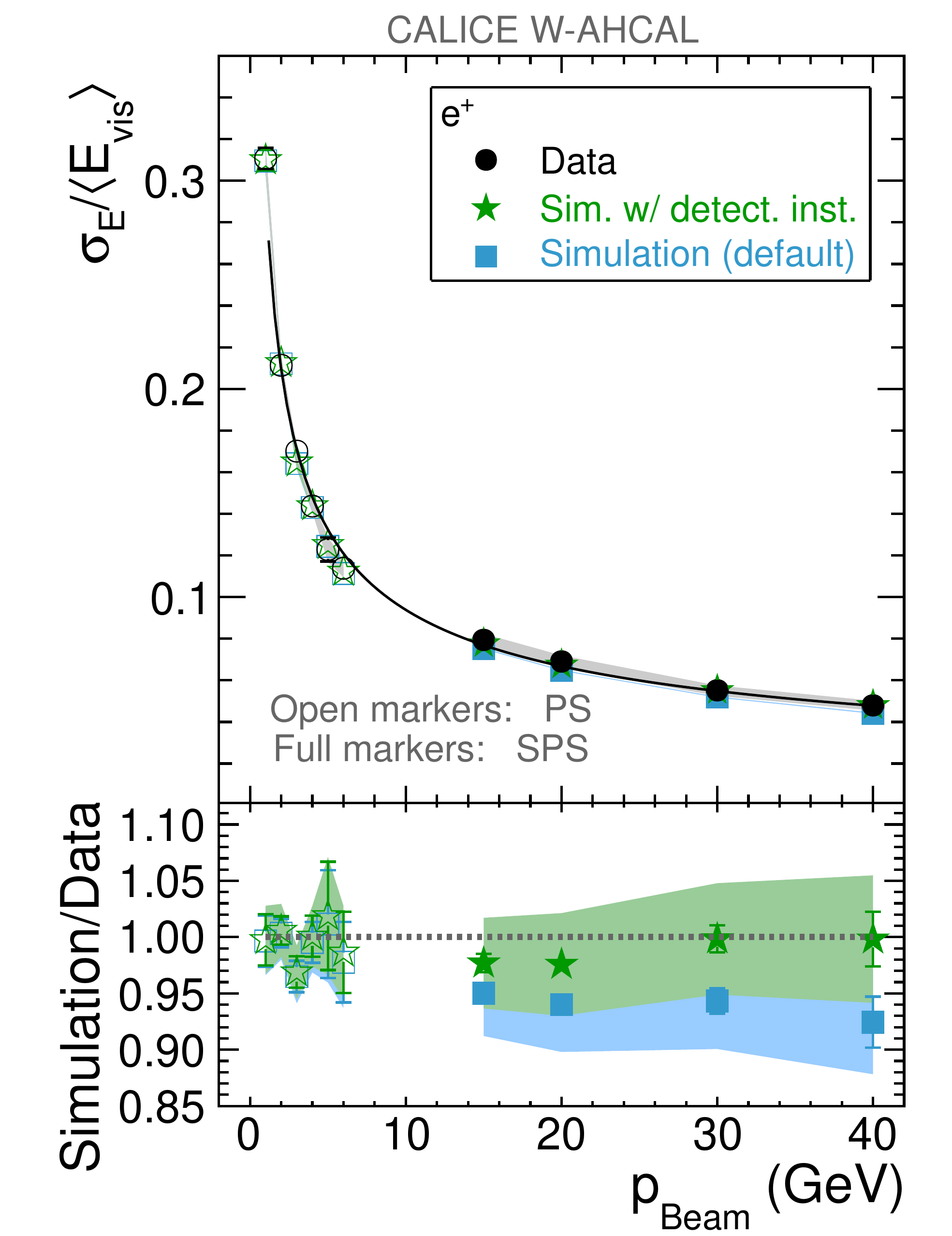}
\end{minipage}
\caption{
Average visible energy (left) and energy resolution (right) as a function of the beam momentum for positron-induced showers for data and simulation. 
The black lines indicate fits to the data points without constraints.
The bands show the overall uncertainty.  
For the energy resolution, re-analysed results from the corresponding PS data analysis \cite{CALICE-WAHCAL-2010} are shown, as well as results from simulations with and without (default) detector instability (see text for definition).
} 
\label{fig:ePlus_linearity_resolution}
\end{figure}

The average $\Pep$ visible energy $\langle E_{\text{vis}} \rangle$ and the width of the energy sum distribution $\sigma_E$ divided by the visible energy, i.e.\ the energy resolution, shown in figure~\ref{fig:ePlus_linearity_resolution}, are obtained by fitting the $\Pep$ energy sum distributions to Gaussian functions.
The fit has been performed in the central region of the energy sum peak defined by the $\text{mean} \pm 1.5\;\sigma$.
The average visible energy is given by the mean of the Gaussian fit function and the width of the energy sum distribution is given by the sigma of the Gaussian fit function.
\\
A minimum $\chi^2$ is used for fitting the calorimeter energy distribution and their energy dependence.
The uncertainties of the parameters are obtained from the covariance matrix after the $\chi^2$~function has been normalised at the fit minimum to $\chi_{\mathrm{min}}^2 = \text{NDF}$ (number of degrees of freedom) by scaling the uncertainties by the factor~$\sqrt{\chi_{\text{not normalised}}^2/\text{NDF}}$ \footnote{
This method may incur problems if applied blindly \cite{Press:1058314};
here it is employed with due attention both for fitting the energy sum distribution and the energy resolution parameters $a$ and $b$, listed in tables \ref{tab:positronResolution}, \ref{tab:pionResolution} and \ref{tab:protonResolution}.
}.
\\
In the left panel of figure~\ref{fig:ePlus_linearity_resolution} the average visible energy is shown as a function of the beam momentum both for data and simulation.  
The systematic uncertainties are evaluated according to table \ref{tab:sys}.
The experimental data for the visible energy agree well with the Monte Carlo simulations within the uncertainties, the deviations being less then 1\%.
The increase of $\langle E_{\text{vis}}\rangle$ with beam momentum is in good agreement with a straight line fit shown for data as a solid line.
The extrapolation of this fit to $p_{\mathrm{Beam}}=0$\,GeV is shown as a dashed line.
\\
The W-AHCAL $\Pep$ energy resolution is displayed in the right panel of figure~\ref{fig:ePlus_linearity_resolution}. 
To better constrain the energy resolution fit results, the low energy data from the PS test beam campaign, described in~\cite{CALICE-WAHCAL-2010}, are included using the SPS selection cuts described in section~\ref{sec:emSelection}.
As discussed in section \ref{sec:simulation}, the simulation includes the effects of intrinsic calorimeter fluctuations as well as instrumental effects such as photo-electron statistics and noise.
The detector instability estimated from the run-per-run variation of the calorimeter response in data (section~\ref{sec:stability}), however, is not included in the simulation.
This is referred to as default simulation in figure~\ref{fig:ePlus_linearity_resolution}.
In order to achieve a more realistic modelling of the data in terms of the energy resolution, the detector instability needs to be accounted for in the simulated energy resolution. 
This can be achieved by adding the observed effect of the detector instability on $\langle E_{\text{vis}}\rangle$ from data in quadrature to the simulated positron energy resolution.
After taking the detector instability into account, the energy resolution measured for positrons is reproduced by the simulations within uncertainties.
\\
The combined data from the PS and the SPS test beam campaigns are fitted to
\begin{equation}
\frac{\sigma_E}{\langle E_{\text{vis}}\rangle} =\frac{a}{\sqrt{E\;[\mathrm{GeV}]}} \oplus b \oplus \frac{c}{E\;[\mathrm{GeV}]},
\label{eq:resFit}
\end{equation}
where $a$ is the stochastic term and $b$ is the constant term, which is dominated by the stability of the calibration;
$c$ is the noise term fixed to the noise in the fiducial volume of the $\Pep$ showers corresponding to 35\,MeV, measured using dedicated random trigger events.
The resulting energy resolution fit for data is displayed as a solid black line in figure \ref{fig:ePlus_linearity_resolution}.
The parameters of the positron energy resolution fits for data and simulation including the detector instability are listed in table~\ref{tab:positronResolution}.
The indicated uncertainties include statistical as well as systematic uncertainties as introduced in section \ref{sec:systematics}.
The stochastic and constant terms derived from data and Monte Carlo agree within the uncertainties. 
The W-AHCAL energy resolution fit results are in agreement with those found using only the PS data sets~\cite{CALICE-WAHCAL-2010}. 
The observed $\Pep$ energy resolution is slightly larger than that found for the Fe-AHCAL with a stochastic term of approximately 22\% \cite{CALICE_AHCAL_emPaper} with beam momenta from 10\,GeV to 50\,GeV.
The larger value of $a$ for the W-AHCAL in comparison to the Fe-AHCAL is expected due to the larger number of $X_0$ per layer in tungsten (2.8\,$X_0$) in comparison to steel (1.2\,$X_0$).

\begin{table}
\centering
\caption{Parameters of the $\Pep$ energy resolution fits for data and simulation using beam momenta from 1\,GeV to 40\,GeV.
The simulated results are obtained after including the detector instability measured in data.
}
\label{tab:positronResolution}
\begin{tabular}{lrr}
\toprule
Parameter                         & Data           & Simulation     \\
\midrule
$a$ ($\%\cdot\sqrt{\text{GeV}}$)  & $29.5 \pm 0.4$ & $28.7 \pm 0.5$ \\
$b$ (\%)                          & $1.1 \pm 1.1$  & $1.7 \pm 0.6$  \\
$c$ (GeV)                         & $0.035$        & $0.035$        \\
\bottomrule
\end{tabular}
\end{table}

In summary, within uncertainties, the electromagnetic data agree well with the simulation for response, linearity, and energy resolution.
For the longitudinal energy profiles layer-by-layer deviations of up to 15\% are observed.

\section{Analysis of hadron data}
\label{sec:hadron}
In the following section, the analysis of hadron-induced showers in the W-AHCAL is discussed and the results are compared to Geant4 simulations.
First, the properties of pion-induced showers are studied, followed by shower properties for protons and kaons.
\\
For brevity, shower properties describing the evolution of hadron showers are presented only for pion-induced showers. 
The shower evolution of pion, kaon, and proton showers are consistent to within 20\%.
The agreement between data and simulation is at the same level for all three particle types.
\\
In order to take particles' mass and decay properties into account, the observables are presented as a function of the available hadron shower energy, $E_{\mathrm{available}}$, given by
\begin{eqnarray}
E_{\mathrm{available}} &=& \sqrt{m_{\text{meson}}^2 + p_{\mathrm{beam}}^2} \quad \quad \text{for \PGpp and \PKp, and}\label{eq:pionavailableE}\\
E_{\mathrm{available}} &=& \sqrt{m_{\text{baryon}}^2 + p_{\mathrm{beam}}^2}-m_{\text{baryon}} \quad \quad \text{for protons},\label{eq:protonavailableE}
\end{eqnarray}
where $m_{\text{meson}}$ and $m_{\text{baryon}}$ are the masses of the corresponding meson or baryon and $p_{\text{beam}}$ is the beam momentum.

\subsection{The pion data}
\label{sec:piAnalysis}
In the following, the calorimeter response, the energy resolution, and variables describing longitudinal and radial shower development of pion-induced showers are studied and compared to Monte Carlo simulations. 
The purity of the pion samples selected based on Cherenkov threshold counters is between 93\% and 100\%~\cite{LCD-Note-2013-006}.

\subsubsection{Calorimeter response and energy resolution}
\begin{figure}[t!]
\centering
\includegraphics[width=0.46\textwidth]{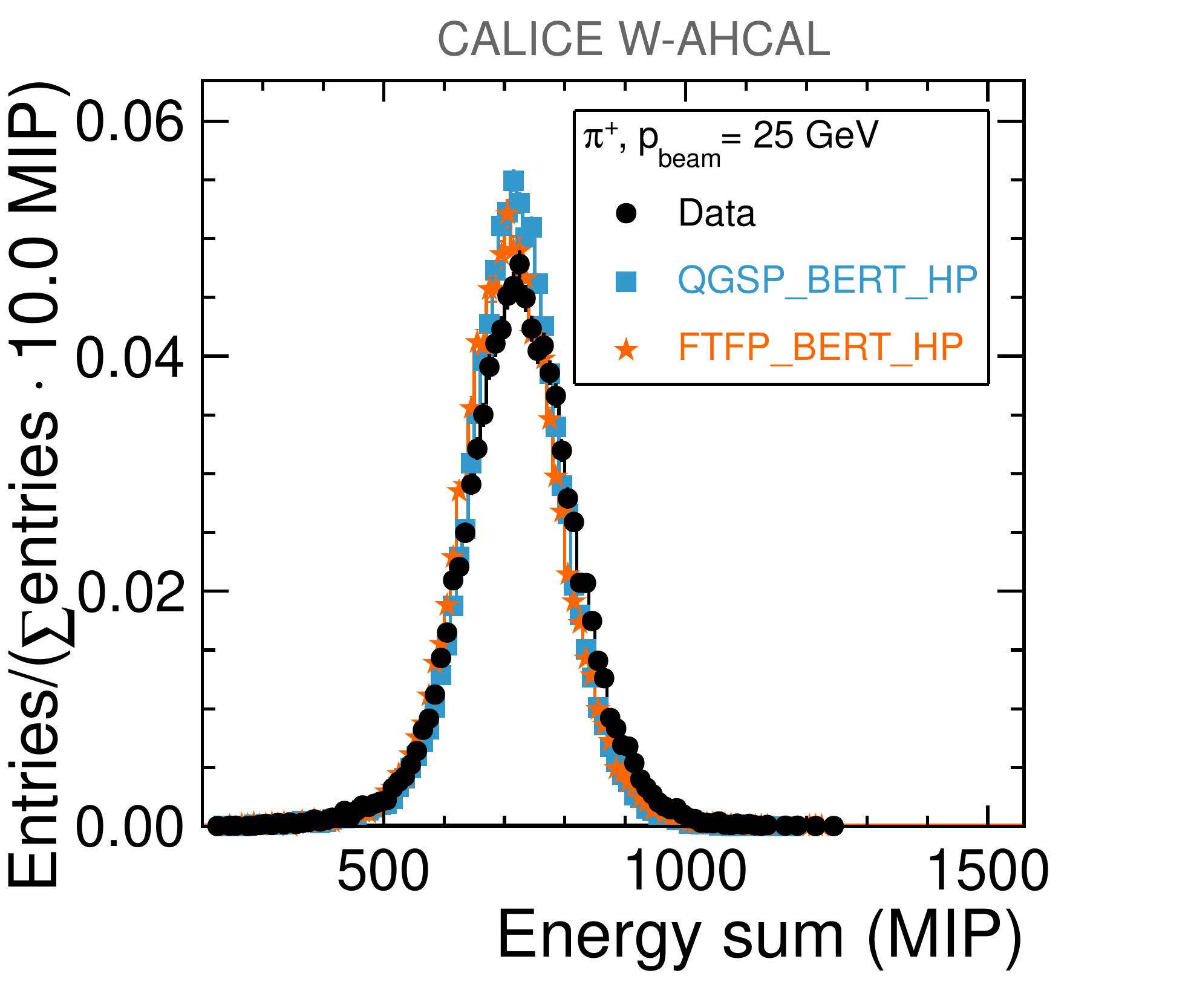}
\hfill
\includegraphics[width=0.46\textwidth]{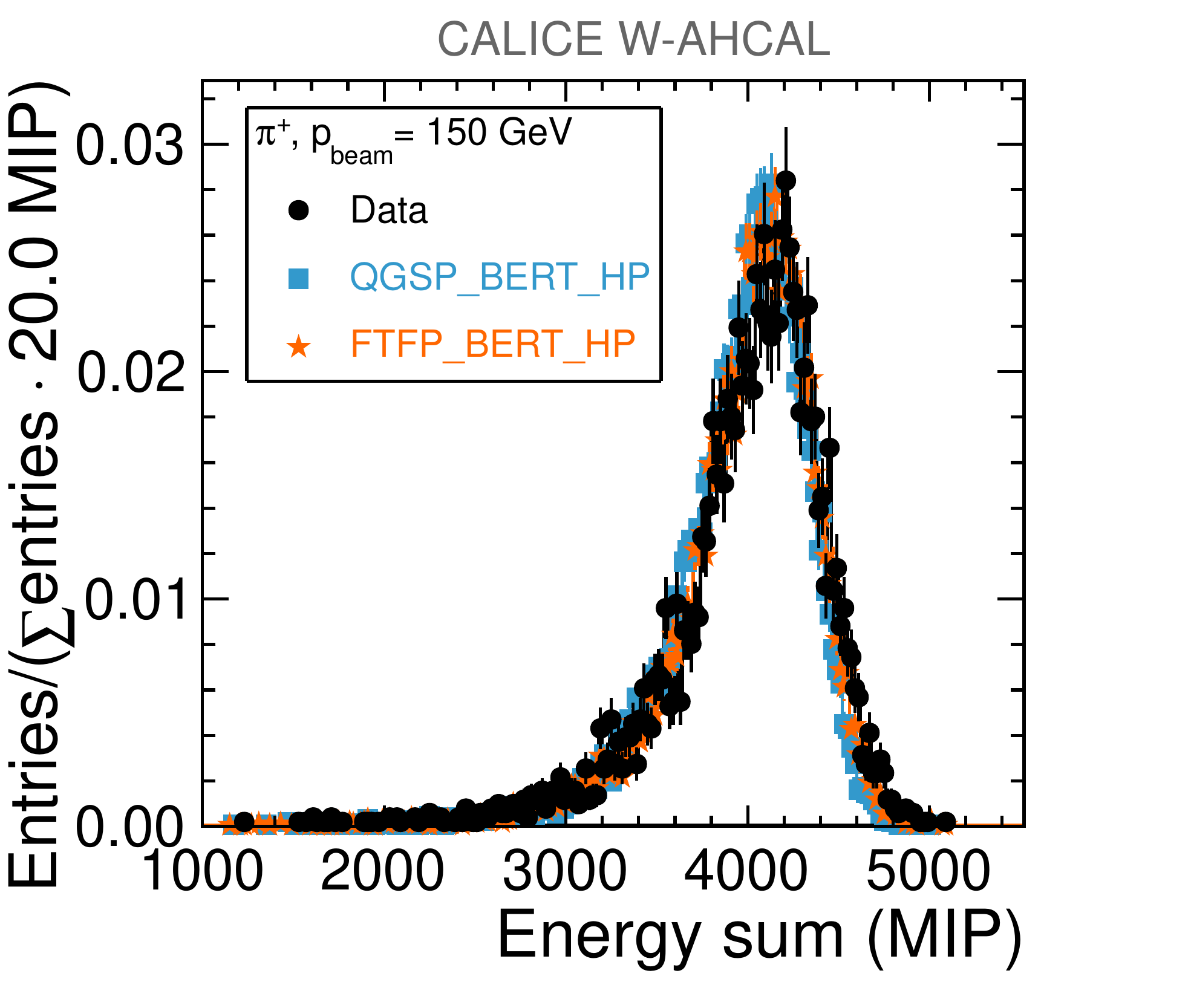}
\caption{Energy sum distributions of \PGpp events at 25\,GeV (left) and 150\,GeV (right).  
The data are compared to selected Geant4 physics lists.
}
\label{fig:piPlusEsumMC}
\end{figure}

In figure~\ref{fig:piPlusEsumMC}, the energy sum distribution for 25\,GeV and 150\,GeV $\PGpp$ events are compared to the Geant4 physics lists QGSP\_BERT\_HP and FTFP\_BERT\_HP. 
Both physics lists describe the shapes fairly well.
In the 150\,GeV distributions of data and simulations, a tail towards lower energies indicates leakage effects in the calorimeter as discussed in section \ref{sec:HadronSelection}.

\begin{figure}[t!]
\begin{minipage}[t]{0.47\linewidth}
\centering
\includegraphics[width=\textwidth]{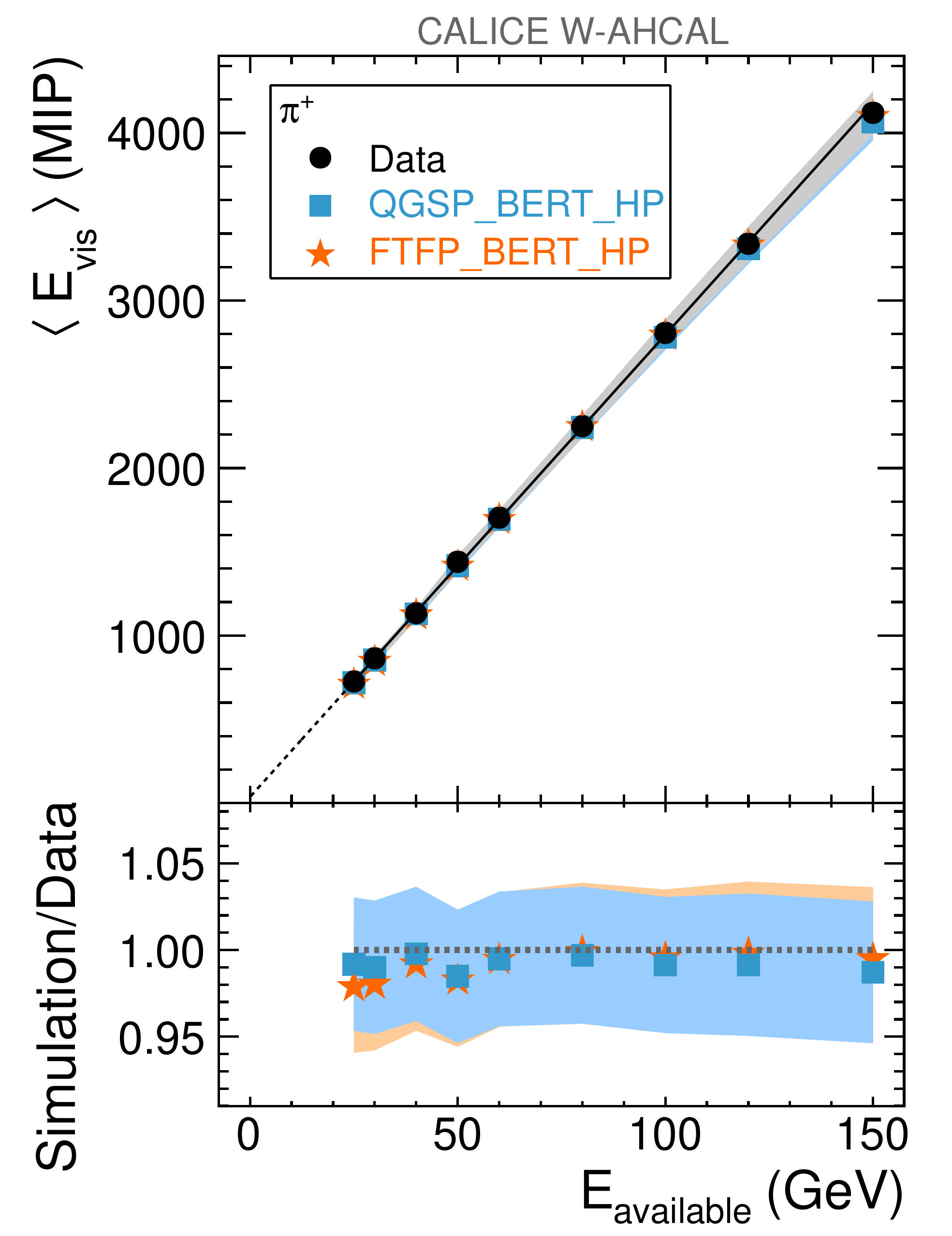}
\end{minipage}
\hfill
\begin{minipage}[t]{0.47\linewidth}
\centering
\includegraphics[width=\textwidth]{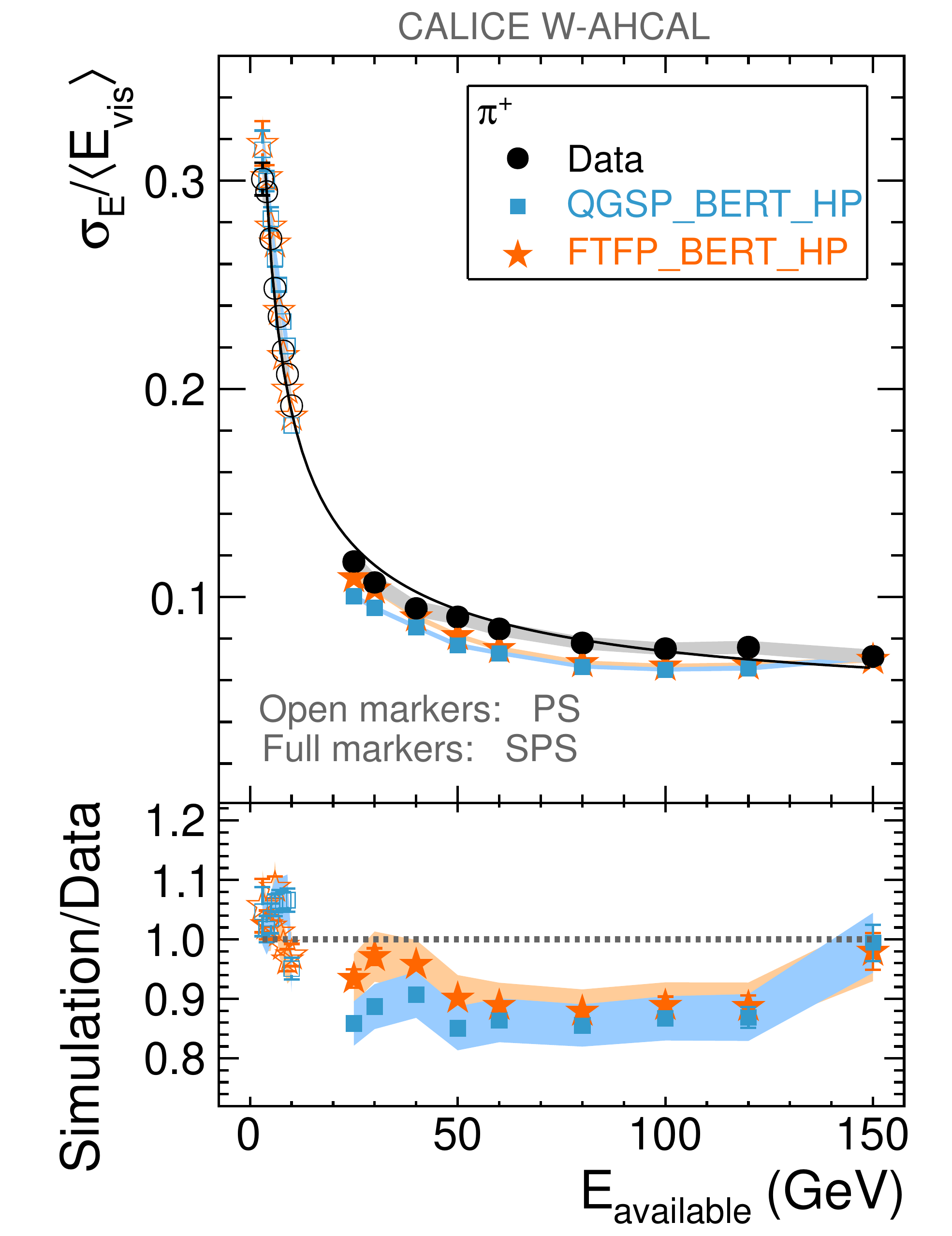}
\end{minipage}
\caption{Average visible energy as a function of the available energy (left) and energy resolution (right) for $\PGpp$ induced showers. 
For the energy resolution, data points from the corresponding PS data analysis \cite{CALICE-WAHCAL-2010} are shown.
The black lines indicate fits to the data points.
The data are compared to selected Geant4 physics lists.
The bands show the overall uncertainties. 
}
\label{fig:piLinearityResolution}
\end{figure}

The average visible energy $\langle E_{\text{vis}} \rangle$ deposited in the calorimeter by \PGpp showers is estimated using Gaussian fits to the energy sum distributions as described in section \ref{sec:positron}.
$\langle E_{\text{vis}} \rangle$ is shown in the range from 25\,GeV to 150\,GeV as a function of the available energy in figure~\ref{fig:piLinearityResolution}~(left) where the available energy for $\PGpp$ showers is given by equation \ref{eq:pionavailableE}.
The solid line indicates a linear fit to the data and shows that the response for pion-initiated showers is linear within the uncertainties. 
The data are compared to Geant4 physics lists. 
A good agreement of $\pm2\%$ between data and QGSP\_BERT\_HP and FTFP\_BERT\_HP is observed which is covered by the systematic uncertainties of the measurement.
At beam momenta below 60\,GeV, however, both Geant4 physics lists tend to slightly underestimate the data.

\FloatBarrier

\begin{table}[t!]
\centering
\caption{
Parameters of the $\PGpp$ energy resolution fits for data and simulations using beam momenta from 3\,GeV to 150\,GeV.
The simulated results are obtained after including the detector instability measured in data.
}
\label{tab:pionResolution}
\begin{tabular}{lrrr}
\toprule
Parameter                        & Data            & QGSP\_BERT\_HP          & FTFP\_BERT\_HP\\
\midrule
$a$ ($\%\cdot\sqrt{\text{GeV}}$) & $57.9\pm 1.1$   & $51.1\pm 2.8$           & $54.6\pm 2.0$ \\
$b$ (\%)                         & $4.6\pm  0.4$   & $3.9\pm 0.7$            & $3.8\pm 0.5$  \\
$c$ (GeV)                        & $0.065$         & $0.065$                 & $0.065$       \\
\bottomrule
\end{tabular}
\end{table}

The relative energy resolution for the $\PGpp$ showers in the energy range from 3\,GeV to 150\,GeV is presented in the right panel of figure~\ref{fig:piLinearityResolution}. 
The simulation includes the detector instability, using the procedure established in section~\ref{sec:positron} and validated with positrons. 
The PS data were selected based on the new selection cuts described in section~\ref{sec:HadronSelection}.
In the combination of the PS and SPS data for the energy resolution fit, the difference of the response between 30 and 38 layers has been taken into account by introducing a systematic uncertainty of 2.0\% in the PS energy range.
In the SPS beam momentum range, except at 150\,GeV, the Monte Carlo simulations underestimate $\sigma_E/\langle E_{\text{vis}}\rangle$ by 3\%--12\% for FTFP\_BERT\_HP and by 10\%--15\% for QGSP\_BERT\_HP.
\\
The energy resolutions for data and simulations are fitted using the energy resolution function defined in equation~\ref{eq:resFit}.
The noise term $c$ has been fixed to 65\,MeV, corresponding to the electronic noise in the full calorimeter volume. 
The resulting parameters of the $\PGpp$ energy resolution fits for all data sets are listed in table \ref{tab:pionResolution}.
The stochastic term found for both physics lists underestimates the value found in data.
The fit results of the stochastic term for the experimental data are slightly lower than those found in the corresponding PS data analysis \cite{CALICE-WAHCAL-2010}.
This is due to the fact that for the PS study, the $\sigma_E/\langle E_{\text{vis}}\rangle$ was estimated using the standard deviation and the mean of the energy sum distribution.
These are more affected by the tails of the distribution than the parameters of the Gaussian fit functions which are used here for both PS and SPS data.
The energy resolution fit results of the W-AHCAL agree with the corresponding fit for Fe-AHCAL combined with a steel tail catcher of in total 12\,$\lambda_{\text{I}}$ obtained with beam momenta from 10\,GeV to 80\,GeV, with a stochastic term of approximately 58\%~\cite{Adloff:2012gv}.

\subsubsection{Spatial development}
\label{sec:had:long}
In order to assess the accuracy of the simulation of the spatial development of hadronic showers, we compare, for data and simulation results, the shower development along the $z$-axis (longitudinal) and in the $xy$-plane (radial).

\paragraph{Longitudinal shower development.}
The development of hadron showers along the beam direction ($z$-axis) is shown in longitudinal profiles, using as a reference the measured shower start.
Thus fluctuations of the shower start are disentangled from the intrinsic longitudinal shower development. 
The layer in which the shower started is identified using the clustering algorithm described in~\cite{Adloff:2013mns}.
We found that data and Monte Carlo at all beam momenta agree in terms of the reconstructed, exponentially falling shower start layer distribution within 2\% for shower starts in the most relevant first 15 layers.
\\
Longitudinal profiles might be particularly sensitive to showers starting close to the entrance of the calorimeter. 
Therefore, an alternative event selection is used here. 
Only events starting in layer four or later are considered in order to ensure that the showers start in the bulk of the calorimeter.
Uncertainties due to layer-by-layer fluctuations of the calorimeter response generated by uncertainties in the calibration and saturation effects are taken into account, in addition to the systematic uncertainties discussed in section~\ref{sec:systematics}.  
The uncertainty due to layer-by-layer fluctuations is estimated by the difference between the effect observed in data and simulations following the procedure described in~\cite{Adloff:2013mns}. 

\begin{figure}[t!]
\centering
\includegraphics[width=0.46\textwidth]{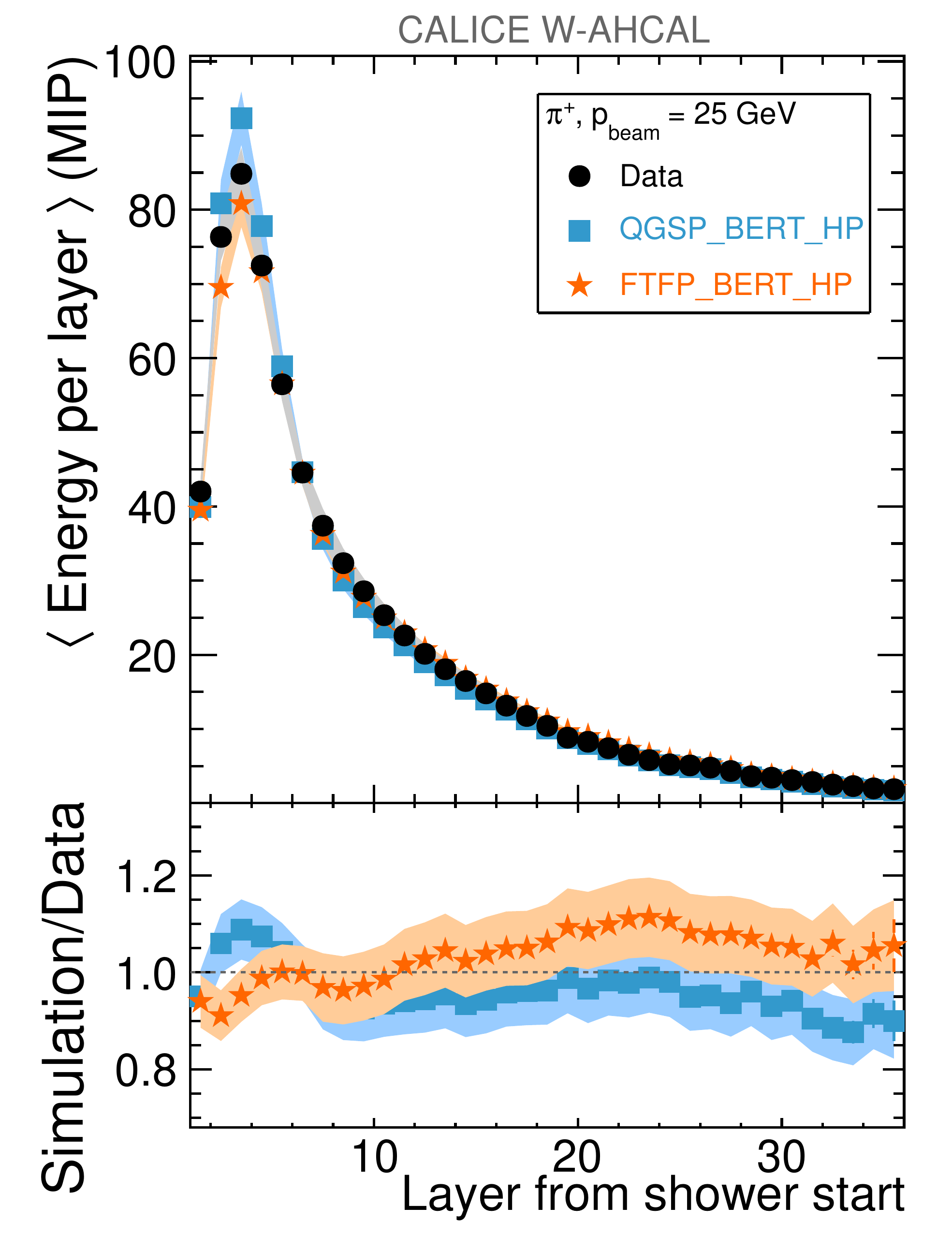}
\hfill
\includegraphics[width=0.46\textwidth]{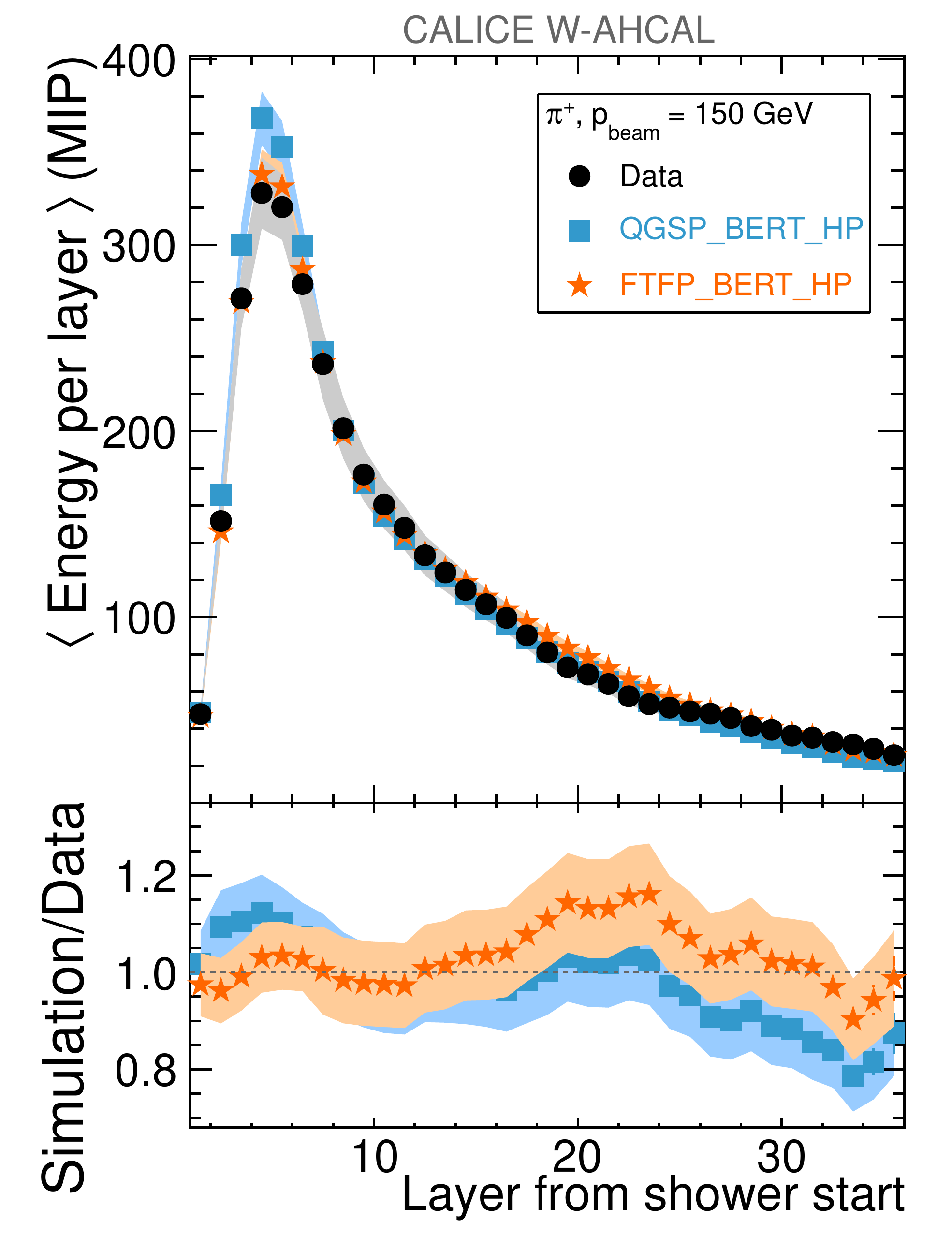}
\caption{Longitudinal profiles from shower start for 25\,GeV (left) and 150\,GeV (right) $\PGpp$.
}
\label{fig:pi_long_profile_mc}
\end{figure}

\begin{figure}[t!]
\centering
\includegraphics[width=0.45\textwidth]{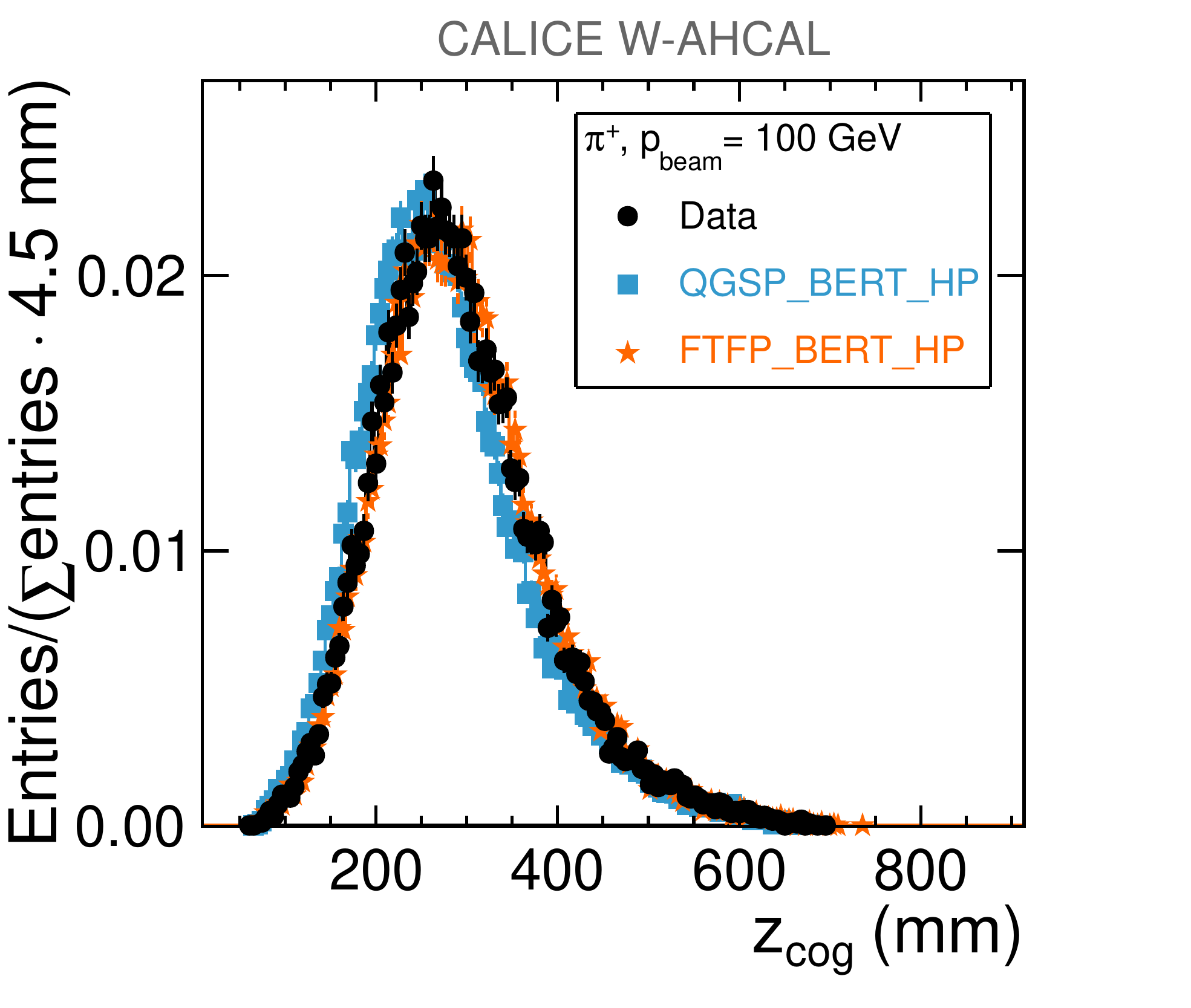}
\hfill
\includegraphics[width=0.45\textwidth]{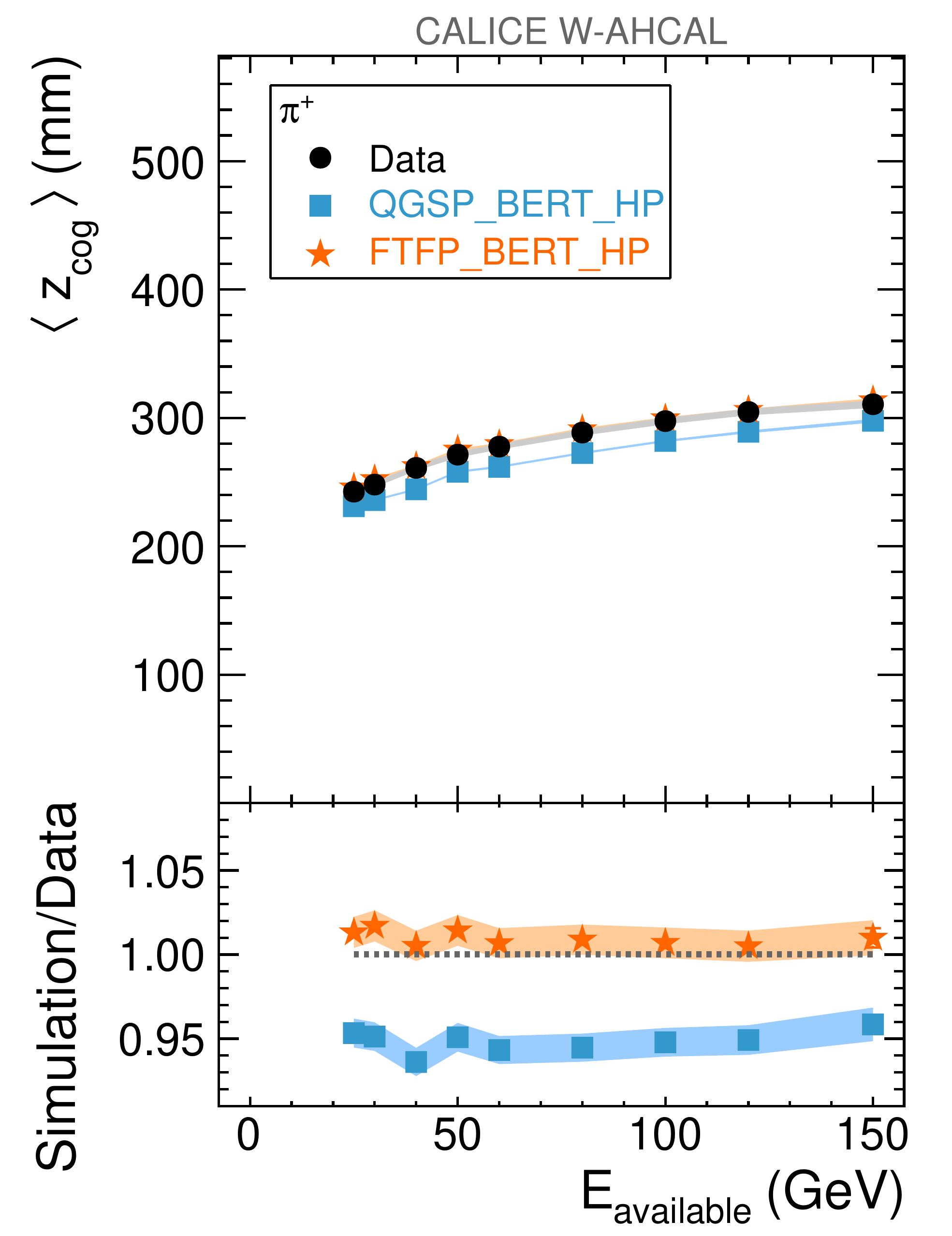}
\caption{Centre-of-gravity distribution in the $z$-direction for pions of 100\,GeV (left).
Dependence of the average centre-of-gravity in the $z$-direction on the available energy (right). 
}
\label{fig:pi_zcog_mc}
\end{figure}

A comparison of the longitudinal profile in data and Monte Carlo for 25\,GeV and 150\,GeV~$\PGpp$ events is shown in figure \ref{fig:pi_long_profile_mc}. 
At both momenta, QGSP\_BERT\_HP overestimates the energy deposition in the first four layers which is compensated by an underestimate of the energy deposition in the subsequent part of the shower, while the opposite behaviour is observed for FTFP\_BERT\_HP at 25\,GeV.  
FTFP\_BERT\_HP gives a good overall description of the longitudinal energy profile at 150\,GeV.
\\
A second characteristic parameter of the shower development along the $z$-axis is the energy weighted centre-of-gravity in the $z$-direction, $z_{\text{cog}}$, as defined in equation \ref{eq:zcog}.
The $z_{\text{cog}}$ distribution is shown in figure~\ref{fig:pi_zcog_mc} (left) for 100\,GeV~$\PGpp$ events.
In the right panel of figure \ref{fig:pi_zcog_mc}, the average centre-of-gravity in the \mbox{$z$-direction} $\langle z_{\text{cog}} \rangle$, estimated as the mean of the $z_{\text{cog}}$ distribution, is shown as a function of the available energy and compared to the Geant4 simulations.
FTFP\_BERT\_HP agrees with the experimental data within 2\% while QGSP\_BERT\_HP shows a 5\% earlier centre-of-gravity in the $z$-direction than observed in data.

\paragraph{Radial shower development.}
The radial energy profile is given by the distribution of the energy density as a function of the radial distance to the shower's $z$-axis.
Here, the radial energy density, $\langle E_{vis}^*\rangle$, is defined as the average energy sum in rings in the projection along the $z$-axis normalised to the ring's surface.
The radial distance to the shower centre is defined as
\begin{equation}
\label{eq:radial_distance}
r_i = \displaystyle\sqrt{(x_i-x_{\mathrm{cog}})^2 + (y_i-y_{\mathrm{cog}})^2},
\end{equation}
where $x_i$ ($y_i$) is the $x$ ($y$) position of the cell $i$, and $x_{\mathrm{cog}}$ ($y_{\mathrm{cog}}$) is the centre-of-gravity in $x$ ($y$)
\begin{equation}
x_{\mathrm{cog}}=\displaystyle \frac{\sum_i E_i \cdot x_i}{\sum_i
  E_i} \; \mathrm{and} \; y_{\mathrm{cog}}=\displaystyle
  \frac{\sum_i E_i \cdot y_i}{\sum_i E_i},
\end{equation}
and $E_i$ is the cell energy.

\begin{figure}[t!]
\centering
\includegraphics[width=0.46\textwidth]{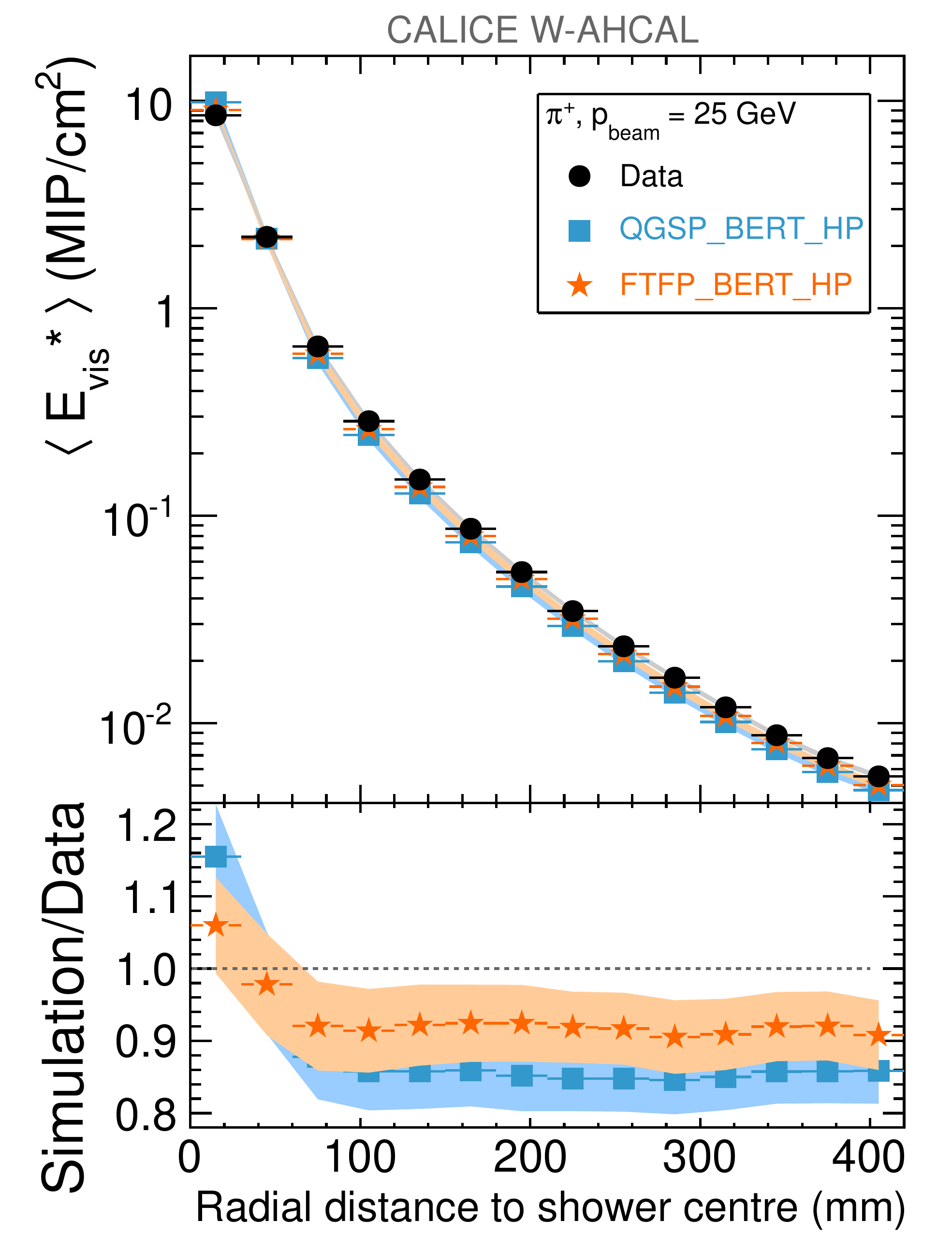}
\hfill
\includegraphics[width=0.46\textwidth]{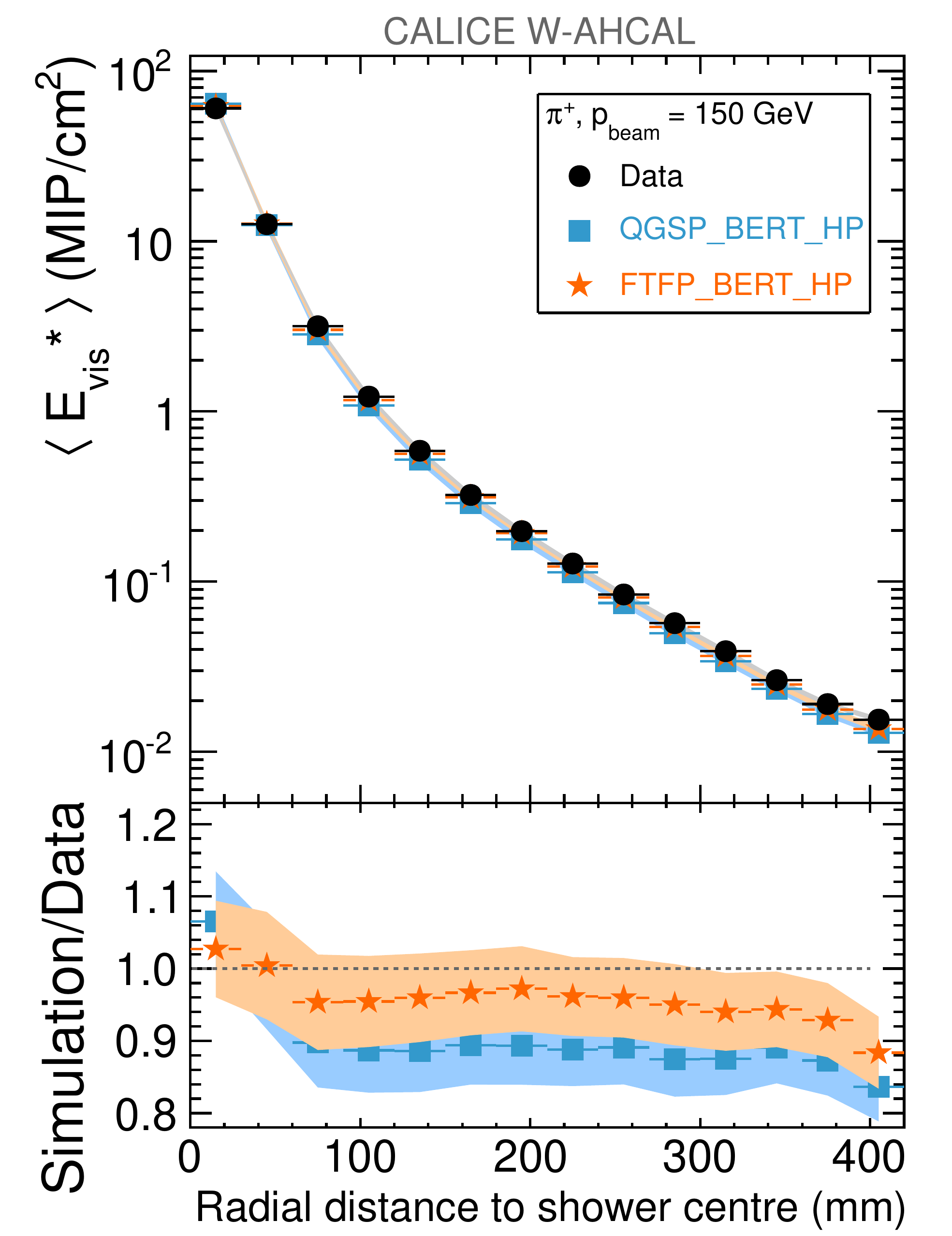}
\caption{The radial profile of 25\,GeV $\PGpp$ (left) and 150\,GeV $\PGpp$ (right). 
Here, the $\langle E_{\text{vis}}^*\rangle$ is visible energy $\langle E_{\text{vis}}\rangle$ normalised to the ring area.}
\label{fig:pi_radialProfile}
\end{figure}

Example radial profiles are displayed in figure~\ref{fig:pi_radialProfile} for $\PGpp$ at 25\,GeV and 150\,GeV.
The bin width of the histogram's horizontal axis corresponds to the dimension of the smallest \mbox{W-AHCAL} tile, i.e. $3\times3$\,cm$^2$.
At both beam momenta, the Geant4 simulations predict a higher density in the core of the shower than observed in data. 
This effect decreases with increasing beam momentum.
The better agreement with data is observed for FTFP\_BERT\_HP at 150\,GeV.

\begin{figure}[t!]
\centering
\includegraphics[width=0.46\textwidth]{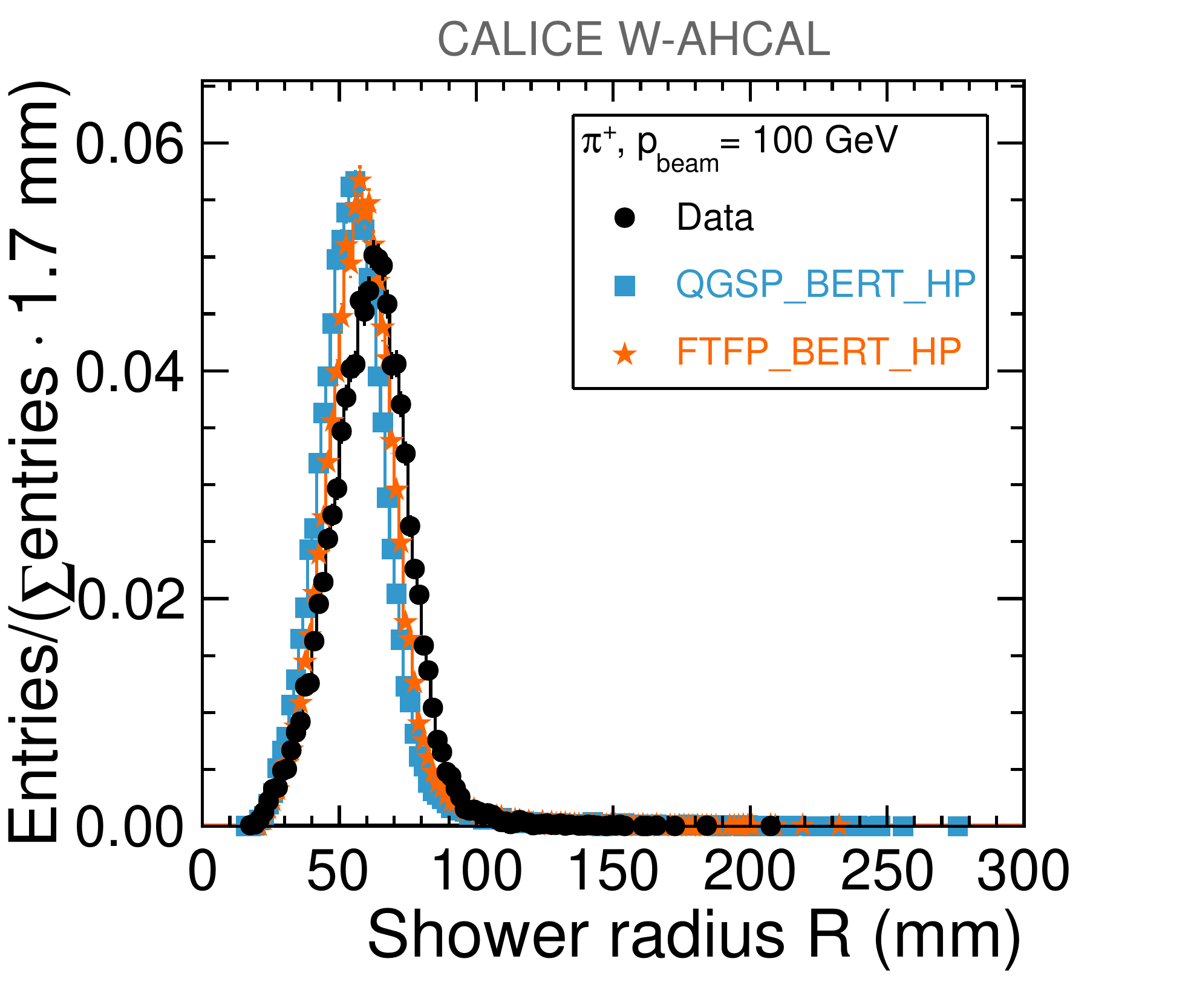}
\hfill
\includegraphics[width=0.46\textwidth]{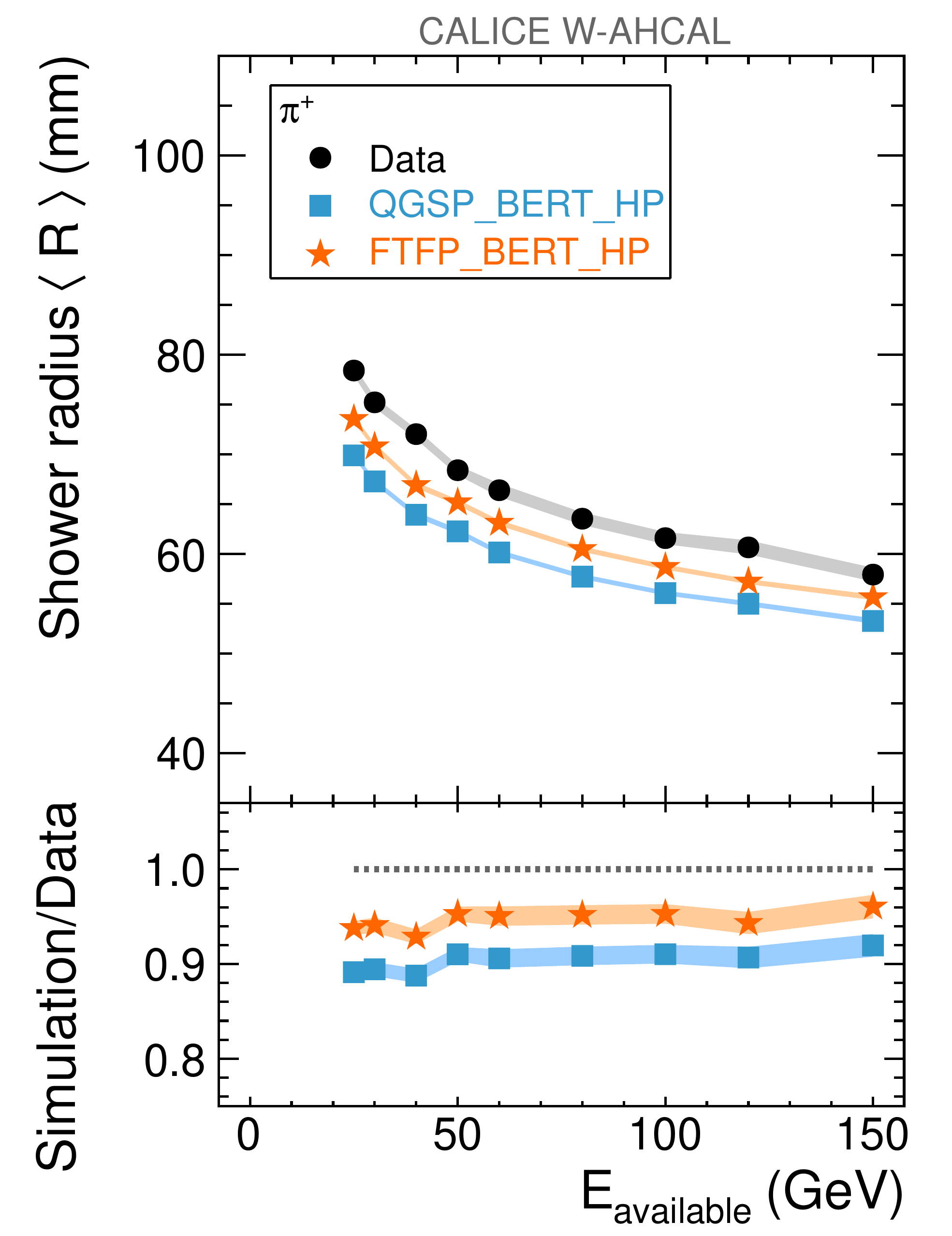}
\caption{Energy-weighted shower radius for pions of 100\,GeV (left).
Dependence of the average shower radius on the available energy (right). 
}
\label{fig:pi_showerRadius}
\end{figure}

\clearpage
To further study the shower development in the radial direction, an energy-weighted shower radius is defined
\begin{eqnarray}
\label{eq:showerRadius}
 R&=\displaystyle\frac{\sum_i E_i \cdot r_i}{\sum_i E_i},
\end{eqnarray}
where $E_i$ is the cell energy, and $r_i$ is defined in equation~\ref{eq:radial_distance}. 
A distribution of the energy-weighted shower radius is shown in the left panel of figure \ref{fig:pi_showerRadius} for 100\,GeV~$\PGpp$.
The mean shower radius $\langle R \rangle$, estimated as the mean of the $R$ distribution, is shown as a function of the available energy in the right panel of figure \ref{fig:pi_showerRadius}. 
The better agreement with data is observed for FTFP\_BERT\_HP, within 7\% or better, however both simulations significantly underestimate the shower radius at all energies.
 
\cleardoublepage
\subsection{The proton data}
The proton events are selected as described in Section~\ref{sec:HadronSelection}. 
The purity of the event samples selected based on Cherenkov threshold counters is between 85\% and 100\%~\cite{LCD-Note-2013-006}.

\begin{figure}[t!]
\begin{minipage}[c]{0.47\linewidth}
\centering
\includegraphics[width=\textwidth]{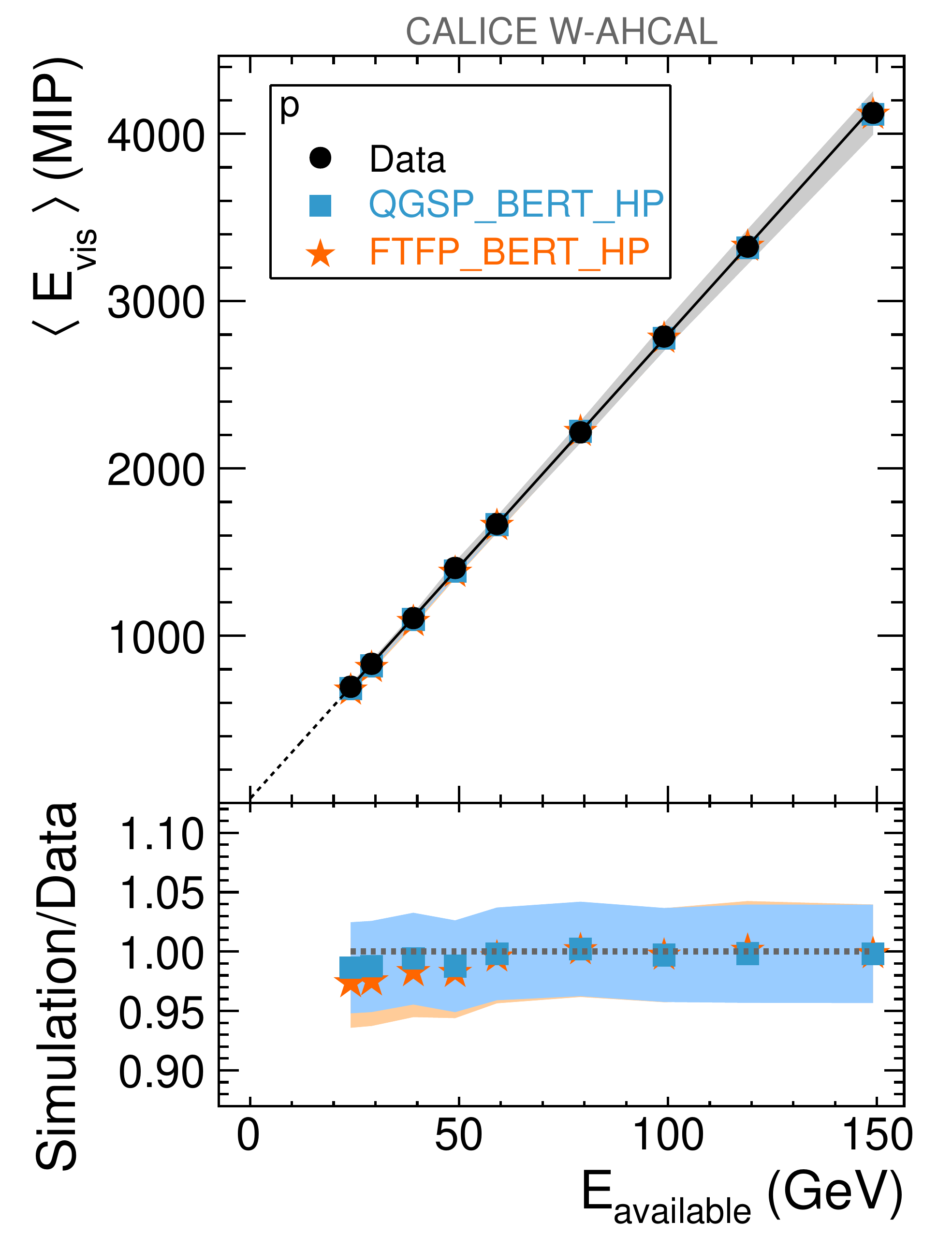}
\end{minipage}
\hfill
\begin{minipage}[c]{0.47\linewidth}
\centering
\includegraphics[width=\textwidth]{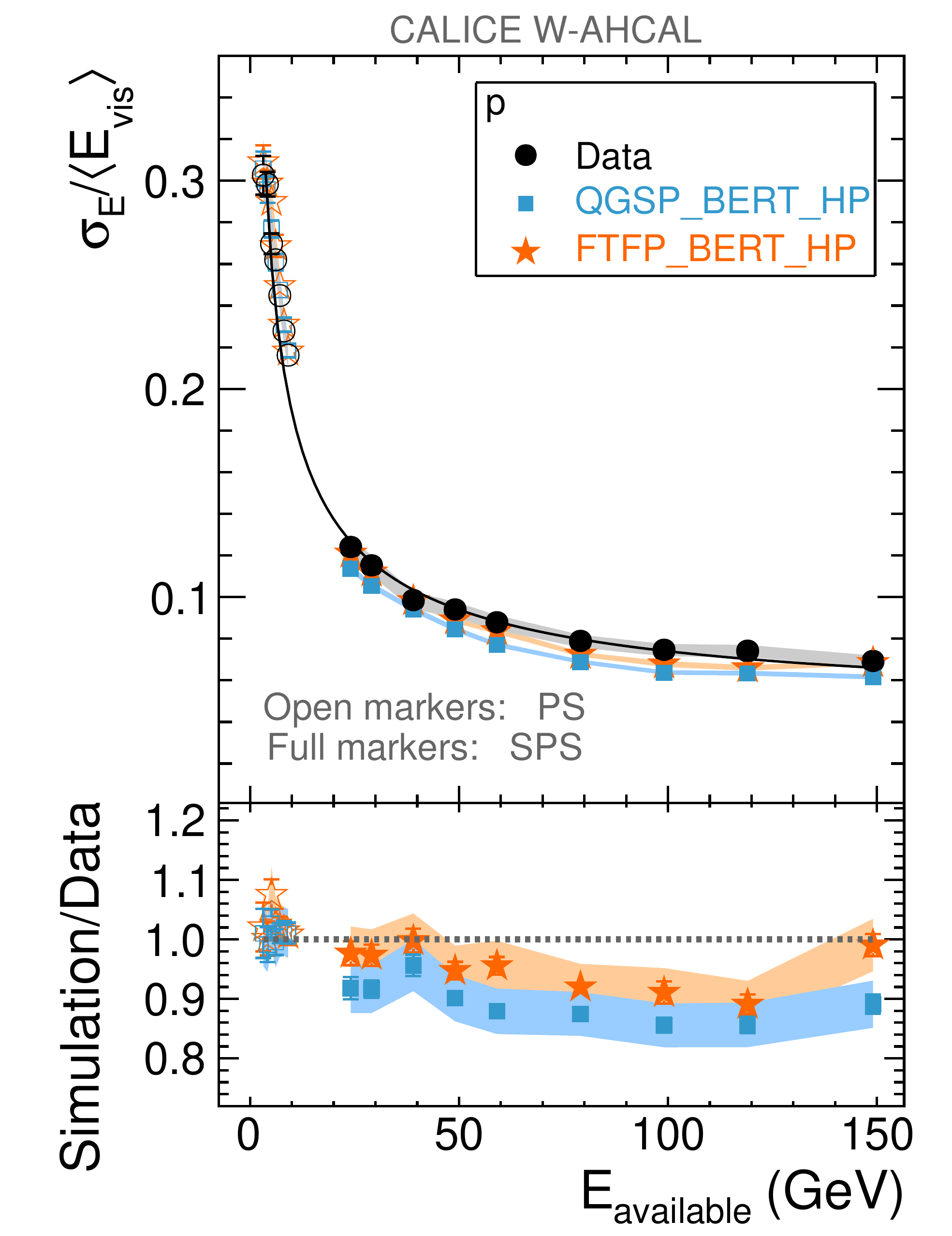}
\end{minipage}
\caption{Average visible energy (left) and energy resolution (right) for proton-induced ($\Pp$) showers as a function of the available energy. 
For the energy resolution, data points from the corresponding PS data analysis \cite{CALICE-WAHCAL-2010} are shown.
The black lines indicate fits to the data points.
The data are compared to selected Geant4 physics lists.
The bands show the overall uncertainties. 
}
\label{fig:proton_linearity_resolution}
\end{figure}

The average visible energy deposited in the calorimeter by proton-induced showers is shown as a function of available energy in the left panel of figure \ref{fig:proton_linearity_resolution} in the range from 25\,GeV to 150\,GeV. 
The data are compared to the selected Geant4 physics lists.
\\
The proton response in this energy range is linear within uncertainties.
Data and both physics lists agree for all analysed energies within uncertainties.
At beam momenta below 60\,GeV, however, both Geant4 physics lists tend to slightly underestimate the data.
\\
The energy resolution for the proton data is presented in the right panel of figure~\ref{fig:proton_linearity_resolution}. 
Data in the PS beam momentum range from 4\,GeV to 10\,GeV~\cite{CALICE-WAHCAL-2010} are included. 
As for the $\Pep$ and $\PGpp$ data, the PS data were selected based on the new selection cuts described in section~\ref{sec:HadronSelection}.
The experimental energy resolution is up to 15\% larger than predicted by the Monte Carlo simulations including the detector instability.
FTFP\_BERT\_HP is closer to data (within 10\%) than QGSP\_BERT\_HP, as already observed for the pion-induced showers.
The parameters of the proton energy resolution fits using equation~\ref{eq:resFit} are listed in table~\ref{tab:protonResolution}.
As also observed for the pion data, the Geant4 physics lists underestimate the stochastic term of the energy resolution fit.
The fit results of the proton data are in agreement with the PS results when taking into account the different methods used to estimate $\sigma_E$ and $\langle E_{\text{vis}} \rangle$, as discussed in section \ref{sec:piAnalysis}.

\begin{table}[!t]
\centering
\caption{
Parameters of the proton energy resolution fits for data and simulations using beam momenta from 4\,GeV to 150\,GeV.
The simulated results are obtained after including the detector instability measured in data.
}
\label{tab:protonResolution}
\begin{tabular}{lrrr}
\toprule
Parameter                          & Data             & QGSP\_BERT\_HP      & FTFP\_BERT\_HP \\
\midrule
$a$ ($\%\cdot\sqrt{\text{GeV}}$)   & $60.7\pm 1.2$    & $56.6\pm 1.7$       & $58.7\pm 1.7$  \\
$b$ (\%)                           & $4.3\pm 0.4$     & $3.1\pm 0.5$        & $3.7\pm 0.5$   \\
$c$ (GeV)                          & $0.065$          & $0.065$             & $0.065$        \\
\bottomrule
\end{tabular}
\end{table}

\subsection{The kaon data}

\begin{figure}[t!]
\centering
\includegraphics[width=0.6\textwidth]{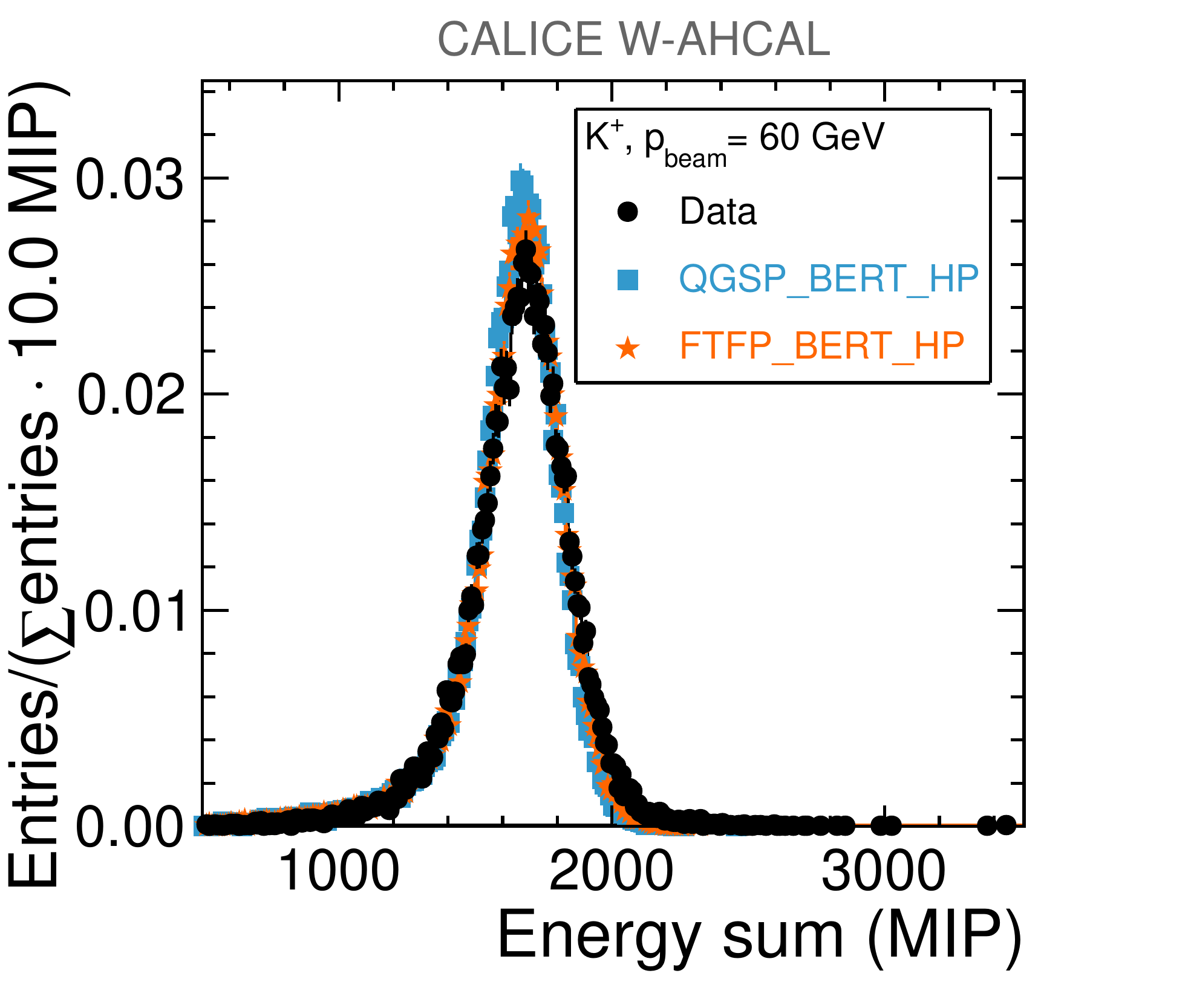}
\caption{Energy sum distributions for 60\,GeV $\PKp$. 
The data are compared to selected Geant4 physics lists. 
}
\label{fig:kaons_esum}
\end{figure}

In the hadron beam, a small fraction of events are expected to have kaon-induced showers. 
Data corresponding to beam momenta of 50\,GeV and 60\,GeV were selected since the purity of the event samples selected based on Cherenkov threshold counters is 89\% at 50\,GeV and 99\% at 60\,GeV~\cite{LCD-Note-2013-006}.
The energy sum distribution for 60\,GeV  $\PKp$ is compared to the simulation in figure~\ref{fig:kaons_esum}. 
A good agreement with both Geant4 physics lists is observed.

\section{Comparison of the response for different particle types}
\label{sec:ComparisonOfResponse}

\begin{figure}[t!]
\centering
\begin{minipage}[t]{0.47\linewidth}
\centering
\includegraphics[width=\textwidth]{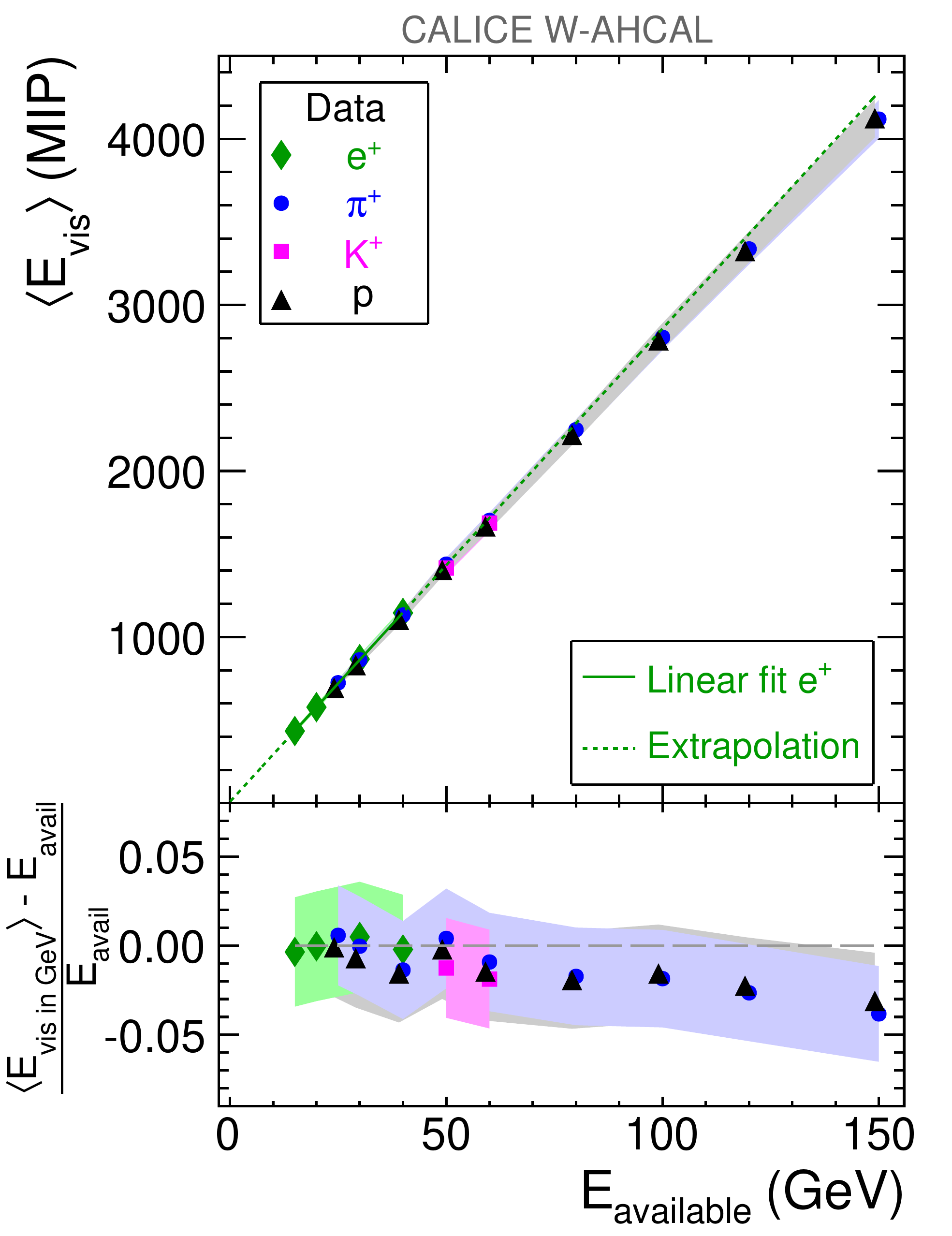}
\end{minipage}\hfill
\begin{minipage}[t]{0.47\linewidth}
\centering
\includegraphics[width=\textwidth]{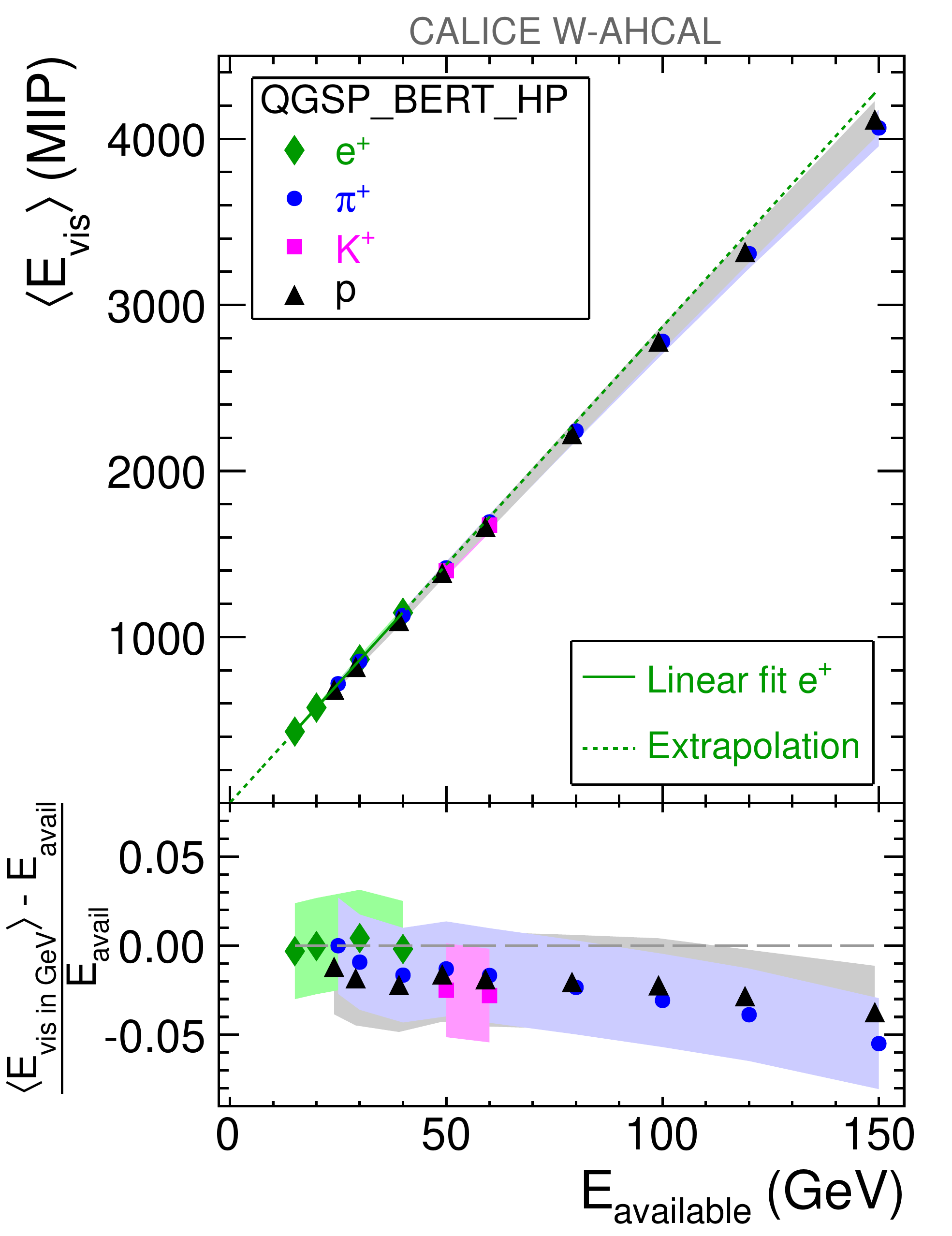}
\end{minipage}
\caption{The response of the CALICE \mbox{W-AHCAL} to different particles as a function of the available energy for data (left) and for
QGSP\_BERT\_HP (right). 
The solid green line indicates a linear fit to the $\Pep$ data. 
The dotted green line indicates the extrapolation of the line to 0\,GeV and 150\,GeV. 
The lower part of the figures shows the comparison of the average visible and the available energy given by $(\langle E_{\text{vis in GeV}}\rangle-E_{\text{avail}})/E_{\text{avail}}$, where $\langle E_{\text{vis in GeV}}\rangle$ corresponds to $\langle E_{\text{vis}}\rangle $ converted from MIP to GeV based on the $\Pep$ fit parameters.
The bands show the overall uncertainties.
}
\label{fig:wahcalResponse_allParticles_Sept2011}
\end{figure}

The average visible energy in $\Pep$, $\PGpp$, $\PKp$, and proton showers are compared in figure~\ref{fig:wahcalResponse_allParticles_Sept2011} for data (left) and Geant4 simulations (right). 
The upper part of the figures shows the average visible energy as a function of the available energy.
The lower part of the figures shows a comparison of the average visible energy to the available energy given by $(\langle E_{\text{vis in GeV}}\rangle-E_{\text{avail}})/E_{\text{avail}}$. 
For this, the average visible energy is converted from MIP to GeV using the $\Pep$ fit results from 15\,GeV to 40\,GeV.
The fit is extrapolated to the full energy range of the data sets.
The bands show the overall uncertainties.
\\
For data, the individual $\Pep$ points agree with the available energy within $\pm0.5\%$.
Up to approximately 60\,GeV, the hadron results agree with the available energy within 2\% which is well within the overall uncertainties of approximately 3.0\%.
For larger beam energies, the reconstructed energies start to deviate from the available energy with values outside the systematic uncertainties of up to 3\% and 4\% at 150\,GeV for protons and pions.
Both longitudinal leakage and saturation effects contribute to the observed deviation from the linear behaviour at high beam energies.
The calorimeter response to $\PGpp$ for beam energies below 80\,GeV is slightly higher than the response to protons, however, the data points agree within the uncertainties of the measurement.
The responses of $\PGpp$ and protons approach each other again with increasing available energy.
In conclusion, given the overall uncertainties in the measurements, the detector response is almost identical for all particle types up to 60\,GeV such that the W-AHCAL is found to be approximately compensating in this energy range.
\\
The observed behaviour is reproduced by the Geant4 simulations as shown in the right panel of  figure~\ref{fig:wahcalResponse_allParticles_Sept2011}.
A similar behaviour was also found for W-AHCAL test beam data and simulations of $\Pep$, $\PGpp$, and proton beams in the energy range from 1\,GeV to 10\,GeV~\cite{CALICE-WAHCAL-2010}.
\section{Summary and conclusions}
\label{sec:summary}
We have presented analysis results of $\Pep$, $\PGpp$, $\PKp$, and proton data with beam momenta from 15\,GeV to 150\,GeV recorded with the CALICE W-AHCAL prototype in 2011 at the CERN SPS.
The measured calorimeter response, energy resolution, and the longitudinal and radial shower \mbox{development} are compared to predictions of the Geant4 physics lists QGSP\_BERT\_HP and \\FTFP\_BERT\_HP as implemented in Geant4 version 9.6.p02. 
The simulations show a reasonable overall agreement with experimental data.
\\
As electromagnetic showers are well understood, a comparison of positron-induced showers between experimental data and simulations is used to validate the detector calibration and the implementation of the detector simulation.
Data and simulation agree with each other within the uncertainties for the detector response and the energy resolution.
For both $\Pep$ data and simulations, the detector response increases linearly with the beam momentum.
The W-AHCAL energy resolution for $\Pep$ data and simulations results in a stochastic term of approximately 29\% and a constant term of approximately 1\%.
\\
The longitudinal energy profiles for $\Pep$ data and simulations agree with each other at the level of 15\%. 
This difference could be due to an uncertainty on the single cell calibration factors, imperfect description of the material in the beam line in simulations, and to uncertainties in the Geant4 modelling.
\\
When comparing the positive hadron data to predictions of the Geant4 physics lists, similar conclusions can be drawn for $\PGpp$, $\PKp$, and protons.
In the case of the calorimeter response to hadrons, a good agreement within $\pm2\%$ between data and simulation is obtained for QGSP\_BERT\_HP and FTFP\_BERT\_HP. 
The W-AHCAL response is the same, within systematic uncertainties, for \Pep, \PGpp, \PKp, and protons for beam momenta up to 60\,GeV above which leakage effects in the hadron showers start to play an increasingly important role.
The level of agreement between the response for the particle types studied is in line with the Geant4 predictions.
\\
For the hadronic energy resolution of the W-AHCAL, a stochastic term of approximately 58\% (61\%) is found for $\PGpp$ (proton) data and
a constant term of approximately 5\% is found for both $\PGpp$ and proton data.
For both $\PGpp$ and proton events, the energy resolution has up to 5\% lower values in FTFP\_BERT\_HP than observed in data, and up to 10\% in QGSP\_BERT\_HP. 
\\
The Geant4 physics lists reproduce the longitudinal shower development of hadron showers within 15\%.
For the radial shower development, both physics lists predict a higher density in the core of the shower than observed in data, with a slightly better agreement, within 10\%, observed for FTFP\_BERT\_HP.\\
\\
In conclusion, we have shown that the Geant4 physics lists QGSP\_BERT\_HP and FTFP\_BERT\_HP reproduce average hadron shower properties recorded with the highly granular CALICE W-AHCAL at the percent level, and spatial shower profiles at the 15\% level or better.
The \mbox{FTFP\_BERT\_HP} physics list gives a slightly better agreement with data for most observables and hadron types than\\ \mbox{QGSP\_BERT\_HP}.
The W-AHCAL response increases linearly as a function of the beam momentum, however at large beam momenta an onset of leakage and saturation effects can be observed, in both data and simulations.
The energy resolution obtained for the \mbox{W-AHCAL} meets the expectations based on experience from the Fe-AHCAL.
The $\Pep$ energy resolution of the W-AHCAL has a slightly larger stochastic term than the \mbox{Fe-AHCAL} energy resolution due to the larger number of $X_0$ per layer in the W-AHCAL.
The hadron energy resolutions for the steel and tungsten absorbers agree with each other since the sampling in terms of nuclear interaction lengths is about the same for steel and tungsten.
Based on the comparison between $\Pep$ and positive hadron data, the W-AHCAL is found to be compensating up to 60\,GeV.

\section*{Acknowledgements}
We gratefully acknowledge the DESY and CERN managements for their support and hospitality, and their accelerator staff for the reliable and efficient beam operation. 
The authors would like to thank the RIMST (Zelenograd) group for their help and sensors manufacturing.
This work was supported 
by the European Commission under the FP7 Research Infrastructures project AIDA, grant agreement no.\ 262025;
by the Bundesministerium f\"{u}r Bildung und Forschung, Germany;
by the DFG cluster of excellence `Origin and Structure of the Universe' of Germany; 
by the Helmholtz-Nachwuchsgruppen grant VH-NG-206;
by the BMBF, grant no.\ 05HS6VH1;
by the Alexander von Humboldt Foundation (including Research Award IV, RUS1066839 GSA);
by the Russian Ministry of Education and Science contracts 4465.2014.2 and 14.A12.31.0006 and the Russian
Foundation for Basic Research grant 14-02-00873A;
by MICINN and CPAN, Spain;
by CRI(MST) of MOST/KOSEF in Korea;
by the US Department of Energy and the US National Science Foundation;
by the Ministry of Education, Youth and Sports of the Czech Republic under the projects AV0 Z3407391, AV0 Z10100502, LG14033  and LA09042;  
and by the Science and Technology Facilities Council, UK.

\bibliographystyle{calice}
\bibliography{bib}

\end{document}